\documentclass[fleqn,usenatbib]{mnras}

\usepackage{mathptmx}

\usepackage[T1]{fontenc}
\usepackage{ae,aecompl}

\usepackage{amsmath}	
\usepackage{amssymb}	
\usepackage{enumitem}   
\usepackage{url}
\usepackage{hyperref}
\usepackage{breakurl}
\usepackage{graphicx}	

\newcommand{\gt}{>}

\title[Observations of a nearby filament]{Observations of a nearby filament of galaxy clusters with the Sardinia Radio Telescope}

\author[V. Vacca et al.]{
Valentina Vacca,$^{1}$\thanks{E-mail: vvacca@oa-cagliari.inaf.it}
M. Murgia,$^{1}$
F. Govoni$^{1}$
F. Loi$^{1,2}$
F. Vazza$^{3,4,5}$ 
A. Finoguenov$^{6}$
\newauthor
E. Carretti$^{1}$
L. Feretti$^{4}$
G. Giovannini$^{3,4}$
R. Concu$^{1}$
A. Melis$^{1}$
C. Gheller$^{7}$
\newauthor
R. Paladino$^{4}$
S. Poppi$^{1}$
G. Valente$^{1,8}$
G. Bernardi$^{3,9,10}$
W. Boschin$^{11,12,13}$
M. Brienza$^{14}$
\newauthor
T.E. Clarke$^{15}$
S. Colafrancesco$^{16}$
T. En{\ss}lin$^{17}$ 
C. Ferrari$^{18}$
F. de Gasperin$^{19}$
\newauthor
F. Gastaldello$^{20,21}$ 
M. Girardi$^{22,23}$ 
L. Gregorini$^{4}$ 
M. Johnston-Hollitt$^{24}$
H. Junklewitz$^{25}$
\newauthor
E. Orr{\`u}$^{14}$
P. Parma$^{4}$
R. Perley$^{26}$
G.B Taylor$^{27}$     
\\
$^{1}$INAF-Osservatorio Astronomico di Cagliari, Via della Scienza 5, I-09047 Selargius (CA), Italy \\
$^{2}$Dipartimento di Fisica, University of Cagliari, Strada Prov.le Monserrato-Sestu Km 0.700,I-09042 Monserrato (CA), Italy\\       
$^{3}$Dip. di Fisica e Astronomia, Universit\`a degli Studi di Bologna, Viale Berti Pichat 6/2, I–40127 Bologna, Italy\\         
$^{4}$INAF - Istituto di Radioastronomia, Via Gobetti 101, I--40129 Bologna, Italy \\
$^{5}$Hamburger Sternwarte, Universit\"at Hamburg, Gojenbergsweg 112, 21029, Hamburg, Germany\\ 
$^{6}$Department of Physics, University of Helsinki, Gustaf H{\"a}llstr{\"op}min katu 2a, 00014 Helsinki, Finland\\
$^{7}$CSCS-ETHZ, Via Trevano 131, Lugano, Switzerland\\
$^{8}$Agenzia Spaziale Italiana (ASI), Roma\\ 
$^{9}$SKA SA, 3rd Floor, The Park, Park Road, Pinelands 7405, The Cape Town, South Africa \\
$^{10}$Department of Physics and Electronics, Rhodes University, PO Box 94, Grahamstown 6140, South Africa \\
$^{11}$Fundaci\'on G. Galilei - INAF TNG, Rambla J. A. Fern\'andez P\'erez 7, E-38712 Bre\~na Baja (La Palma), Spain\\
$^{12}$Instituto de Astrof\'isica de Canarias, C/V\'ia L\'actea s/n, E-38205 La Laguna (Tenerife), Spain\\
$^{13}$Dep. de Astrof\'isica, Univ. de La Laguna, Av. del Astrof\'isico Francisco S\'anchez s/n, E-38205 La Laguna (Tenerife), Spain\\
$^{14}$ASTRON, the Netherlands Institute for Radio Astronomy, Postbus 2, 7990 AA, Dwingeloo, The Netherlands\\ 
$^{15}$Naval Research Laboratory, Washington, District of Columbia 20375, USA\\
$^{16}$School of Physics, University of the Witwatersrand, Private Bag 3, 2050, Johannesburg, South Africa\\
$^{17}$Max Planck Institut f\"{u}r Astrophysik, Karl-Schwarzschild-Str.1, 85740 Garching, Germany\\
$^{18}$Laboratoire Lagrange, UCA, OCA, CNRS, Blvd de l'Observatoire, CS 34229, 06304 Nice cedex 4, France\\ 
$^{19}$University of Leiden, Rapenburg 70, 2311 EZ Leiden, the Netherlands \\
$^{20}$INAF - IASF Milano, Via Bassini 15, I-20133 Milano, Italy \\ 
$^{21}$Dep. of Physics and Astronomy, University of California at Irvine, 4129 Frederick Reines Hall, Irvine, CA 92697-4575, USA \\
$^{22}$Dip. di Fisica dell'Universit\`a degli Studi di Trieste - Sezione di Astronomia, via Tiepolo 11, I-34143 Trieste, Italy  \\
$^{23}$INAF - Osservatorio Astronomico di Trieste, via Tiepolo 11, I-34143 Trieste, Italy \\
$^{24}$International Centre for Radio Astronomy Research, Curtin University,
Bentley, WA 6102, Australia\\
$^{25}$Argelander-Institut f\"{u}r Astronomie, Auf dem H\"{u}gel 71 D-53121 Bonn, Germany \\
$^{26}$National Radio Astronomy Observatory, P.O. Box O, Socorro, NM 87801, USA \\
$^{27}$Department of Physics and Astronomy, University of New Mexico, Albuquerque NM, 87131, USA 
}

\date{Accepted 2018 April 22. Received 2018 April 13; in original form 2018 January 04}

\pubyear{2015}

\begin{document}
\label{firstpage}
\pagerange{\pageref{firstpage}--\pageref{lastpage}}
\maketitle

\begin{abstract}
We report the detection of diffuse radio emission
which might be connected to a large-scale filament of the cosmic web covering a 8$^{\circ}\times$8$^{\circ}$ area in the sky, likely associated with a z$\approx$0.1 over-density traced by nine massive galaxy clusters. In this work, we present radio observations of this region taken with the Sardinia Radio Telescope. Two of the clusters in the field host a powerful radio halo  sustained by violent ongoing mergers and provide direct proof of intra-cluster magnetic fields. In order to investigate the presence of large-scale diffuse radio synchrotron emission in and beyond the galaxy clusters in this complex system, we combined the data taken at 1.4\,GHz obtained with the Sardinia Radio Telescope with higher resolution data taken with the NRAO VLA Sky Survey. We found 28 candidate new sources with a size larger and  X-ray emission fainter than known diffuse large-scale synchrotron cluster sources for a given radio power. This new population is potentially the tip of the iceberg of a class of diffuse large-scale synchrotron sources associated with the filaments of the cosmic web. In addition, we found in the field a candidate new giant radio galaxy.
\end{abstract}




\section{Introduction}

One of the major challenges of the new generation of astronomical instruments
is the detection of the magnetic cosmic web. According to cosmological
simulations, half of the baryons expected in the Universe from the cosmic
microwave background observations should be located in the filaments
connecting galaxy clusters, in the form of a diffuse plasma \citep[with median
  over-densities of about 10-30 and temperatures $10^5-10^7$\,K,
  see][]{Cen1999,Dave2001}.  Owing to the low density and intermediate
temperature range, the thermal component of the cosmic web is hardly
detectable in mm/sub-mm through the Sunyaev-Zel'dovich (SZ) effect and in
X-rays via bremsstrahlung emission.  Thermal emission from plasma associated
with filaments of the cosmic web has not yet been firmly detected.  By using
data from the Planck satellite of the region between the galaxy cluster couple
A399-A401, the \cite{Planck2013} provided evidence of a SZ signal from the
medium between the two clusters within their virial radii.  Later, X-ray
observations by \cite{Eckert2015} indicated structures coherent over scales of
8\,Mpc, associated with the galaxy cluster A2744 and with the same spatial
location of galaxy over-densities and dark matter.  Recently, a signature of
inter-galactic filaments between close galaxy couples have been statistically
reported using Planck data \citep{DeGraaff2017}, while a statistical study based
on a sample of 23 massive galaxy clusters conducted by \cite{Heines2017}
showed that half of the expected cluster mass growth rate is due to X-ray
groups in the infall region when clusters between the present epoch and
z$\sim$0.2 are considered.

A non-thermal component is expected to be present as well and to emit in the
radio band via the synchrotron mechanism. This component is even more
difficult to observe due to the expected very low density of cosmic rays and
the weakness of the magnetic fields involved \citep[e.g.][]{Vazza2015}.
Magnetic fields in the large-scale structure of the Universe have been
discovered only in the second half of the last century through the observation
of large-scale diffuse synchrotron sources both at the center and in the
periphery of galaxy clusters, called respectively radio halos and relics
\citep[see, e.g.,][]{Feretti2012}.  The properties of these sources in total
intensity and polarization indicate magnetic fields, tangled on scales from a
few kpc up to a few hundred kpc, characterized by central $\mu$G strengths
that decline towards the periphery as a function of the thermal gas density of
the intra-cluster medium
\citep[e.g.,][]{Murgia2004,Govoni2006,Vacca2010,Govoni2017}.

Numerical simulations suggest that cosmological filaments host weak magnetic
fields \citep[with strengths $\gtrsim$\,nG,][]{Brueggen2005,Ryu2008,Vazza2014},
whose properties may reflect the primordial magnetic field strength and
structure. Beyond a few Mpc from the cluster center, only hints of the
presence of non-thermal emission are available.  Bridges of diffuse radio
emission have been found to connect the galaxy cluster center and periphery,
as in the case of the Coma cluster \citep[see][and references
  therein]{Kronberg2007}, A2744 \citep{Orru2007}, and A3667
\citep{Carretti2013}.  Extended emission regions at a projected distance of
2\,Mpc from the cluster center and probably connected with the large-scale
structure formation processes, have been detected in the galaxy cluster A2255
\citep{Pizzo2008}. At larger distances from the cluster center, diffuse radio
emission associated with the optical filament of galaxies ZwCl\,2341.1+0000
has been observed by \cite{Bagchi2002} and \cite{Giovannini2010},
corresponding to a linear size larger than a few Mpc.  Later, by using optical
data, \cite{Boschin2013} found that this cluster belongs to a complex system
undergoing a merger event, posing new questions on the nature of the observed
diffuse emission that could be alternatively due to a two relics plus halo
system.  Diffuse emission along an optical filament has been recently
detected also in the A3411 - A3412 system \citep{Giovannini2013} that could be
powered by accretion shocks as material falls along the filament. According to
an alternative explanation, the observed emission is evidence of
re-acceleration of fossil electrons \citep{vanWeeren2013,vanWeeren2016}.
However, no clear and firm association of the filaments of the cosmic web with
a radio signal has been found to date. The major limit to observation of the
magnetization of large-scale structure is the scanty sampling of short
\emph{uv}-spacing in present interferometers: the Square Kilometre Array
(SKA), and its precursors and path-finders can overcome this limit and have
the capabilities to detect radio emission from the cosmic web at frequencies
below 1.4\,GHz, if the magnetization of the medium is at least about 1\% of
the energy in the thermal gas \citep{Vazza2015}.

Waiting for the SKA, a valuable contribution to the study of the magnetization of the
cosmic web through its synchrotron radio emission can be given by the analysis
of radio emission with simultaneously interferometric and single-dish
telescopes.  By combining single-dish data collected with the 305\,m Arecibo
telescope with interferometric data collected with the Dominion Radio
Astrophysical Observatory (DRAO) at 0.4\,GHz, \cite{Kronberg2007} imaged a
region of the sky of 8$^{\circ}$ around the Coma cluster.  They detected
large-scale synchrotron emission in the direction of the cluster,
corresponding to a linear size of about 4\,Mpc if located at the same redshift
as Coma.  A possible inter-cluster source associated with the filament between
A2061 and A2067 has been claimed by \cite{Farnsworth2013}, after subtracting
point sources from 100\,m Green Bank Telescope (GBT) observations at 1.4 GHz
by using interferometric data.  Extragalactic sources are typically observed
with interferometers to reach a suitable spatial resolution.  Nevertheless,
full-synthesis interferometric observations do not recover information on
angular scales larger than those corresponding to their minimum baseline and
do not have sensitivity on such large scales.  On the contrary, single-dish
telescopes reveal emission over scales as large as the size of the scanned
region. These characteristics make single-dishes and interferometers
complementary instrument for the study of large-scale synchrotron sources. The
ability to reconstruct emission from large-scale structures of single-dishes
combined with the high spatial resolution of interferometers, allows us to
recover the emission of sources covering angular size larger than a few tens
of arc-minutes and to separate this emission from that of possible embedded
compact sources.

Great effort in the last years have been devoted to constrain the magnetic
field strength beyond the cluster volume both on the basis of radio
observations alone and with multi-wavelength data.  \cite{Brown2017},
constrained the magnetic field in the cosmic web to be in the range
0.03--0.13\,$\mu$G (3$\sigma$) and the primordial magnetic field less than
1\,nG by cross-correlating the radio emission observed with the \emph{S-band
  Polarization All Sky Survey} radio survey at 2.3 \,GHz with a simulated
model of the local synchrotron cosmic web. By cross-correlating radio
observations at 180\,MHz obtained with the Murchison Widefield Array and
galaxy number density estimates from the \emph{Two Micron All-Sky Survey} and
the \emph{Wide-field InfraRed Explorer redshift catalogues},
\cite{Vernstrom2017} put limits on the magnetic field strength of the cosmic
web of 0.03--1.98\,$\mu$G (1$\sigma$).

In this paper, we present a study of the radio emission of a new filament of
the cosmic web by using new single-dish observations and interferometric data.
The paper is organized as follows. In \S\,\ref{filament} a description of the
system is given. In \S\,\ref{observations} we summarize the radio
observations.  In \S\,\ref{combination}, we present the combination of single
data with interferometric snapshot observations. In
\S\,\ref{x-ray} and in \S\,\ref{sz} we report the X-ray and sub-millimetre
properties of the sources in this region. In \S\,\ref{Results} and
\S\,\ref{Discussion}, we present and discuss our results and in \S\,\ref{conclusions} we present
our conclusions. In Appendix\,\ref{pscat} and in Appendix\,\ref{RRG} we check
radio properties of a sample of compact sources and of interesting radio
galaxies in the field.

In the following, we consider a flat Universe and adopt a $\Lambda$CDM
cosmology with $H_0=$67.3\,km\,s$^{-1}$\,Mpc$^{-1}$, $\Omega_{\rm m}=$0.315, and
$\Omega_{\Lambda}=$0.685 \citep{Planck2014}.

\section{The filaments of the cosmic web}
\label{filament}

\begin{table*}
        \caption{Clusters in the field with known redshift.}
        \begin{tabular}{ccccccc}
          \hline
          \hline
  Cluster          &   RA (J2000) & Dec (J2000) & z         & $^{\prime\prime}$/kpc & $D_{\rm c}$  & Reference             \\
                   &   h:m:s      & $^{\circ}$:$^{\prime}$:$^{\prime\prime}$       &           &                    &  Mpc       &                        \\
       \hline

A523               & 04:59:06.2   & +08:46:49   & 0.104           &  1.98              &  452       & \cite{Girardi2016} \\
A525               & 04:59:30.9   & +08:08:27   & 0.1127           &  2.13              &  608       & \cite{Chow2014}\\        
MACS\,J0455.2+0657 & 04:55:17.4   & +06:57:47   & 0.425             &  5.77              &  1696      &   \cite{Cavagnolo2008}\\       
ZwCl\,0510.0+0458A & 05:12:39.4   & +05:01:31   & 0.027            &  0.56&120&      \cite{Baiesi1984}\\
ZwCl\,0510.0+0458B & 05:12:39.4   & +05:01:31   & 0.015             & 0.32& 67& \cite{Baiesi1984}\\
RXC\,J0503.1+0607  & 05:03:07.0   & +06:07:56   & 0.088            &  1.71              &  384       & \cite{Crawford1995} \\
A529               & 05:00:40.7   & +06:10:22   & 0.1066           &  2.03              &  463       &  \cite{Chow2014}        \\
A515               & 04:49:34.5   & +06:10:09   & 0.1061           &  2.02              &  461       & \cite{Chow2014}        \\ 
A526               & 04:59:51.8   & +05:26:26   & 0.1068           &  2.03              &  464       &  \cite{Chow2014}        \\
RXC\,J0506.9+0257  & 05:06:54.1   & +02:57:08   & 0.1475           &  2.68              &  634       & \cite{Bohringer2000}\\
A520               & 04:54:19.0   & +02:56:49   & 0.199            &  3.41              &  844       & \cite{Struble1999} \\
A509               & 04:47:42.2   & +02:17:16   & 0.0836           &  1.63              &  365       & \cite{Struble1999} \\
A508               & 04:45:53.9   & +02:00:24   & 0.1481          &  2.69              &  635       &  \cite{Chow2014} \\
            \hline
            \hline

        \end{tabular}
\label{tab:A}\\
       Col 1: Cluster name; Col 2 \& 3: cluster coordinates; Col 4: redshift;
       Col 5: Linear scale conversion factor; Col 6: co-moving radial distance; Col 7: reference for the redshift. 
    \end{table*}

\begin{figure}
\includegraphics[width=0.5\textwidth]{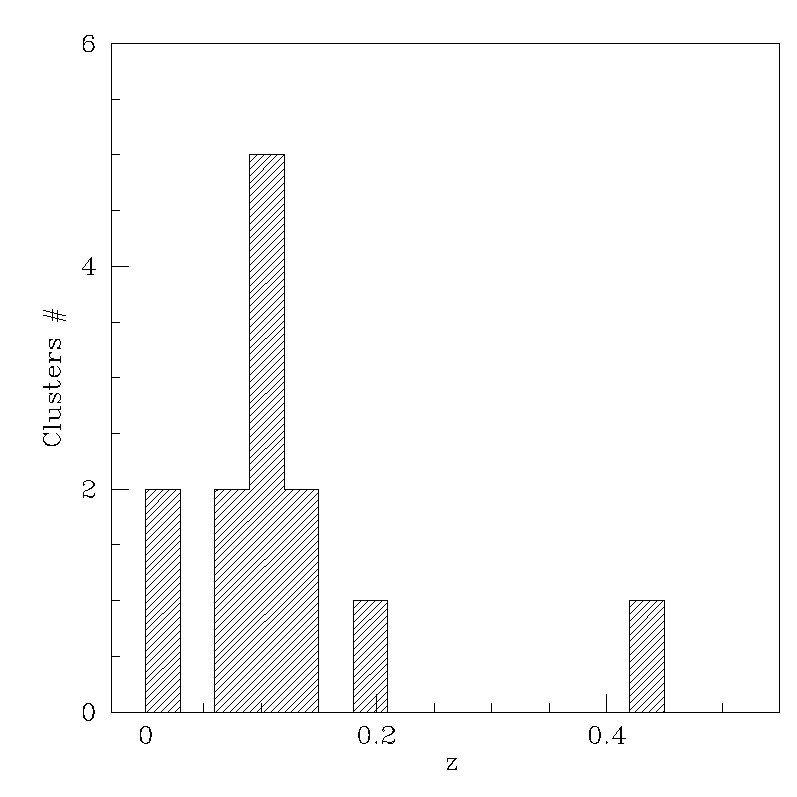}
\caption{Redshift distribution of clusters in the field of view with a redshift identification.}
\label{histo_z} 
\end{figure}

Clusters form at the intersection of filaments of the cosmic web through which
they accrete material during the processes that cause their formation.  We
identified a rich region in the radio sky that covers a size of
8$^{\circ}\times$8$^{\circ}$ and hosts 43 clusters.  Among these, 13 have
either a spectroscopic or photometric redshift identification (see
Table\,\ref{tab:A}).  These redshifts range from 0.015 to 0.425 as shown in
the histogram in Fig.\,\ref{histo_z}, which clearly shows a predominant peak
around $z=0.11$ with standard deviation 0.02. This peak is caused by nine
clusters with redshift in the range $0.08\lesssim z\lesssim 0.15$, i.e. A523,
A525, RXC\,J0503.1+0607, A529, A515, A526, RXC\,J0506.9+0257, A508, and A509.
Two clusters have a redshift smaller than 0.08: ZwCl\,0510.0+0458\,A and
ZwCl\,0510.0+0458\,B, located along the same line of sight, at RA J2000
05:12:39.4 and Dec J2000 +05:01:31, at a distance of 8\,Mpc.  Two clusters
have redshift larger than 0.12: A520, and MACS\,J0455.2+0657.
MACS\,J0455.2+0657 is the most distant system in the field. It is a very
luminous, massive, and distant galaxy cluster, part of the MAssive Cluster
Survey (MACS), see \cite{Mann2012}. According to these authors, this is a
pronounced cool core, characterized by a single brightest cluster galaxy (BCG)
and by a perfect alignment between the X-ray peak and the BCG.  In addition,
we can identify in the field 29 additional clusters belonging to the Zwicky
Clusters of Galaxies catalogue \citep{Zwicky1965} and one belonging to the Hubble
Space Telescope Medium Deep Survey Cluster Sample \citep{Ostrander1998} but,
no redshift information is available in the literature for them.  The full
list of these clusters along with their coordinates is given in
Table\,\ref{tab:noz}.

Some among the clusters in the field have been already found to trace the
cosmic web structure. \cite{Einasto2001} identify the super-clusters SCL\,061
and SCL\,062 that include respectively A509-A526 and A515-A523-A525-A529-A532
(A532 is outside our field of view). Later, a new classification has been
performed by \cite{Chow2014} who identify the super-cluster MSCC\,145
consisting of the clusters A515-A526-A525-A529. These systems are poorly known
in radio, optical, and X-rays, with the exception of A523 and A520, which are the only
two systems in the field known to host diffuse large-scale synchrotron
emission (see \S\,\ref{A520} and \S\,\ref{A523}).

\begin{figure*}
\begin{center}
\includegraphics[width=1.\textwidth, angle=0]{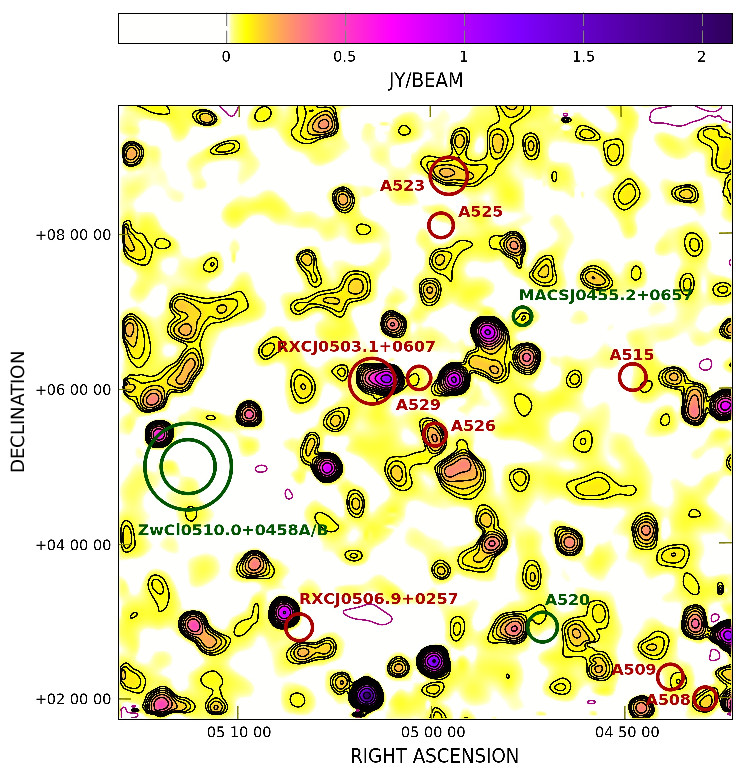}
\caption{SRT image in colours and contours in the frequency range 1.3-1.8\,GHz. The contours are -3$\sigma$, 3$\sigma$ ($\sigma=20$\,mJy/beam) and the remaining scale by a factor $\sqrt{2}$. Negative contours are represented in magenta, while positive contours are in black. 
The angular resolution of the images is 13.9$^{\prime}\times$12.4$^{\prime}$ and BPA=-58.5$^{\circ}$ at the center of the observing band. The circles and labels identify the position of the galaxy clusters in the field of view with known redshift. The size of the circles is proportional to $R_{200}$ of the corresponding clusters (see Table\,\ref{xraytab}), large circles are close-by clusters while small circles are distance clusters. For systems with an upper limit for $R_{200}$, we used the value of the upper limit. We highlight in red clusters with redshift between 0.08 and 0.15, in green clusters with redshift outside this range. }
\label{FULL_optical}
\end{center}
 \end{figure*}

\subsection{Abell 520}
\label{A520}
This galaxy cluster is a dynamically young system, undergoing a strong merger
event in the NE-SW direction \citep{Govoni2004}.  In radio, it is famous for
hosting an extended synchrotron source with a largest linear size of 1.4\,Mpc,
originally classified as a radio halo \citep{Giovannini1999,Govoni2001}, with
the south-west edge of the radio emission coincident with a bow shock, as
found by \cite{Markevitch2005}. A recent re-analysis of the radio emission in
the cluster revealed that the radio halo has a flux density at 1.4\,GHz of
$S_{\rm 1.4\,GHz}=(16.7\pm0.6)$\,mJy \citep{Vacca2014}. The peculiar
morphology and properties of the source do not reflect the typical properties
of radio halos and raise new questions about its nature, as noted already by
\cite{Govoni2001}.  By using optical observations from the \emph{Telescopio
  Nazionale Galileo} and the \emph{Isaac Newton Telescope} facilities,
\cite{Girardi2008} found evidence that the cluster formation is taking place
at the crossing of three filaments: one in the north-east/south-west
direction, one in the east-west direction, and one almost aligned with the
line of sight.

\subsection{Abell 523}
\label{A523}

According to \cite{Chow2014}, A523 is an isolated galaxy cluster. It consists
of an irregular south/south-west sub-cluster strongly interacting with a more
compact sub-cluster on the north/north-east direction \citep{Girardi2016}. The
system has been discovered to host large-scale diffuse radio emission
\citep{Giovannini2011} in the form of a radio halo located at the center of
the cluster with a flux $S_{\rm 1.4\,GHz}=(59\pm5)\,$mJy and a largest linear
angular size of 12$^{\prime}$.  According to a re-analysis of the radio data
by \cite{Girardi2016}, this source has a flux density at 1.4\,GHz of $S_{\rm
  1.4\,GHz}=(72\pm3)$\,mJy, a largest linear size of 1.3\,Mpc and it is
characterized by a polarized signal. Polarization is uncommon among radio
halos. It has been detected in only other two cases, A2255 \citep{Govoni2005}
and MACS\,J0717+3745 \citep{Bonafede2009}, and indicates an intra-cluster
magnetic field fluctuating on scales as large as hundreds of kpc. Compared to
the other radio halos, this system is under-luminous in X-rays with respect to radio, opening new questions on the formation of radio halos and their link with merger events.

\begin{table}
        \caption{Clusters in the field without redshift information.}
       \begin{tabular}{cccccccc}
          \hline
          \hline
  Cluster          &   RA (J2000) & Dec (J2000)   \\
                   &   h:m:s      & $^{\circ}$:$^{\prime}$:$^{\prime\prime}$                            \\
       \hline
ZwCl\,0440.1+0514  &         04:42:45.5 & +05:19:37\\
ZwCl\,0440.3+0547  &         04:42:58.2 & +05:52:36\\
ZwCl\,0440.9+0407  &         04:43:32.3 & +04:12:34\\
ZwCl\,0441.7+0423  &         04:44:20.6 & +04:28:30\\
ZwCl\,0444.7+0828    &       04:47:25.2 & +08:33:18\\
ZwCl\,0445.1+0223  &         04:47:42.4 & +02:28:16\\
ZwCl\,0445.6+0539  &         04:48:16.1 & +05:44:14\\
ZwCl\,0446.2+0235  &         04:48:48.6 & +02:40:12\\
ZwCl\,0446.6+0150   &        04:49:11.8 & +01:55:10\\
ZwCl\,0448.2+0919  &         04:50:56.2 & +09:24:03\\
ZwCl\,0448.5+0551  &         04:51:10.3 & +05:56:02\\
ZwCl\,0451.3+0159    &       04:53:54.0 & +02:03:51\\
ZwCl\,0452.1+0627  &         04:54:47.0 & +06:31:47\\
HSTJ\,045648+03529   &       04:56:48.3 & +03:52:57\\
ZwCl\,0454.3+0534   &        04:56:58.0 & +05:38:38\\
ZwCl\,0455.3+0155  &         04:57:53.9 & +01:59:34\\
ZwCl\,0455.2+0746  &         04:57:54.5&  +07:50:34\\
ZwCl\,0457.0+0511  &         04:59:39.6 & +05:15:26\\
ZwCl\,0458.2+0137  &         05:00:47.6 & +01:41:22\\
ZwCl\,0458.5+0102   &        05:01:04.9 & +01:06:20\\
ZwCl\,0458.5+0536  &         05:01:10.1 & +05:40:20\\
ZwCl\,0459.2+0212  &         05:01:48.2 & +02:16:17\\
ZwCl\,0459.6+0606  &         05:02:16.7 & +06:10:15\\
ZwCl\,0459.8+0943  &         05:02:32.8 & +09:47:14\\
ZwCl\,0502.0+0350  &         05:04:38.1 & +03:54:05\\
ZwCl\,0502.1+0201   &        05:04:42.0 & +02:05:05\\
ZwCl\,0503.1+0751  &         05:05:48.7 & +07:55:01\\
ZwCl\,0504.9+0417  &         05:07:32.6 & +04:20:53\\
ZwCl\,0505.1+0655  &         05:07:47.6 & +06:58:52\\
ZwCl\,0508.8+0241  &         05:11:24.8 & +02:44:36\\

            \hline
            \hline
        \end{tabular}\\
        \label{tab:noz}
     Col 1: Cluster name; Col 2 \& 3: cluster coordinates.
    \end{table}

\section{Radio Observations}
\label{observations}

In order to look for the presence of diffuse emission from low density
environments connecting galaxy clusters, we observed with the SRT a region of
the sky hosting nine clusters with redshift z$\approx$0.1: A523, A525,
RXC\,J0503.1+0607, A529, A515, A526, RXC\,J0506.9+0257, A509, and A508, see
Table\,\ref{tab:A}. 
The radio observations presented in this paper were done in the context of the SRT Multi-frequency Observations of Galaxy
Clusters program \citep[SMOG, PI. Matteo Murgia, see][for a description of the
  project]{Govoni2017,Loi2017}.  We observed an area of $8^{\circ}\times
8^{\circ}$ centred at right ascension (RA) 05h:00m:00.0s and declination
(Dec) +05$^{\circ}$:48$^{\prime}$:00.0$^{\prime \prime}$, for a total exposure
time of about 18 hours and three smaller fields of view centred respectively
on the galaxy clusters A520, A526, and A523 over a region of $3^{\circ}\times
3^{\circ}$ of about 1.5 hour each. The observations were taken with
the L- and P-band dual-frequency coaxial receiver for primary-focus operations
\citep{Valente2010}. We used the L-band component only covering the frequency
range 1.3-1.8\,GHz. The data stream was recorded with the SArdinia
Roach2-based Digital Architecture for Radio Astronomy
\cite[SARDARA][]{Melis2017} back-end. The configuration used had 1500\,MHz of
bandwidth and 16384 channels $\sim$90\,kHz each, full-Stokes parameters, and
sampling at 10 spectra per second. We opted for the on-the-fly (OTF) mapping
strategy scanning the sky alternatively along the RA and Dec direction with a
telescope scanning speed of 6\,$^{\prime}$/s and an angular separation of 3\,$^{\prime}$ between
scans, this choice guarantees a proper sampling of the SRT beam whose full
width at half maximum (FWHM) is 13.9\,$^{\prime}\times$12.4\,$^{\prime}$ at 1550\,MHz, the
central frequency of these observations. Since the sampling time is 100\,ms, 
the individual samples
are separated in the sky by 36\,$^{\prime\prime}$ along the scanning direction.  A detailed
summary of the SRT observations is given in Table\,\ref{tab:B}.

The calibration and the imaging of the data were performed with the proprietary
Single-dish Spectral-polarimetry Software \citep[SCUBE,][]{Murgia2016}. We
used 3C\,147 as band-pass and flux density calibrator and the flux density
scale of \cite{Perley2013} was assumed.  We flagged about 23\% of the data
because of radio frequency interference (RFI). After calibration, a baseline
was fitted to the data in order to subtract any emission not related to
the target (due e.g, to our Galaxy).  The subtraction of the baseline has
been performed as follows. First, we convolved the NVSS image of this region
to the resolution of the SRT. We used this image to identify empty regions in
the SRT image.  The linear fit of the emission of these regions was then
subtracted from the data.  The baseline-subtracted data were then projected in
a regular three-dimensional (RA - Dec - observing frequency) grid with a
spatial resolution of 3\,$^{\prime}$/pixel in order to guarantee a beam FWHM
sampling of four pixels. These steps were performed in each frequency
channel.  The resulting frequency cube was stacked and de-stripped by mixing
the Stationary Wavelet Transform (SWT) coefficients \cite[see][for
  details]{Murgia2016}.  This image was then used to produce a new mask and
the procedure was repeated. This procedure was performed for each dataset
separately and the images produced for each dataset were averaged.

Finally, we applied a hard threshold denoising algorithm on scales below two
pixels (1\,pixel=3\,$^{\prime}$) as the signal amplitude is expected to be
negligible with respect to the noise amplitude on these angular scales.  The
denoised image obtained by the combination of all the SRT data is shown in
Fig.\,\ref{FULL_optical} in colours and contours.  The noise level 1$\sigma$ of
the radio image is 20\,mJy/beam, comparable with the confusion noise for this
frequency band at the SRT resolution.  The zero-level of the image is
$\sim$25\% of the noise level. A detailed description of the calibration and
imaging procedures of the SRT data are given in \cite{Murgia2016} and
\cite{Govoni2017}.

The circles and labels in Fig.\,\ref{FULL_optical} indicate the position in
the field of the clusters with a redshift identification. This figure reveals
that most of the nine clusters with redshift $\approx$ 0.1 are all located
along a line crossing the field of view from north to south and connecting the
galaxy clusters A523 and A508, passing through A525, RXC\,J0503.1+0607, A529,
A526, and A509, while RXC\,J0506.9+0257 and A515 are respectively slightly
south-east and slightly west of the filament. A520 has a redshift higher than
the average redshift of this group. A possible explanation is that A520
appears to belong to this region of the sky only because of projection
effects.  Alternatively, the filament could be partially located in the plane
of the sky and partially oriented along the line of sight in the direction of
A520.  This is supported by the findings of \cite{Girardi2008} which suggest
the hypothesis that A520 could be connected with the filament at z$\approx$0.1
examined in this work. However, as shown in Table\,\ref{tab:A}, the redshift of
A520 implies a distance along the line of sight of about 410 Mpc, therefore the
hypothesis of a connection is still controversial.

\subsection{SRT beam and calibration scale}
\label{fluxscale}
We cross-checked the calibration procedures by comparing the SRT flux scale
with the NVSS flux scale. To perform the comparison, we produced SRT images in
the same frequency range as the NVSS images, we tailored from the whole SRT
bandwidth (1.3-1.8 GHz) two smaller sub-bands centred at the same
frequencies and with the same width of the NVSS observations (1364.9 - 1414.9
MHz and 1435.1 - 1485.1 MHz) and then binned them in one image.

\begin{figure*}

\begin{center}
\includegraphics[width=0.8\textwidth]{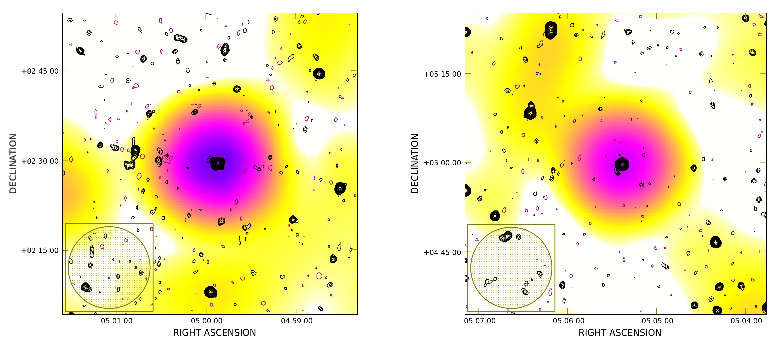}
\caption{Sources used for the beam measurement (left panel) and for the comparison of the NVSS and SRT flux density scales (right panel). The SRT image in the same frequency range as the NVSS is shown in colours, while the NVSS emission is shown in contours. The contour levels are -1.35, 1.35\,mJy/beam (with a beam of 45$^{\prime \prime}$) and increase by a factor $\sqrt{2}$, positive contours shown in black and negative contours shown in magenta. The beam in the bottom left corner represents the SRT spatial resolution of 13.66$^{\prime}$ as found in \S\,\ref{fluxscale}. The colorbar is the same as Fig.\,\ref{FULL_optical}.}
\label{test_source}
\end{center}
 \end{figure*}

As a first step, we measured the beam of the SRT images by fitting the emission
of the brightest isolated point source in the field with the software SCUBE
\citep{Murgia2016}.  This compact source is a quasar with redshift $z=2.384$
\citep{Xu2014} and is located at RA J2000 04h:59m:53.94s and Dec J2000
02$^{\circ}$:29$^{\prime}$:39.72$^{\prime\prime}$ (see
Fig.\,\ref{test_source}, left panel). By fitting the peak emission with a
circular 2-d Gaussian with four free parameters (peak, $x$ and $y$ coordinates
of the centre, and FWHM), we obtained for the SRT beam a FWHM of
$\theta=(820\pm 1)\,^{\prime\prime}$.

We then compared the SRT and NVSS flux scales by fitting the peak brightness
of the brightest non-variable isolated point-source.  The source is located at
RA J2000 05h:05m:24.21s and Dec J2000
04$^{\circ}$:59$^{\prime}$:48.44$^{\prime\prime}$ (see
Fig.\,\ref{test_source}, right panel).  During the fitting procedure the beam FWHM was kept fixed to the value derived above. The position of the peak as
measured by the SRT and by the NVSS agrees within 15$^{\prime\prime}$.  The
peak in the two images are in agreement within about 10\%, with a value
respectively of ($0.96\pm 0.03$)\,Jy/beam and
($0.86\pm0.03$)\,Jy/beam. Note that the source used for the fit is not
the brightest source in the field but one among the brightest isolated
sources. Indeed, we prefer to discard non-isolated sources because of possible
blending of sources.  Moreover, we note that the flux scale of the NVSS and of
SRT observations are based respectively on the scales by \cite{Baars1977} and
\cite{Perley2013}, and they agree within 1\%.

\section{Combination of single-dish and interferometric data}

\label{combination}

In this section, we present the combination of our single-dish data with
mosaic interferometric observations taken from the NVSS, following the approach
described by \cite{Loi2017}. 
Single-dish observations can recover a maximum angular scale corresponding
to the angular size of the scanned region but they miss high spatial
resolution. Interferometers can reach high spatial resolutions but
are limited by the maximum angular scale corresponding to their minimum
baseline. In the case of a full-synthesis Very Large Array (VLA) observation
this translates to $\sim$16$^{\prime}$ for the most compact configuration at L-band. For
snapshot observations as the NRAO VLA Sky Survey \citep[NVSS,][]{Condon1998}
this number should be divided by
two\footnote{\burl{https://science.nrao.edu/facilities/vla/docs/manuals/oss/performance/resolution}}
but we consider the full angular size since the NVSS image we used has been obtained
with a mosaic.
By combining the SRT and NVSS data, we retain the ability to observe
large-scale structures up to 8$^{\circ}$ and, at the same time, the resolution
of the VLA in D configuration at 1.4\,GHz, i.e. 45$^{\prime\prime}$.  The
combination has been performed in the Fourier domain with the SCUBE software
package \citep{Murgia2016}.  The SRT has a size of 64\,m, while the VLA in its
most compact configuration has a minimum baseline of 35\,m, this means that
there is a common range of spacing and consequently of wave-numbers $k$ in
the Fourier space\footnote{For practical purposes, in our algorithm we define
  the wave number $k=0.5\frac{\Delta\times N_{\rm pix}}{\theta_{\rm A}}$,
  where $\Delta=15^{\prime \prime}$/pixel is the angular size of the pixel of
  the image, $N_{\rm pix}=2048$ is the number of pixels in the image along one
  side, and $\theta_{\rm A}$ is the angular scale in $^{\prime\prime}$.}.  We
combined the NVSS image with the SRT image obtained in the same frequency
range and select in this image the same region of the sky as the NVSS image,
so that they should contain the same radio power. 
\begin{figure}
\begin{center}
\includegraphics[width=0.5\textwidth]{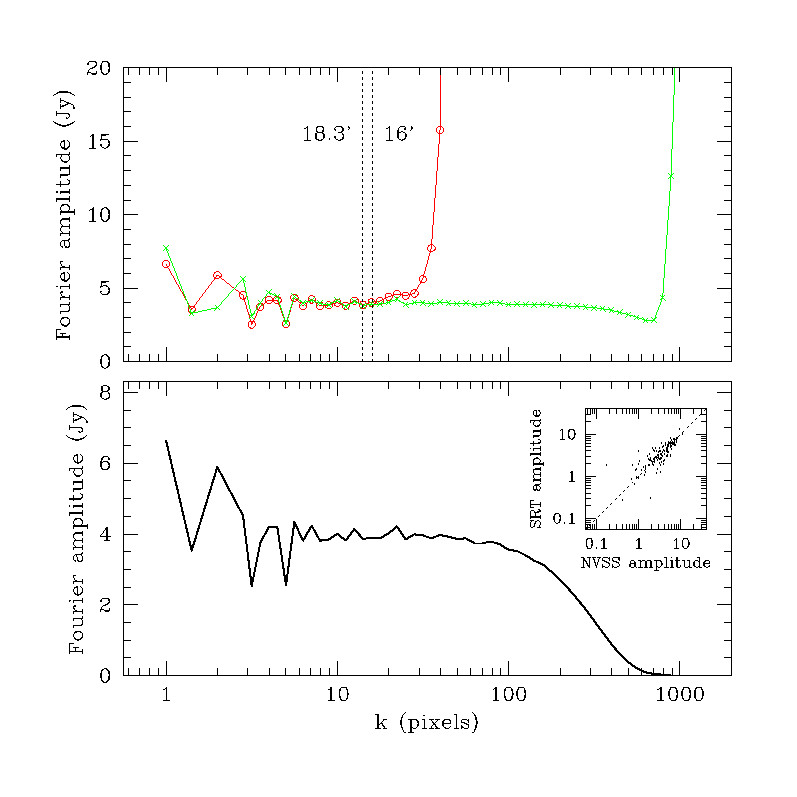}
\caption{\emph{Top panel}: Power spectrum of the SRT data (red) and power spectrum of the NVSS data (green). The dashed vertical black lines identify the overlapping region used to adjust the flux density scales. \emph{Bottom panel}: Tapered combined power spectrum. In the inset the comparison between the amplitude of the SRT and of the NVSS data in the region of overlapping (18.3$^{\prime}$-16$^{\prime}$) in the Fourier space is shown.} 
\label{combo}

\end{center}
 \end{figure}
In order to combine the two images we required that the two power spectra have
the same power spectral density in the inner portion of the overlapping region
in Fourier-space, corresponding to angular scales between 16$^{\prime}$ and
18.3$^{\prime}$.  This requirement translates into a scaling factor of the SRT
power spectrum normalization of $\approx$1.003.  In \S\,\ref{fluxscale}, by
comparing the flux of a single source in the two images, we found that the two
flux density scales agree within 10\%.  In this section we make a more robust
statistical comparison, by using all the sources available in the field.  This
agreement confirms the accuracy of the SRT flux density scale calibration.

The two power spectra were then merged with a weighted sum: the single-dish
data-weights were set to zero for angular scales smaller than 16$^{\prime}$,
to 1 for angular scales larger than 18.3$^{\prime}$, and with a weight that
linearly varies from 0 to 1 in between. Vice-versa for the interferometric
data-weights.  The final power spectrum was tapered with the interferometric
beam and the data were back-transformed to obtain the combined image.  The
interferometric and single-dish power spectra are shown in the top panel of
Fig.\,\ref{combo}, after the beam deconvolution (they have been divided in
Fourier space by the Fourier transform of the corresponding beam).  The merged
power spectrum is shown in the bottom panel of Fig.\,\ref{combo}: in the inset
the SRT amplitude versus the NVSS amplitude is shown for measurements
corresponding to angular scales in the overlapping region.  The average of the
amplitude of the points in the inset corresponds to the value shown in the
large panels.  The combined image has a beam of 45$^{\prime \prime}$ and a
noise 0.45\,mJy/beam, dominated by the NVSS noise. To better highlight the
presence of diffuse large-scale sources, we convolved to a beam of
$3.5^{\prime}\times 3.5^{\prime}$. The resulting combined image is shown in
Fig.\,\ref{srt_all} (grey colours, top panels).

 \begin{table*}
        \caption{Data}
        \begin{tabular}{ccccccc}
          \hline
          \hline
            Date &  Receiver            &   Target &  FOV  & OFT scan axis & Calibrators & Time on source   \\
            \hline
               14Jul2016       & L-band  & A520-A526-A523   &  $8^{\circ}\times 8^{\circ}$     & 1$\times$(RA$+$Dec)& 3C84,3C147, 3C138& 10\,h\\
               23Jul2016       & L-band  & A520-A526-A523   &  $8^{\circ}\times 8^{\circ}$     & 1$\times$(RA$+$Dec)& 3C286,3C147, 3C138&9\,h \\
               22Jul2016       & L-band  & A520             &  $3^{\circ}\times 3^{\circ}$     & 1$\times$(RA$+$Dec) & 3C84,3C147, 3C138& 1\,h\,20\,min\\
               22Jul2016       & L-band  & A523             &  $3^{\circ}\times 3^{\circ}$     & 1$\times$(RA$+$Dec)& 3C84,3C147, 3C138& 1\,h\,40\,min \\
               22Jul2016       & L-band  & A526             &  $3^{\circ}\times 3^{\circ}$     & 1$\times$(RA$+$Dec)& 3C84,3C147, 3C138& 1\,h\,20\,min\\

            \hline
            \hline
        \end{tabular}
     
        \label{tab:B}
 \end{table*}

\subsection{Point source subtraction}
\label{ps_sub}
To investigate the presence of large-scale diffuse structures associated with the
galaxy clusters reported in Table\,\ref{tab:A} and with the low-density
environments connecting them, we performed a subtraction of the point sources
with the SCUBE software package \citep{Murgia2016} in the combined image at
$45^{\prime\prime}$ resolution. 

The subtraction is done as follows:
\begin{itemize}
\item[(i)] the strongest point source in the image is identified and fitted with a non-zero baseline 2-d elliptical Gaussian;
\item[(ii)] the model is then subtracted from the image; 
\item[(iii)] the two steps above are repeated until an user-defined threshold is reached. 
\end{itemize}

To take into account the possibility that each source is embedded into
large-scale diffuse emission, we model the source with a 2-d elliptical
Gaussian sitting on a plane. Overall, the free parameters of the fit are nine:
the $x,y$ coordinates in the sky of the centre of the Gaussian, the FWHM along
the two axis, the position angle, the amplitude, and the three components of
the direction normal to the plane.  Only in the case of sources weaker than a
user-defined-threshold (in this case 10$\sigma$) is the FWHM is forced to assume
the value of the beam size.  This algorithm has been tested in the context of
\cite{Fatigoni2017}.

As outputs, the task produces a residual image.  We repeated the procedure two
times. At first, we used a threshold of about 7$\sigma$=3\,mJy/beam, where
$\sigma$=0.45\,mJy/beam is the noise of the image. As a second step, we again run
the algorithm on the residual image, with a lower threshold of
4$\sigma$=1.8\,mJy/beam. This conservative threshold has been chosen to avoid
over-subtraction. In total, 3872 sources have been subtracted.  In Fig.\,\ref{srt_all}
(top right and bottom left panels), the resulting residual image is shown in blue
contours.  In Appendix\,\ref{pscat} we use a sub-sample of the radio galaxies
in the field to cross check the flux scale after the combination of
single-dish and interferometric data.

\section{X-ray emission}
\label{x-ray}

The X-ray properties of the majority of the clusters in the sample are
unknown.  Information for only a few of the clusters in Table\,\ref{tab:A} can
be found in the literature.  The X-ray luminosity of A523, RXC\,J0506.9+0257, and
RXC\,J0503.1+0607 are reported in the catalogue by \cite{Bohringer2000}, while
A520 and MACS\,J0455.2+0657 can be found in \cite{Mahdavi2007} and
\cite{Mann2012}, respectively.  To our knowledge, no information is available
for the remaining clusters.

By applying the approach proposed by \cite{Finoguenov2007}, we derived the
flux $S_{\rm X}$ in the 0.5-2\,keV energy band and the luminosity $L_{\rm X,
  0.1-2.4\,keV}$ within an extraction area of radius $R_{200}$\footnote{This
  radius is defined as the radius that encompasses the matter with density 200
  times the critical density.} for the clusters in Table\,\ref{tab:A}, by using
data from the \emph{ROSAT All-Sky Survey} \citep[RASS,][]{Voges1999}.  The
fluxes of the clusters have been corrected for the aperture and for the Point
Spread Function and the luminosities have been corrected for the dimming \cite[for details about the procedure please refer
  to][]{Finoguenov2007}. The values of flux and luminosity of these clusters
are reported in Table\,\ref{xraytab}. For the sources with significance of the count rate below 1$\sigma$, the values we provide for all the properties are 2$\sigma$ upper limits.

\begin{table*}
        \caption{X-ray fluxes of galaxy clusters}
\begin{center}
        \begin{tabular}{cccccccc}
          \hline
          \hline
Cluster & z & $n_{\rm H}$ & $R_{200}$&$M_{200}$& $S_{\rm X}$ & $D_{\rm L}$ & $L_{\rm X, 0.1-2.4\,keV}$ \\
& &$10^{20}$\,cm$^{-2}$ & $^{\circ}$& $10^{14}M_{\odot}$&$10^{-12}$erg\,s$^{-1}$\,cm$^{-2}$ & Mpc & $10^{43}$erg\,s$^{-1}$ \\
\hline
A523 & 0.104 & 10.6 & 0.192 & 2.91 & 1.6 $\pm$ 0.3 & 499&7.3 $\pm$ 1.6\\
A525 & 0.1127 & 9.98 & 0.128 & 1.08 & 0.3 $\pm$ 0.1 & 543&1.6 $\pm$ 0.8\\
MACSJ0455.2+0657 & 0.425 & 8.41 & 0.096 & 12.87 & 1.3 $\pm$ 0.2 & 2416&116.1 $\pm$ 21.6\\
ZwCl0510.0+0458A & 0.027 & 12.0 & < 0.28 & < 0.19 & < 0.4 & 123&< 0.1\\
ZwCl0510.0+0458B & 0.015 & 12.0 & < 0.448 & < 0.14 & < 0.7 & 68& < 0.1\\
RXCJ0503.1+0607 & 0.088 & 8.61 & 0.235 & 3.39 & 2.9 $\pm$ 0.4 &418& 9.0 $\pm$ 1.4\\
A529 & 0.1066 & 8.25 & < 0.121 & < 0.79 & < 0.2 & 512 &< 0.9\\
A515 & 0.1061 & 8.21 & 0.137 & 1.12 & 0.3 $\pm$ 0.2 &509 &1.6 $\pm$ 0.8\\
A526 & 0.1068 & 7.81 & 0.117 & 0.72 & 0.2 $\pm$ 0.1 & 513&0.8 $\pm$ 0.6\\
RXCJ0506.9+0257 & 0.1475 & 8.57 & 0.14 & 2.91 & 0.8 $\pm$ 0.2 & 727 &7.7 $\pm$ 2.1\\
A520 & 0.199 & 5.66 & 0.155 & 8.64 & 2.6 $\pm$ 0.4 & 1011&45.0 $\pm$ 6.1\\
A509 & 0.0836 & 8.88 & < 0.133 & < 0.53 & < 0.2 &395 &< 0.5\\
A508 & 0.1481 & 9.85 & 0.118 & 1.75 & 0.4 $\pm$ 0.2 & 730&3.5 $\pm$ 1.5\\
           \hline
            \hline
        \end{tabular}
\label{xraytab}\\
\end{center}
Col 1: Cluster name; Col 2: redshift; Col 3: cluster Hydrogen column density;
Col 4: $R_{200}$ of the system;  Col 5: X-ray flux in the 0.5-2\,keV energy band within $R_{200}$; Col 6: Mass of the galaxy cluster within $R_{200}$; Col 7: Luminosity distance; Col 8: X-ray luminosity in the 0.1-2.4\,keV energy band within $R_{200}$.    \\
\normalsize

    \end{table*}

\section{millimetre and sub-millimetre emission}
\label{sz}
To investigate the properties of the cosmic web in this region of the sky, we
superimposed the SRT contours on the \emph{Planck} image, see
Fig.\,\ref{srt_all} (colours, bottom left panel).  The Compton parameter map (hereafter $Y$-map) has been
obtained by using the data from the full mission and by applying the Modified
Internal Linear Combination Algorithm \citep[MILCA][]{Hurier2013}
method\footnote{\burl{https://www.cosmos.esa.int/web/planck/pla}}. As described
in the work by \cite{Planckcoll2016a}, this method removes the signal of the
Cosmic Microwave Background taking into account its spectral properties.
 
Most of the clusters in Table\,\ref{tab:A} show a counterpart in the
\emph{Planck} image.  However, in the catalogue by \cite{Planckcoll2016b} only
an association for A520 and one for A539 (RA 05:16:37.3, Dec +06:26:16,
z=0.0284) within 4$^{\prime}$ are present. The
strongest signal is detected in the direction of A523 and A520, where it
extends well beyond the cluster outskirts. A signal appears to be present also in the
direction of A525, A529, RXC\,J0503.1+0607, MACS\,J0455.2+0657,
RXC\,J0506.9+0257, A515, A508, A509, and A539 but, unfortunately, the galaxy cluster A539 is at the edge of the SRT field of view. On
the contrary, for A526 and ZwCl\,0510.0+0458\,A/B there is no evidence of a
signal above the noise level.

\section{Results}
\label{Results}

\begin{figure}
\begin{center}
\includegraphics[width=0.5\textwidth, angle=0]{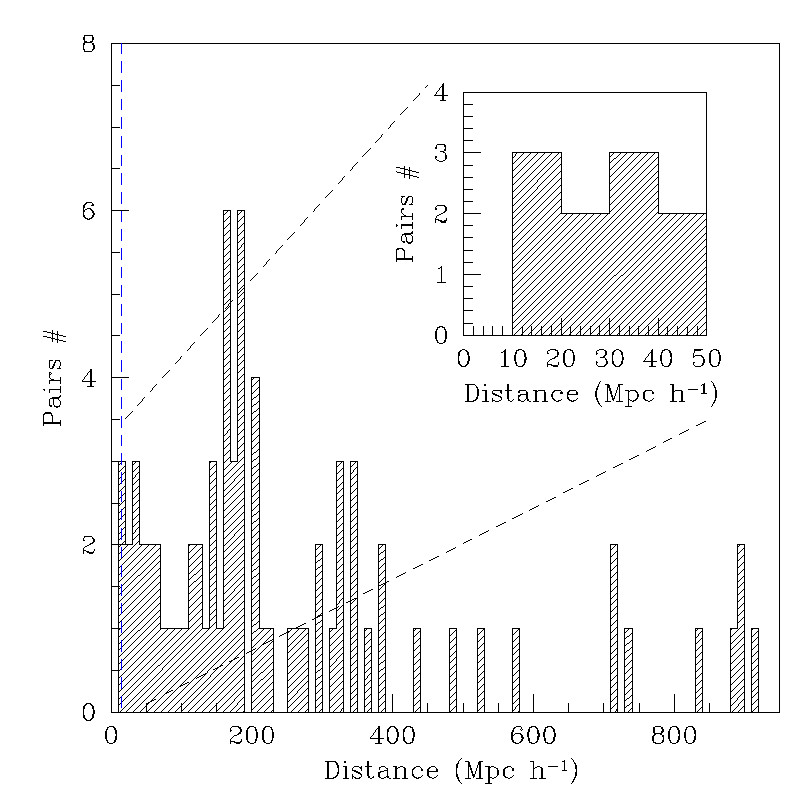}\\
\end{center}
\caption{Distribution of 3-d distances between all the pairs of clusters with known redshift in the field. The dashed vertical blue line marks a 3-d distance of 15 Mpc$h^{-1}$. In the inset a zoom of the region between 0 and 50\,Mpc$h^{-1}$ is shown. }
\label{colberg}
 \end{figure}

The SRT contours reveal a field rich of radio sources.  This region of the sky
is populated by several galaxy clusters starting from the north with the
complex A523-A525 down to the south with the complex A508-A509. A large
fraction of them are approximately at the same redshift ($\approx$0.1, see
Fig.\,\ref{histo_z}), and possibly form a filament of the cosmic web sitting
in the plane of the sky. This hypothesis is supported by the fact that some
of them have been observed to be undergoing accretion processes as
confirmed by the diffuse large-scale synchrotron emission observed at their
centre. This is the case of A523 and A520, even if it is not clear whether A520 belongs to the filament because of its large distance along the line of sight. In the CDM cosmology,
clusters of galaxies with a mass exceeding $10^{14}$ $M_{\odot}$ should have a
80\% probability of being connected to a neighbouring cluster by a filamentary
joint of dark and gas matter, in case they are less than 15 Mpc\,$h^{-1}$ apart
\citep{Colberg2005}. Based on the X-ray data (Table\,\ref{xraytab}) we can estimate that at least eight of the clusters (A523, A525, MACS\,J0455.2+0657, RXC\,J0503.1+0607, A515, RXC\,J0506.9+0257, A520, and A508) in the field have a mass $M_{\rm 200} \ge 10^{14} M_{\odot}$.  Therefore, several of these
objects might be connected by filaments difficult to observe through the X-ray/SZ
effect but possibly detectable in the radio window.  In Fig.\,\ref{colberg},
we show the distribution of 3-d distances between all the pairs of clusters
with known redshift in the field. Since an absolute error in the redshift
estimate of 0.003 (3\%) translates in an error in the 3-d distance of about
13\,Mpc$h^{-1}$, we looked for pairs with 3-d distance in the range
0--30\,Mpc$h^{-1}$. We found five pairs within this range of 3-d distance 
(A526 and RXC\,J0503.1+0607, A526 and A529, A509 and RXCJ0503.1+0607, A509 and
A529 and A509 and A526), suggesting that this field is likely to host
filaments of dark and gas matter connecting galaxy clusters.

To identify possible diffuse radio emission in the field and better understand
its nature, we subtracted the point sources and compared the residual radio
emission with the signal observed at other wavelengths.  In the following we
present the results and describe the properties of the diffuse radio sources
we found.

\subsection{Diffuse emission}
\label{diff}

\begin{figure*}
\begin{center}
\includegraphics[width=1.0\textwidth, angle=0]{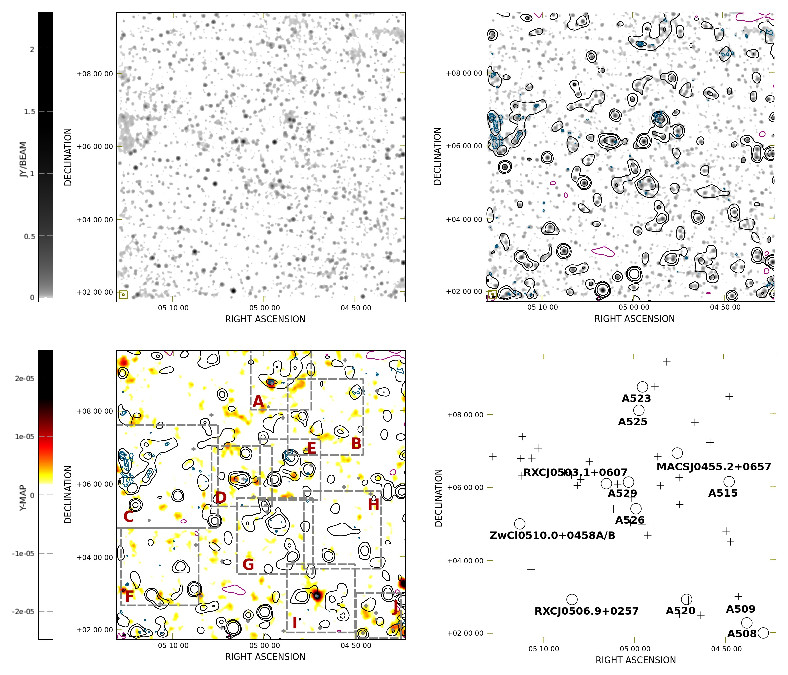}\\
\end{center}
\caption{\emph{Top left panel}: SRT+NVSS image in colours (FWHM 3.5$^{\prime}$). \emph{Top right panel}: contours of the SRT image (-60\,mJy/beam, 60\,mJy/beam, 150\,mJy/beam, 500\,mJy/beam, negative in magenta and positive in black; resolution 13.9\,$'\times$12.4\,$'$) and of the SRT+NVSS image after compact sources subtraction (starting at 3$\sigma$ and increasing by a factor $\sqrt{2}$, in blue; $\sigma$=2.5\,mJy/beam, resolution 3.5$^{\prime}\times$3.5$^{\prime}$) overlaid on the SRT+NVSS image in colours (FWHM 3.5$^{\prime}$). \emph{Bottom left panel}: SRT and SRT+NVSS after point-source subtraction in contours overlaid on the \emph{Planck} Y-map (see text) in colours. The contours from SRT and from the residual SRT+NVSS image are the same as in the top right panel. The resolution of the \emph{Planck} Y-map is 10$^{\prime}$. Dots indicate the position of clusters with redshift identification (see Table\,\ref{tab:A}), while crosses indicate clusters without redshift identification (see Table\,\ref{tab:noz}). \emph{Bottom right panel}: Location of clusters with known redshift (circles and labels) and of the sources discussed in this paper (crosses), see Table\,\ref{tab_diff}. }
\label{srt_all}
 \end{figure*}

In this section we inspect the SRT, the SRT+NVSS and the residual SRT+NVSS
after compact source subtraction 
images at 1.4\,GHz (see Fig.\,\ref{srt_all}, top right panel) and compare them with X-ray ROSAT data in the 0.1--2.4\,keV
from the RASS and the SZ Compton parameter from millimetre/sub-millimetre
observations obtained with the \emph{Planck} satellite.

\begin{figure*}
\begin{center}
\includegraphics[width=0.6\textwidth, angle=0]{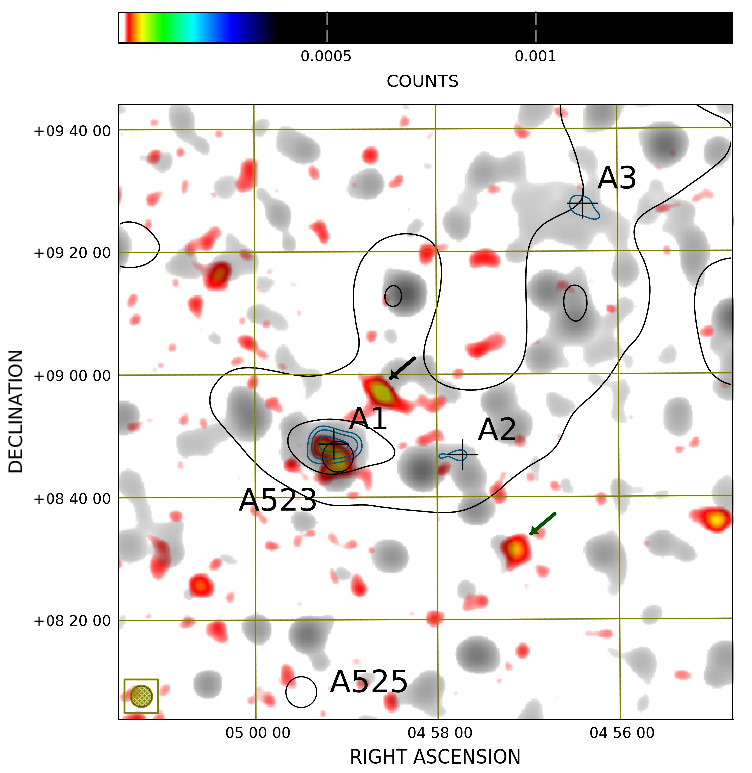}
\includegraphics[width=0.35\textwidth, angle=0]{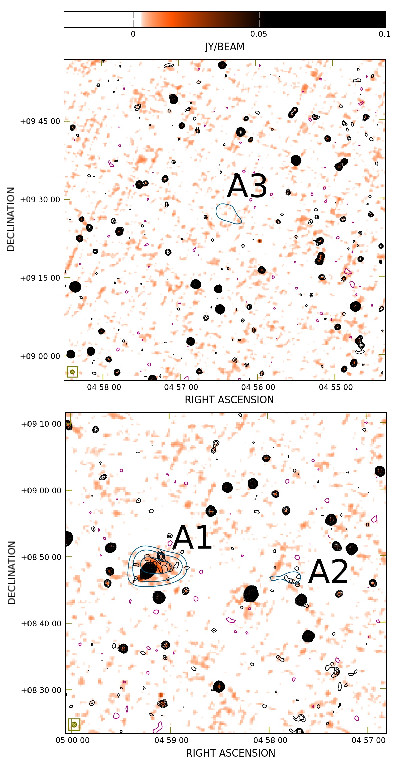}\\
\end{center}
\caption{{\bf Region A.} \emph{Left panel:} SRT contours (-60\,mJy/beam,
  60\,mJy/beam, 150\,mJy/beam, 500\,mJy/beam, negative in magenta and positive
  in black; resolution 13.9\,$'\times$12.4\,$'$) and SRT+NVSS after compact sources subtraction contours (in blue, starting from 3$\sigma$ and the remaining increasing by a factor $\sqrt{2}$; $\sigma$=2.5\,mJy/beam, resolution 3.5$^{\prime}\times$3.5$^{\prime}$) overlaid on X-ray emission from the RASS in
  red colours and radio emission from the SRT+NVSS (resolution $3.5^{\prime}$)
  in gray colours. \emph{Right panels:} NVSS contours -3$\sigma$ (magenta, $\sigma$=0.45\,mJy/beam),
  3$\sigma$, and increasing by a factor $\sqrt{2}$ (black) overlaid on the
  TGSS in colours. The black and green arrows mark regions where a strong X-ray
  emission is detected, as described in the text.}
\label{im1}
 \end{figure*}

\begin{figure*}
\begin{center}
\includegraphics[width=0.6\textwidth, angle=0]{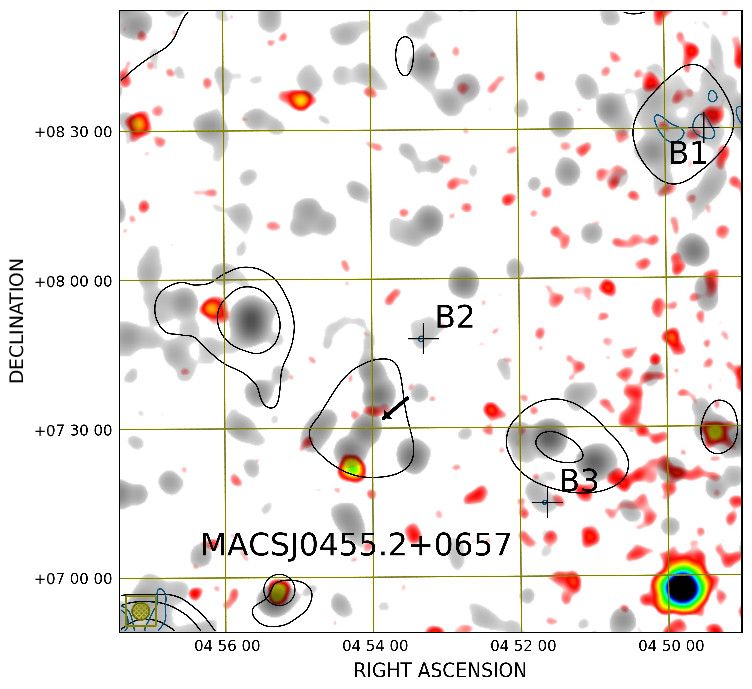}
\includegraphics[width=0.35\textwidth, angle=0]{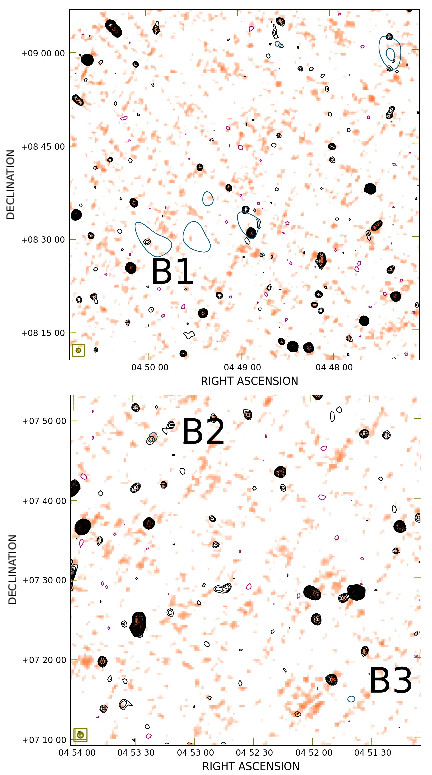}\\
\end{center}
\caption{{\bf Region B.} \emph{Left panel:} SRT contours (-60\,mJy/beam,
  60\,mJy/beam, 150\,mJy/beam, 500\,mJy/beam, negative in magenta and positive
  in black; resolution 13.9\,$'\times$12.4\,$'$) and SRT+NVSS after compact sources subtraction contours (in blue, starting from 3$\sigma$ and the remaining increasing by a factor $\sqrt{2}$; $\sigma$=2.5\,mJy/beam, resolution 3.5$^{\prime}\times$3.5$^{\prime}$) overlaid on X-ray emission from the RASS in
  red colours and radio emission from the SRT+NVSS (resolution $3.5^{\prime}$)
  in gray colours. \emph{Right panels:} NVSS contours -3$\sigma$ (magenta, $\sigma$=0.45\,mJy/beam),
  3$\sigma$, and increasing by a factor $\sqrt{2}$ (black) overlaid on the
  TGSS in colours. The black arrow marks the region shown in the zoom of Fig.\,\ref{SRT_TGSS_cluster}. The colorbars are the same as in Fig.\,\ref{im1}.}
\label{im4}
 \end{figure*}

\begin{figure}
\begin{center}
\includegraphics[width=0.4\textwidth, angle=0]{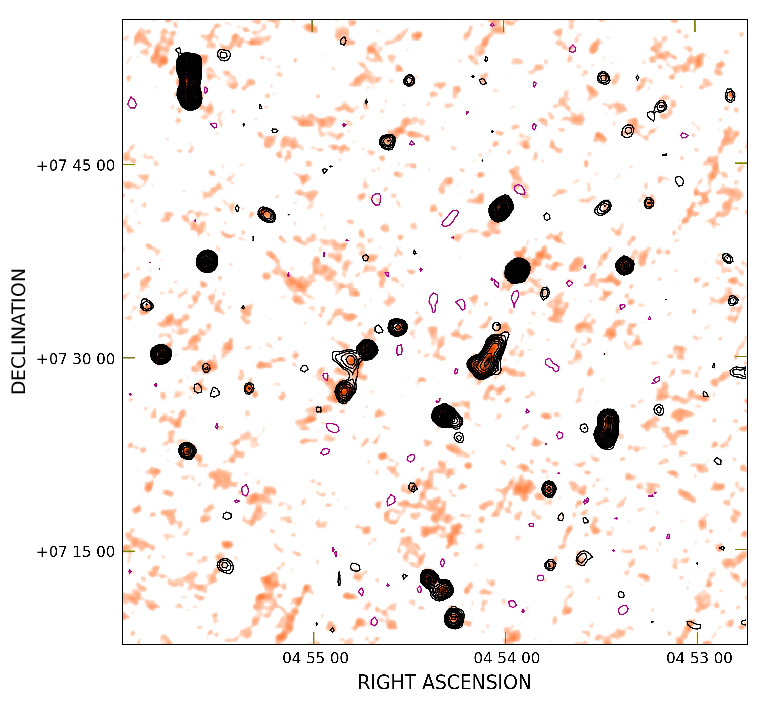}\\
\end{center}
\caption{Zoom of the central portion of Region B (see arrow in Fig.\,\ref{im4}). The image shows the NVSS contours (-1.35\,mJy/beam, 1.35\,mJy/beam, and the remaining scaled by a factor $\sqrt{2}$, negative in magenta and positive in black) overlaid on the TGSS colours. The colorbar is the same as in the right panels of Fig.\,\ref{im1}. }
\label{SRT_TGSS_cluster}
 \end{figure}

After the subtraction of the compact sources, 35 patches of diffuse synchrotron
emission survive, whose locations are shown in Fig.\,\ref{srt_all} along with
the location of the clusters with redshift identification (bottom right
panel). The sources are mainly located along the group of clusters with
redshift z$\approx$0.1 (A523, A525, RXC\,J0503.1+0607, A529, A515, A526,
RXC\,J0506.9+0257, A508, and A509), and in the region around A539 (middle left
of the image). We identified eight regions of interest where these 35
candidate sources are located (gray boxes in the bottom left panel of
Fig.\,\ref{srt_all}).  Due to the large size of the field of view, to better
visualize these diffuse radio sources, we present a zoom of each region in figures from
Fig.\,\ref{im1} to Fig.\,\ref{im7}, labelled respectively with capital letters from A to J (see
Fig.\,\ref{srt_all}). In each Figure, on the left panel, we overlay the SRT
and the SRT+NVSS contours after the point-source subtraction on the X-ray and
SRT+NVSS images in colours. On the right panel(s), we show the NVSS contours
overlaid on the TIFR GMRT Sky Survey Alternative Data Release \citep[TGSS
  ADR,][]{Intema2017} in colours.  We indicate the position of the clusters with redshift identifications with circles and labels and mark all the spots
of diffuse emission with a cross and the letter corresponding to the region of interest
plus progressive numbers.  These sources are those visible in the residual
image after a 3$\sigma$ cut in radio brightness ($\sigma=2.5$\,mJy/beam). For
these sources, we give an estimate of the flux ($S_{\rm residual}$) and the
largest linear size (LLS) in Table\,\ref{tab_diff}.  In order to evaluate the
integrated flux and the size of these sources, we blanked the convolved
residual image at its 2$\sigma$ level. A summary of the radio and X-ray properties
of these sources is given in Table\,\ref{tab_diff} and
Table\,\ref{tab_diff_xray} respectively. Sources at the edge of the image
have not been considered because they could be artefacts of the imaging or
point-source subtraction process. The only exception is the patch of diffuse
emission close-by the galaxy cluster A539.  In the following we report a description of the emission observed
in each region.

\subsubsection{Region A}
\label{RegionA}

\begin{figure*}
\begin{center}
\includegraphics[width=0.6\textwidth, angle=0]{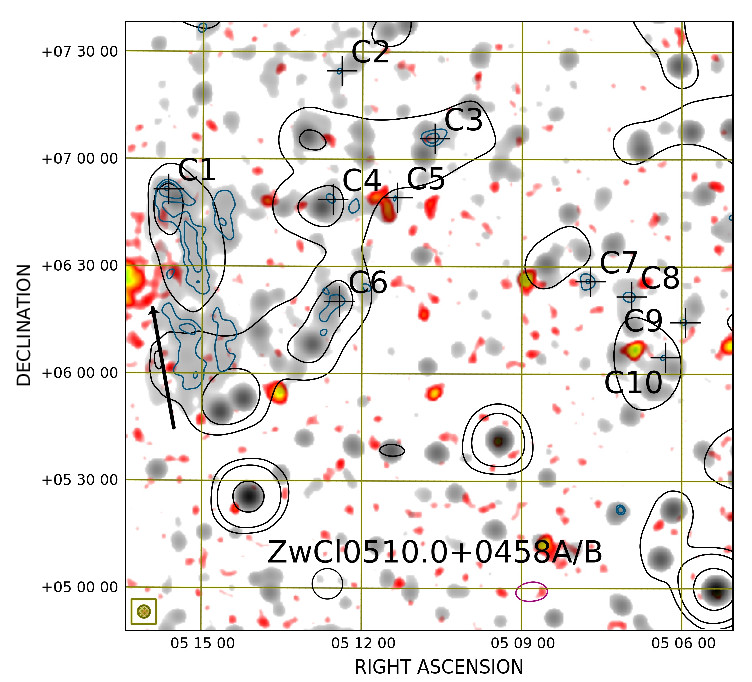}
\includegraphics[width=0.35\textwidth, angle=0]{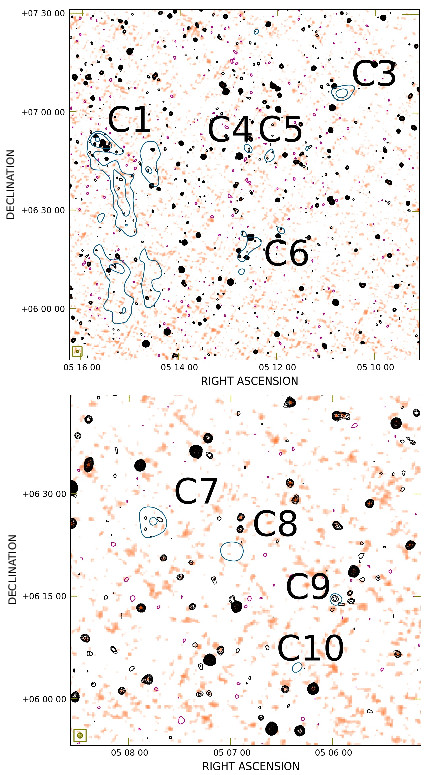}\\
\end{center}
\caption{{\bf Region C.} \emph{Left panel:} SRT contours (-60\,mJy/beam,
  60\,mJy/beam, 150\,mJy/beam, 500\,mJy/beam, negative in magenta and positive
  in black; resolution 13.9\,$'\times$12.4\,$'$) and SRT+NVSS after compact sources subtraction contours (in blue, starting from 3$\sigma$ and the remaining increasing by a factor $\sqrt{2}$; $\sigma$=2.5\,mJy/beam, resolution 3.5$^{\prime}\times$3.5$^{\prime}$) overlaid on X-ray emission from the RASS in
  red colours and radio emission from the SRT+NVSS (resolution $3.5^{\prime}$)
  in gray colours. \emph{Right panels:} NVSS contours -3$\sigma$ (magenta, $\sigma$=0.45\,mJy/beam),
  3$\sigma$, and increasing by a factor $\sqrt{2}$ (black) overlaid on the
  TGSS in colours. The black arrow marks the position of A539. The colorbars are the same as in Fig.\,\ref{im1}.}
\label{im8}
 \end{figure*}
 
Moving from north to south of the full field of view, the first object is A523
(Fig.\,\ref{im1}), which hosts a powerful radio halo \citep{Giovannini2011}
also visible in the NVSS.  This system is known to be over-luminous in radio
with respect to the X-rays.  The SRT data indicate a large-scale emission with
an elongation to the west of A523. After subtracting the point sources in the
combined SRT+NVSS image, a residual diffuse emission emerges at the position
of A523 (source A1) and another 12$^{\prime}$ (source A2), west of A1. A
few patches of diffuse emission can be seen also on the north-west (source
A3), where hints of radio emission are already visible in the combined SRT+NVSS
image.  At the location of sources A2 and A3 no point source emission is
detected in the NVSS image.  X-ray emission is clearly visible toward A523
and to the north-west (see black arrow in the image), while only hints of emission
are present at the location of A2 and A3. South-west of A2, an X-ray signal is
detected (green arrow), whose diffuse nature is uncertain. By inspecting the high resolution NVSS and TGSS
images corresponding to the X-ray signal, we find hints of diffuse emission. In
the SRT+NVSS image after point source subtraction, a diffuse patch at
2$\sigma$ of significance is detected at the same spatial location.  

In the residual image, we measure a
flux of the radio halo in A523 of (44$\pm$5)\,mJy and a largest linear size
(LLS) of about 1.2\,Mpc (10$^\prime$).  As a comparison, \cite{Giovannini2011}
give a flux $S_{\rm 1.4\,GHz}=(59\pm5)\,$mJy, while \cite{Girardi2016} report
$S_{\rm 1.4\,GHz}=(72.3\pm0.6)\,$mJy.  The integrated flux measured in this
work is slightly weaker than {that} estimated by \cite{Giovannini2011}.  This
difference can be explained by the different procedures of point-sources
subtraction. The corresponding radio power at 1.4\,GHz is
1.33$\times$10$^{24}$W/Hz.  The shape and the size of the radio halo in the
galaxy cluster A523 from our images are very similar to those previously derived
from interferometric observations only. The source appears quite roundish with
a slight elongation in the direction perpendicular to the merger axis
(SSW-NNE), as derived by X-ray and optical data \citep{Girardi2016}.

\subsubsection{Region B}

\begin{figure*}[t]
\begin{center}
\includegraphics[width=0.6\textwidth, angle=0]{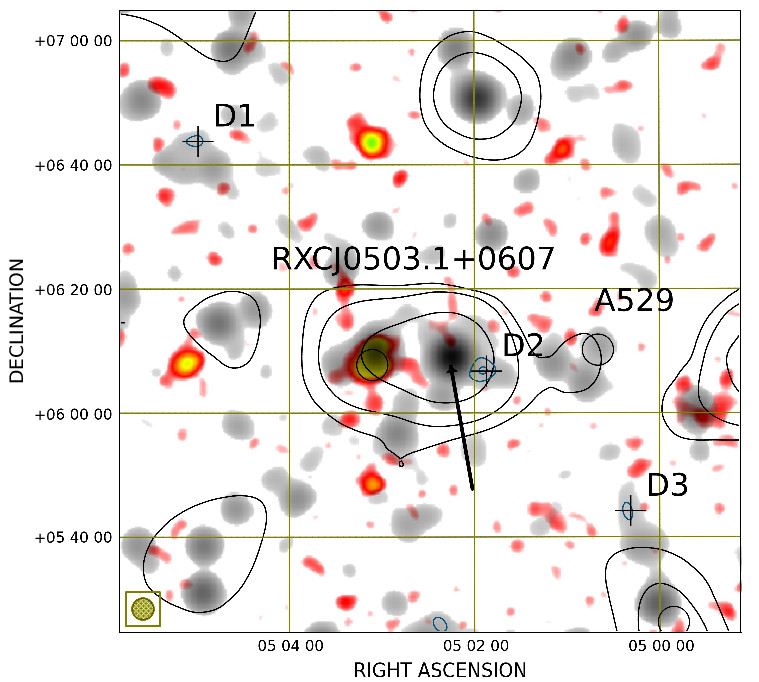}
\includegraphics[width=0.35\textwidth, angle=0]{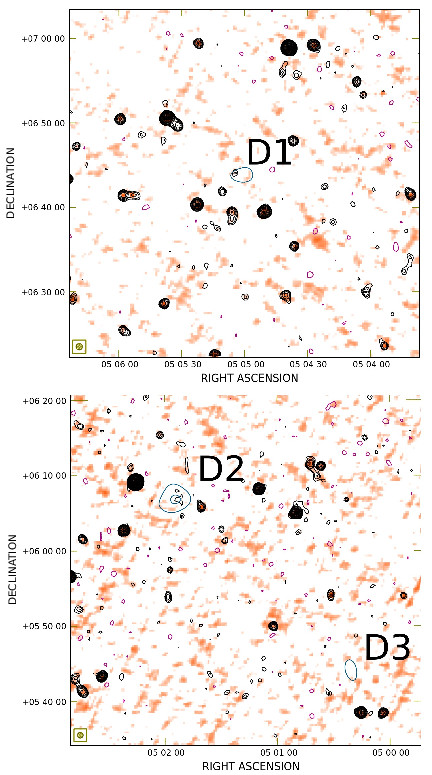}\\
\end{center}
\caption{{\bf Region D.} \emph{Left panel:} SRT contours (-60\,mJy/beam,
  60\,mJy/beam, 150\,mJy/beam, 500\,mJy/beam, negative in magenta and positive
  in black; resolution 13.9\,$'\times$12.4\,$'$) and SRT+NVSS after compact sources subtraction contours (in blue, starting from 3$\sigma$ and the remaining increasing by a factor $\sqrt{2}$; $\sigma$=2.5\,mJy/beam, resolution 3.5$^{\prime}\times$3.5$^{\prime}$) overlaid on X-ray emission from the RASS in
  red colours and radio emission from the SRT+NVSS (resolution $3.5^{\prime}$)
  in gray colours. \emph{Right panels:} NVSS contours -3$\sigma$ (magenta, $\sigma$=0.45\,mJy/beam),
  3$\sigma$, and increasing by a factor $\sqrt{2}$ (black) overlaid on the
  TGSS in colours. The black arrow marks the position of the quasar
  PKS\,0459+060. The colorbars are the same as in Fig.\,\ref{im1}.}
\label{im6}
\end{figure*}
In Fig.\,\ref{im4}, we show the region to the south-west of A523.  In the top
right corner of the image, an excess of radio emission is detected by the SRT
(source B1).  Some point sources can be identified in the outskirts of it but not embedded in the emission, excluding the possibility that this source is a
blending of point-like sources.  Two other diffuse patches of emission can be
identified in the field (source B2 and B3), not clearly associated with
point-like sources but rather located at positions where an excess in radio
emission is observed in the combined SRT+NVSS image \citep[refer
  to][for a similar case]{Macario2014}. At the location of Source B2 a hint
of diffuse emission is observed also in the NVSS and TGSS images.

According to the literature, no galaxy cluster is present in the field shown
in Fig.\,\ref{im4}. However, high resolution NVSS and TGSS data
(Fig.\,\ref{SRT_TGSS_cluster}) show that the emission detected by the SRT at the
centre of the panel, south-east of B2, is a blending of compact radio sources
that might be part of a galaxy cluster.  In a region of radius 20$^{\prime}$
centred on the radio galaxy indicated by the arrow in Fig.\,\ref{im4}, only the source
2MASX\,J04534530+0719303 located at RA 04h53m45.3s and Dec +07d19m31s has a
spectroscopic redshift measurement \citep[z=0.104,][]{Rines2003}. These
galaxies could be part of a galaxy cluster or a group of galaxies at the same
redshift of the filament.

\subsubsection{Region C}
\begin{figure*}
\begin{center}
\includegraphics[width=0.6\textwidth, angle=0]{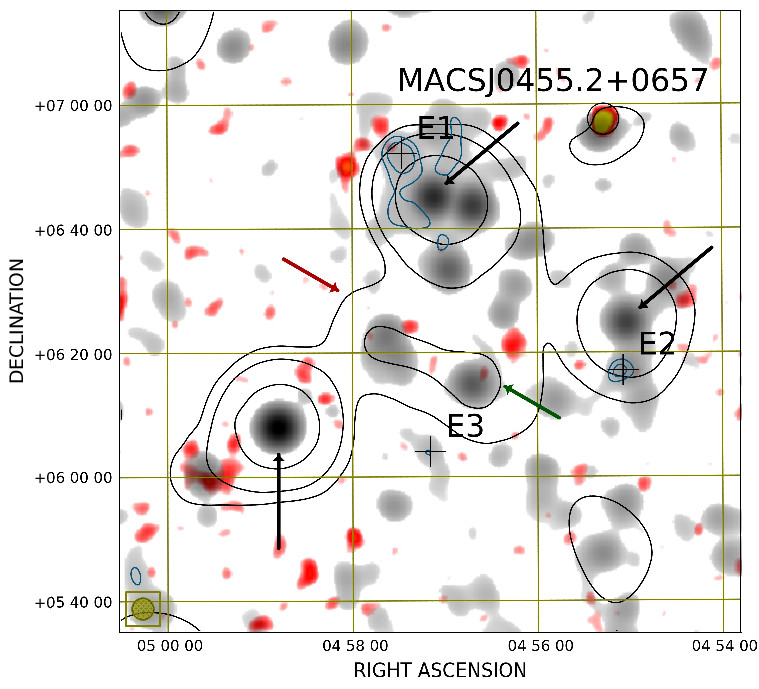}
\includegraphics[width=0.35\textwidth, angle=0]{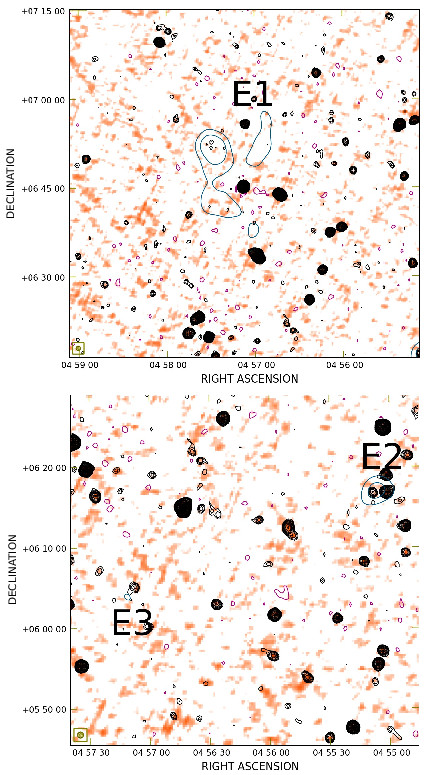}\\
\end{center}
\caption{{\bf Region E.} \emph{Left panel:} SRT contours (-60\,mJy/beam,
  60\,mJy/beam, 150\,mJy/beam, 500\,mJy/beam, negative in magenta and positive
  in black; resolution 13.9\,$'\times$12.4\,$'$) and SRT+NVSS after compact sources subtraction contours (in blue, starting from 3$\sigma$ and the remaining increasing by a factor $\sqrt{2}$; $\sigma$=2.5\,mJy/beam, resolution 3.5$^{\prime}\times$3.5$^{\prime}$) overlaid on X-ray emission from the RASS in
  red colours and radio emission from the SRT+NVSS (resolution $3.5^{\prime}$)
  in gray colours. \emph{Right panels:} NVSS contours -3$\sigma$ (magenta, $\sigma$=0.45\,mJy/beam),
  3$\sigma$, and increasing by a factor $\sqrt{2}$ (black) overlaid on the
  TGSS in colours. The black arrows mark the position of 4C\,+06.21,
  PKS\,0456+060, and LQAC\,073+006\,001, and the red arrow
  the faint large-scale emission SRT emission in between these three
  bright compact sources. The green arrow marks the location where signal is
  clearly visible in the SZ and  X-ray images while in the SRT+NVSS after
  compact sources subtraction is is detected only with a 2$\sigma$ significance. The colorbars are the same as in Fig.\,\ref{im1}.}
\label{im2}
 \end{figure*}

East of the filament, another region of interest can be identified (see
Fig.\,\ref{im8}) on the west of the galaxy cluster A539, whose position is
indicated by the arrow in the same Figure.  The cluster appears to be an
emitter in the X-ray and millimetre/sub-millimetre domain but, unfortunately,
it is outside the field of view of our SRT observations and only the regions to
its west has been mapped. In the SRT image, we see a large arm extending from
the cluster westward, still visible in the residual image as a large diffuse structure.  This structure can not be explained as the blending
of radio galaxies in the field, so it is likely related to a large-scale
diffuse source.  After the subtraction of point sources, ten patches (Sources
from C1 to C10) remain.  Apart from C1, C6 and C9 could be at least
partially the leftovers of the point-like source subtraction process, the
remaining sources seem to be large-scale diffuse synchrotron sources.  Among
these, the largest and more interesting is C1. Hints of diffuse large-scale
emission at the same spatial location are present in the SRT+NVSS and in the
X-ray image. A slight excess of large-scale radio emission can be identified
as well at higher resolution in the NVSS. However, since it is located at the
edge of the image, more observations are required to confirm it and
investigate its nature.

\subsubsection{Region D}
\begin{figure*}
\begin{center}
\includegraphics[width=0.6\textwidth, angle=0]{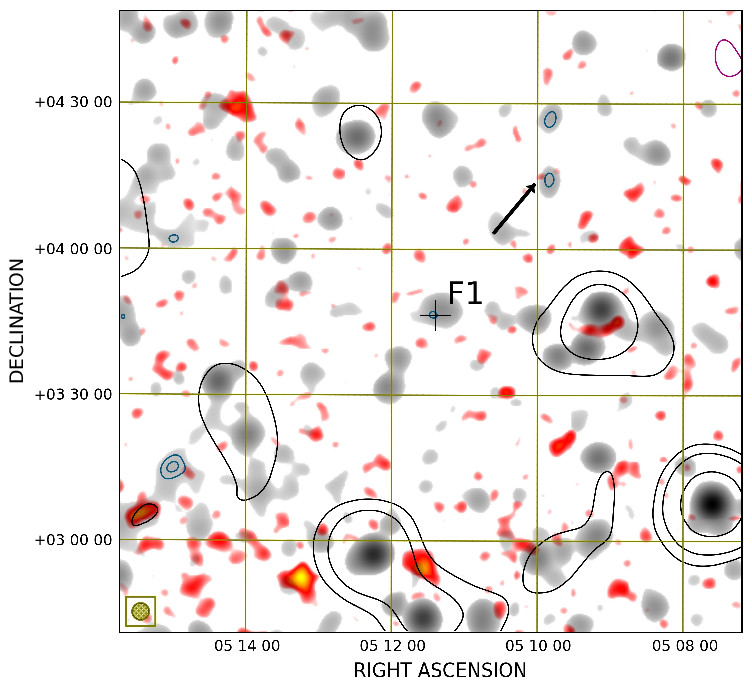}
\includegraphics[width=0.35\textwidth, angle=0]{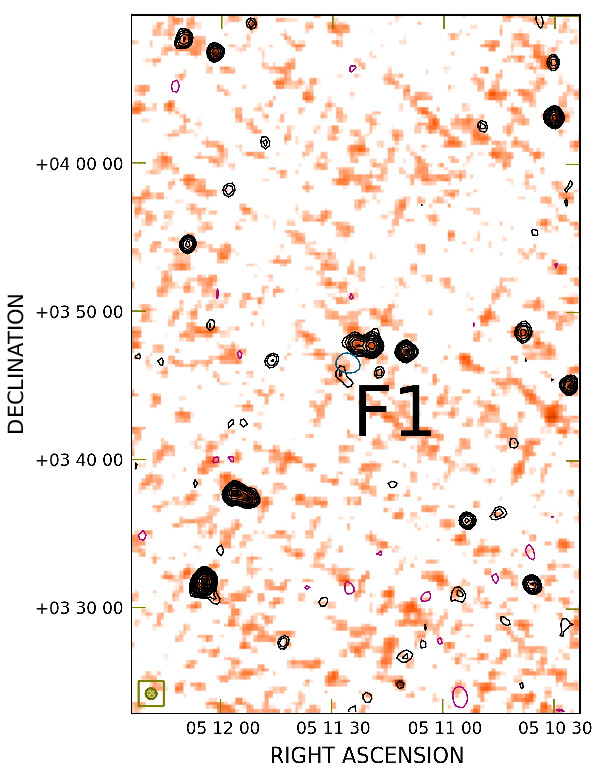}\\
\end{center}
\caption{{\bf Region F.} \emph{Left panel:} SRT contours (-60\,mJy/beam,
  60\,mJy/beam, 150\,mJy/beam, 500\,mJy/beam, negative in magenta and positive
  in black; resolution 13.9\,$'\times$12.4\,$'$) and SRT+NVSS after compact sources subtraction contours (in blue, starting from 3$\sigma$ and the remaining increasing by a factor $\sqrt{2}$; $\sigma$=2.5\,mJy/beam, resolution 3.5$^{\prime}\times$3.5$^{\prime}$) overlaid on X-ray emission from the RASS in
  red colours and radio emission from the SRT+NVSS (resolution $3.5^{\prime}$)
  in gray colours. \emph{Right panel:} NVSS contours -3$\sigma$ (magenta, $\sigma$=0.45\,mJy/beam),
  3$\sigma$, and increasing by a factor $\sqrt{2}$ (black) overlaid on the
  TGSS in colours. The black arrow marks the position of the candidate giant
  radio galaxy we found (see \S\,\ref{RRG}). The colorbars are the same as in Fig.\,\ref{im1}.}
\label{imf}
 \end{figure*}
The complex RXC\,J0503.1+0607 - A529 is located at the centre of the panel
shown in Fig.\,\ref{im6} and the galaxy cluster ZwCl\,0459.6+0606 sits between
them (see Table\,\ref{tab:noz}). In the bottom right corner of the image, the
galaxy cluster A526 is also partially visible.  In the X-rays, only
RXC\,J0503.1+0607 shows a signal above the noise level. In radio, both
RXC\,J0503.1+0607 and A529 can be seen: RXC\,J0503.1+0607 is a strong
emitter, while A529 is characterized by a faint signal. The peak observed in
the SRT image between the two clusters is approximately at the same angular
location as ZwCl\,0459.6+0606 (see arrow in Fig.\,\ref{im6}) and is due to the
quasar PKS\,0459+060 (RA 05h02m15.4s, Dec +06d09m07s, z=1.106). The
superposition of the SRT and the SRT+NVSS images, reveals the presence of
several point sources in the field, as well as hints of large-scale emission
in between the two clusters. This indication is confirmed by the residual
image, where patches of large-scale diffuse emission are present (D1, D2, and
D3).  Source D1 in the top left corner is located in a region where the
SRT+NVSS image reveals a large-scale patch of radio emission.  Source D2 does
not show a connection with point-like sources in the field and appear located
at the periphery of ZwCl\,0459.6+0606. Indications of the presence of a
large-scale signal at locations of D1 and D2 can be seen in the NVSS
image.  A puzzling patch of diffuse emission is observed to the south-west
(source D3), in the outskirts of A526, without a clear association with any point
sources but rather corresponding to hints of diffuse emission in
the SRT+NVSS images.

\subsubsection{Region E}
\begin{figure*}
\begin{center}
\includegraphics[width=0.6\textwidth, angle=0]{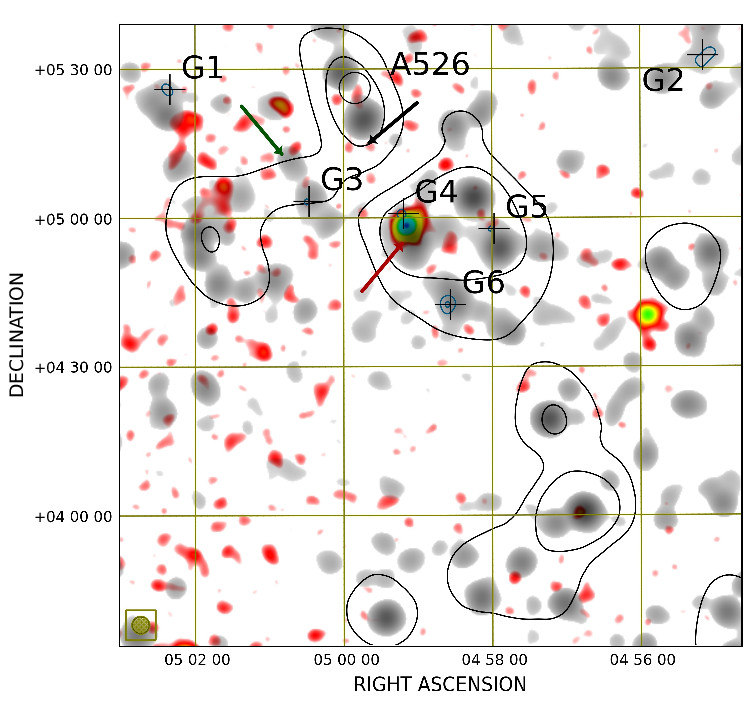}
\includegraphics[width=0.35\textwidth, angle=0]{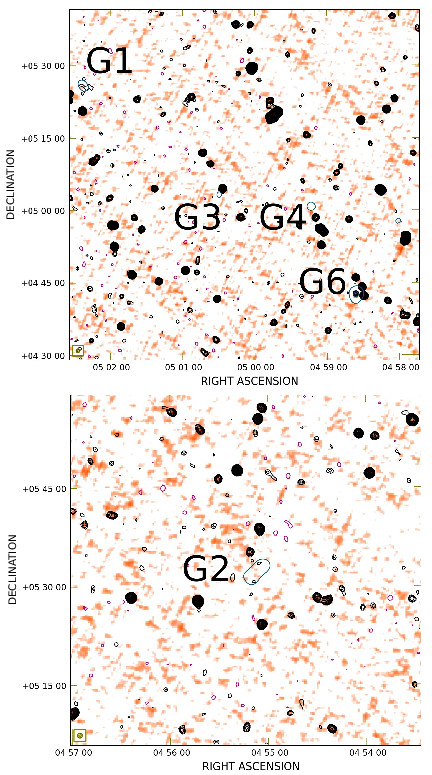}\\
\end{center}
\caption{{\bf Region G.} \emph{Left panel:} SRT contours (-60\,mJy/beam,
  60\,mJy/beam, 150\,mJy/beam, 500\,mJy/beam, negative in magenta and positive
  in black; resolution 13.9\,$'\times$12.4\,$'$) and SRT+NVSS after compact sources subtraction contours (in blue, starting from 3$\sigma$ and the remaining increasing by a factor $\sqrt{2}$; $\sigma$=2.5\,mJy/beam, resolution 3.5$^{\prime}\times$3.5$^{\prime}$) overlaid on X-ray emission from the RASS in
  red colours and radio emission from the SRT+NVSS (resolution $3.5^{\prime}$)
  in gray colours. \emph{Right panels:} NVSS contours -3$\sigma$ (magenta, $\sigma$=0.45\,mJy/beam),
  3$\sigma$, and increasing by a factor $\sqrt{2}$ (black) overlaid on the
  TGSS in colours. The black and red arrow mark respectively the locations of
  the galaxy cluster ZwCl\,0457.0+0511 and of the compact source
  PMN\,J0459+0455, while the green arrow a bridge of radio emission detected
  by SRT. The colorbars are the same as in Fig.\,\ref{im1}.}
\label{im5}
 \end{figure*}
In Fig.\,\ref{im2}, another interesting complex can be identified 2$^{\circ}$
south of A523, south-east of the galaxy cluster MACS\,J0455.2+0657. In this
region, MACS\,J0455.2+0657 is the only galaxy cluster with a spectroscopic
redshift measurement ($z=0.425$). In the field of view of Fig.\,\ref{im2}, the
SRT image reveals three bright peaks (see black arrows in Fig.\,\ref{im2}) and
a faint large-scale emission in between them (red arrow). At higher
resolution, several point sources can be identified. Three of them have the
same spatial location as the three peaks detected by the SRT: 4C\,+06.21
\cite[RA 04h57m07.7099s, Dec +06d45m07.260s, z=0.405,][]{Drinkwater1997},
PKS\,0456+060 \cite[RA 04h58m48.782s, Dec
  +06$^{\circ}$08$^{\prime}$04.26$^{\prime\prime}$, z= 1.08,][]{Veron1996},
and LQAC\,073+006\,001 (RA 04h55m03.1s, Dec
+06$^{\circ}$24$^{\prime}$54$^{\prime\prime}$, no redshift identification).
After point source subtraction, three interesting large-scale patches
emerge. One off the bright radio source 4C\,+06.21 (source E1), one some
10$^{\prime}$ south of LQAC\,073+006\,001 (source E2), and one south of the
central radio emission (source E3), where no emission appears in the SRT image
above the noise level.  The source E1 does not appear to be a blending of
point-like sources as shown by the superposition with higher resolution data
in Fig.\,\ref{im2} and hints of large-scale diffuse emission are already
present in the NVSS image. E1 is characterized by an average brightness of
$\approx$\,0.13\,$\mu$Jy/arcsec$^2$, its integrated flux density is (109 $\pm$
10)\,mJy, and its largest angular size is 25$^{\prime}$. If we assume that this
emission is at the same redshift as the radio source 4C\,+06.21 (z=0.405), its
LLS is 8.7\,Mpc and its radio power at 1.4\,GHz is
6.8$\times$10$^{25}$\,W/Hz. However, if we assume that this source is located at
the average redshift of the filament ($z\approx 0.1$), its LLS is about 3\,Mpc
and its radio power at 1.4\,GHz is 3.0$\times$10$^{24}$\,W/Hz.  
No strong X-ray emission is present in the field except that seen to the north-east of the radio source E1.
Source E3 is
located at position where the combined NVSS indicates the presence of an
excess of large-scale diffuse emission, while the source E2 could be the
residual of point-like emission.    
At the
centre of the panel, at coordinates RA 04h56m25.24s, Dec
+06$^{\circ}$13$^{\prime}$29$^{\prime\prime}$ (see green arrow), a
2$\sigma$-level signal is observed in the residual radio image. Since it is
below the 3$\sigma$ significance threshold, we did not consider it in our
analysis. However, at the same spatial location, a faint X-ray emission and a
SZ signal (see Fig.\,\ref{srt_all}) are present, suggesting that these
emissions could be real and not a statistical fluctuation.

\subsubsection{Region F}
\begin{figure*}
\begin{center}
\includegraphics[width=0.6\textwidth, angle=0]{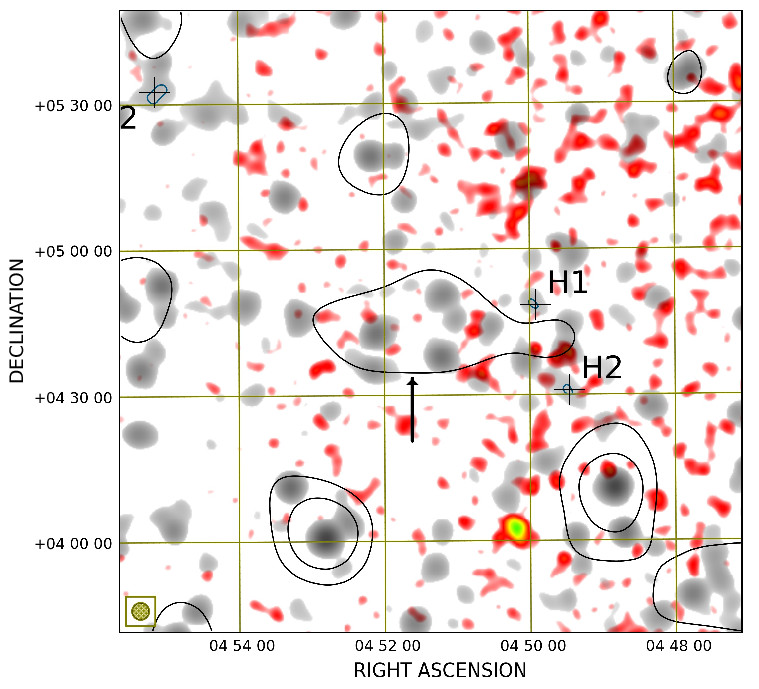}
\includegraphics[width=0.35\textwidth, angle=0]{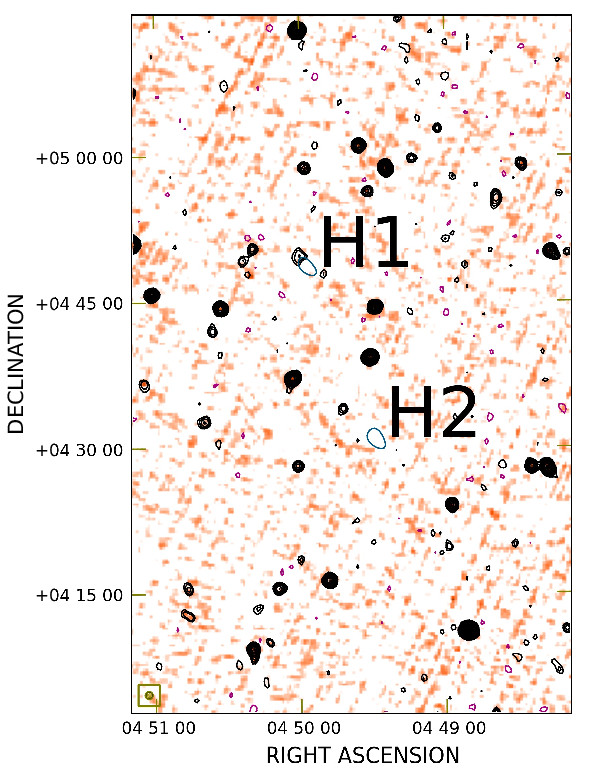}\\
\end{center}
\caption{{\bf Region H.} \emph{Left panel:} SRT contours (-60\,mJy/beam,
  60\,mJy/beam, 150\,mJy/beam, 500\,mJy/beam, negative in magenta and positive
  in black; resolution 13.9\,$'\times$12.4\,$'$) and SRT+NVSS after compact sources subtraction contours (in blue, starting from 3$\sigma$ and the remaining increasing by a factor $\sqrt{2}$; $\sigma$=2.5\,mJy/beam, resolution 3.5$^{\prime}\times$3.5$^{\prime}$) overlaid on X-ray emission from the RASS in
  red colours and radio emission from the SRT+NVSS (resolution $3.5^{\prime}$)
  in gray colours. \emph{Right panel:} NVSS contours -3$\sigma$ (magenta, $\sigma$=0.45\,mJy/beam),
  3$\sigma$, and increasing by a factor $\sqrt{2}$ (black) overlaid on the
  TGSS in colours. The black arrow marks a blending of point-like sources. The colorbars are the same as in Fig.\,\ref{im1}.}
\label{im9}
 \end{figure*}
 
In the region shown in Fig.\,\ref{imf} no galaxy cluster is identified in
{\b the} literature. The field is rich in point-like radio sources as shown by the
SRT+NVSS image. The SRT+NVSS image after point source subtraction shows five
patches of diffuse emission. The two on the left side are not considered here,
since they are at the edge of the image and might be artefacts. The two spots
in the top right region reveal an extended radio galaxy (see arrow in the
image) that we classify as a candidate new giant radio galaxy and discuss in Appendix\,\ref{RRG}. Our interest here is the source F1 located at the centre of the image. This source is located
close-by a point-like source, spatially coincident with a fainter large-scale
region visible also in the NVSS image.

\subsubsection{Region G}

\begin{figure*}
\begin{center}
\includegraphics[width=0.6\textwidth, angle=0]{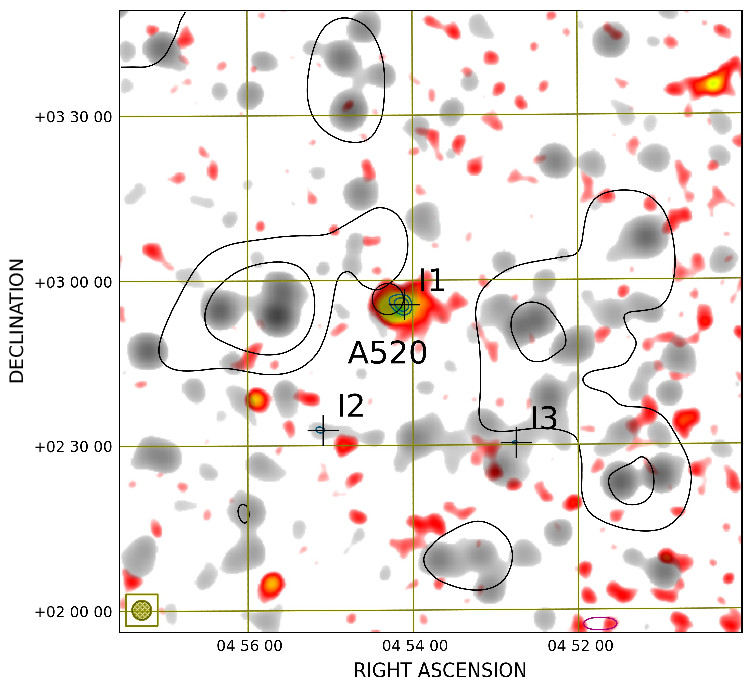}
\includegraphics[width=0.35\textwidth, angle=0]{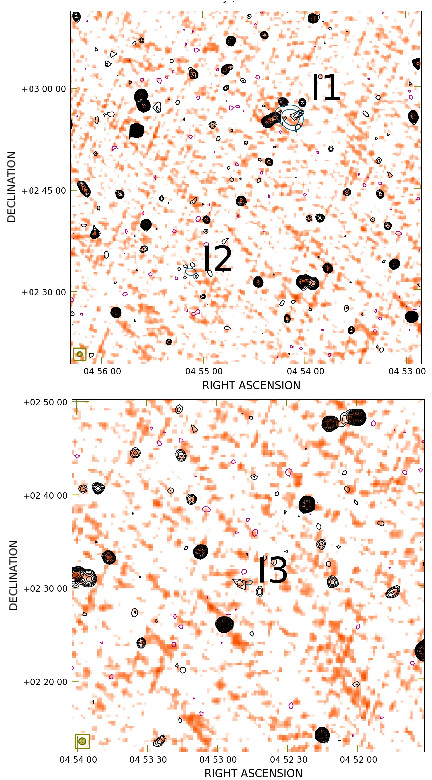}\\
\end{center}
\caption{{\bf Region I.} \emph{Left panel:} SRT contours (-60\,mJy/beam,
  60\,mJy/beam, 150\,mJy/beam, 500\,mJy/beam, negative in magenta and positive
  in black; resolution 13.9\,$'\times$12.4\,$'$) and SRT+NVSS after compact sources subtraction contours (in blue, starting from 3$\sigma$ and the remaining increasing by a factor $\sqrt{2}$; $\sigma$=2.5\,mJy/beam, resolution 3.5$^{\prime}\times$3.5$^{\prime}$) overlaid on X-ray emission from the RASS in
  red colours and radio emission from the SRT+NVSS (resolution $3.5^{\prime}$)
  in gray colours. \emph{Right panels:} NVSS contours -3$\sigma$ (magenta, $\sigma$=0.45\,mJy/beam),
  3$\sigma$, and increasing by a factor $\sqrt{2}$ (black) overlaid on the
  TGSS in colours. The colorbars are the same as in Fig.\,\ref{im1}.}
\label{im3}
 \end{figure*}

The galaxy cluster A526 is shown in the top of Fig.\,\ref{im5}. The cluster
is visible in radio and X-rays, but shows no counterpart at
millimetre/sub-millimetre wavelengths.  To the south-west, large-scale emission is
detected with the SRT. No galaxy clusters have been identified in this region
apart from the galaxy cluster ZwCl\,0457.0+0511 (see black arrow in the
image), south-west of A526 at an angular offset of about 10$^{\prime}$. Two
strong X-ray sources are seen: one south of A526, approximately
coincident with the radio source PMN\,J0459+0455 (RA 04h59m04.7s, Dec
+04d55m59s, see red arrow) and the other in the west with no obvious radio
counterpart.  After subtracting the point source emission, several patches of
diffuse emission emerge. Source G1 and G2, respectively in the top left and
right corners, have the same spatial location of an excess in the NVSS image.
Another patch of residual radio emission is present in the north-east of the image (source
G3), close-by to a point source surrounded by a fainter large-scale signal and
where the SRT image shows a bridge of radio emission (see green arrow)
connecting A526 with an over-density of radio galaxies south-east of A526.  At
the centre of the panel, three patches of diffuse emission are present corresponding to
 the large-scale central SRT radio emission: sources G4, G5,
G6. Sources G4 and G5 appear to be diffuse sources, while source G6 could be a
residual of the point source located below.

\subsubsection{Region H}
This region (Fig.\,\ref{im9}) is full of point-like sources as shown by the
SRT+NVSS image. In the SRT image a central patch likely due to the blending of
point-like sources is present (see arrow in the image). Above and below this
emission, two patches of residual radio signal (sources H1 and H2) are
identified that do not appear to be directly linked to point-like sources. In
particular, at the location of H1, an indication of diffuse emission is detected
also in the NVSS image.

\subsubsection{Region I}
\begin{figure*}
\begin{center}
\includegraphics[width=0.6\textwidth, angle=0]{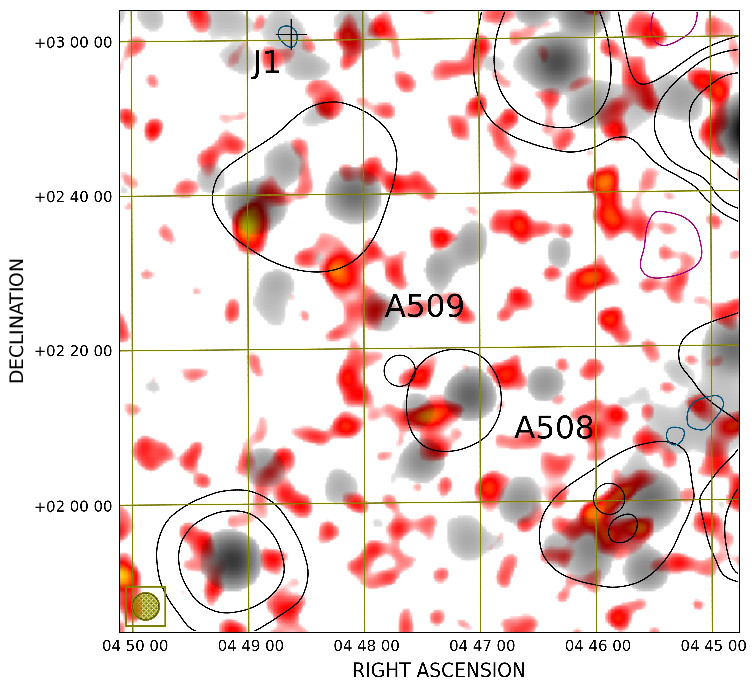}
\includegraphics[width=0.35\textwidth, angle=0]{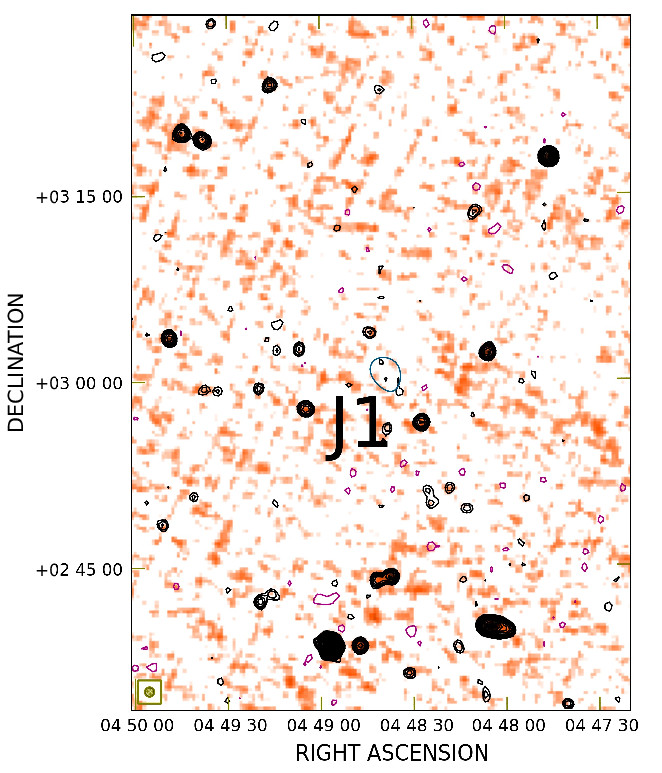}\\
\end{center}
\caption{{\bf Region J.} \emph{Left panel:} SRT contours (-60\,mJy/beam,
  60\,mJy/beam, 150\,mJy/beam, 500\,mJy/beam, negative in magenta and positive
  in black; resolution 13.9\,$'\times$12.4\,$'$) and SRT+NVSS after compact sources subtraction contours (in blue, starting from 3$\sigma$ and the remaining increasing by a factor $\sqrt{2}$; $\sigma$=2.5\,mJy/beam, resolution 3.5$^{\prime}\times$3.5$^{\prime}$) overlaid on X-ray emission from the RASS in
  red colours and radio emission from the SRT+NVSS (resolution $3.5^{\prime}$)
  in gray colours. \emph{Right panel:} NVSS contours -3$\sigma$ (magenta, $\sigma$=0.45\,mJy/beam),
  3$\sigma$, and increasing by a factor $\sqrt{2}$ (black) overlaid on the
  TGSS in colours. The colorbars are the same as in Fig.\,\ref{im1}.}
\label{im7}
 \end{figure*}

\begin{figure*}
\begin{center}
\includegraphics[width=0.45\textwidth, angle=0]{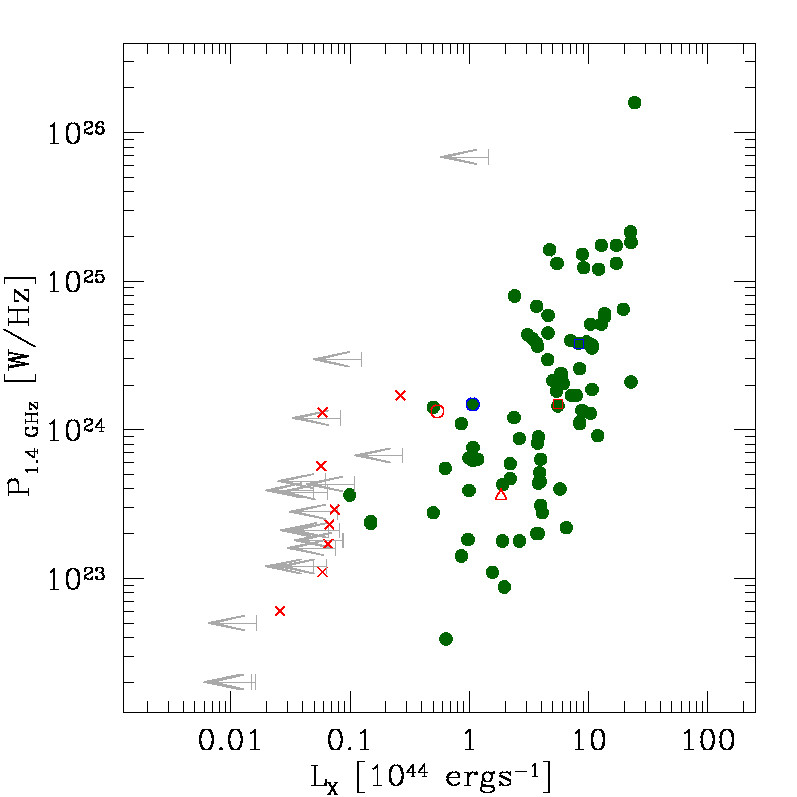}\includegraphics[width=0.45\textwidth, angle=0]{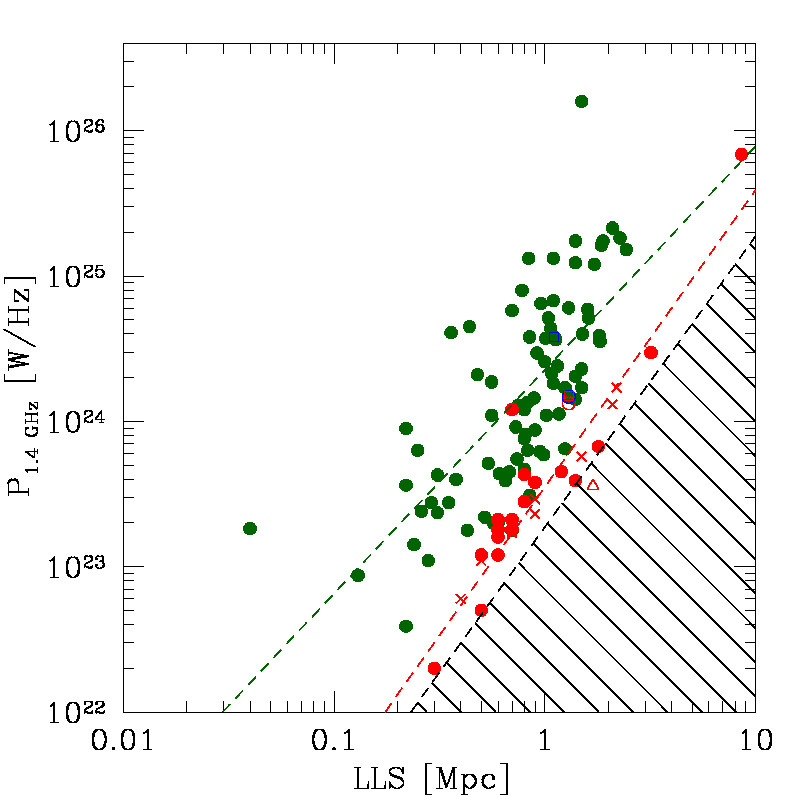}\\
\end{center}
\caption{\emph{Left panel}: Radio power at 1.4\,GHz $P_{\rm 1.4\,GHz}$ versus X-ray luminosity in the energy band 0.1-2.4\,keV $L_{\rm X, 0.1-2.4\,keV}$. \emph{Right panel}: Radio power $P_{\rm 1.4\,GHz}$ versus LLS both at 1.4\,GHz. 
Green dots and blue open symbols are from \citet{Feretti2012}. Red symbols are 
measurements from this work: open symbols identify A523 (circle), A520
(square), G4 (triangle), while crosses identify the remaining sources with an
X-ray identification. Grey arrows are upper limits from this work. The shaded area in the
right panel indicates the region of the ($P_{\rm 1.4\,GHz}$, LLS) plane that
cannot be accessed considering the 2$\sigma$ ($\sigma$=2.5\,mJy/beam) sensitivity of the SRT+NVSS image after point source subtraction. Dashed lines represent a linear fit in logarithmic scale of Eq.\,\ref{corr} for cluster sources (green) and for the new sources (red).}
\label{halos}
 \end{figure*}

\begin{figure}
\begin{center}
\includegraphics[width=0.45\textwidth, angle=0]{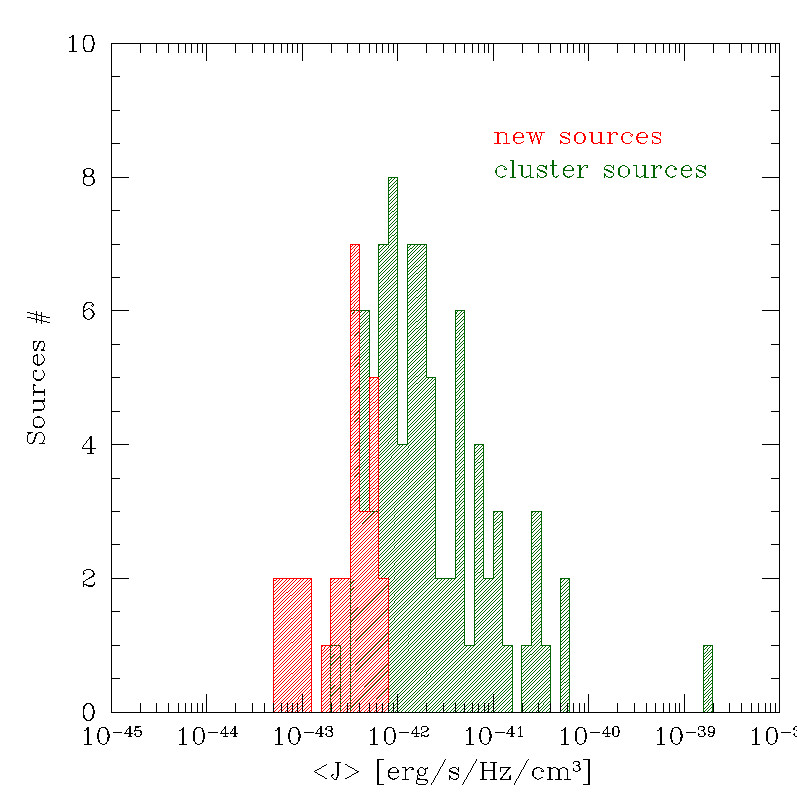}
\end{center}
\caption{Histogram in logarithmic scale of the mean emissivity of the new sources presented in this paper (red) compared to the cluster diffuse radio sources from \citet{Feretti2012} (green).}
\label{emissivity}
 \end{figure}
At the centre of Fig.\,\ref{im3} the galaxy cluster A520 is present, located
in the south of the full field of view. Strong X-ray emission is spatially
coincident with this cluster, that appears to be very bright at
millimetre/sub-millimetre wavelengths (see Fig.\,\ref{srt_all}).  Hints of
diffuse emission are present in the NVSS in the direction of A520 (where
indeed a radio halo is present), east, south and south-west of the cluster.
After the point source subtraction in the SRT+NVSS image, patches of diffuse
large-scale emission remain at these locations and the radio halo in A520
(source I1) is now clearly visible.  In the residual image blanked at the
2$\sigma$ level, we measure an integrated flux of (12 $\pm$ 3)\,mJy and a LLS
of about 1\,Mpc for this radio halo. The corresponding radio power at 1.4\,GHz
is 1.48$\times$10$^{24}$\,W/Hz. As a comparison, the values reported in
the literature are (16.7$\pm$0.6)\,mJy and 1.05\,Mpc \citep{Govoni2001,Vacca2014}
which are very consistent with ours.  South of A520, the sources I2 and I3 can be
observed. They do not show a clear link with radio galaxies in the field, on
the contrary they are located where hints of
diffuse radio emission are present in the NVSS image, so they could be real
large-scale diffuse synchrotron sources. The present observations of the radio
halo in the galaxy cluster A520 confirm what was suggested by the data at
higher resolution.  The source is quite complex and with a strong elongation
in the NE-SW direction. The peak of the emission is located in the SW at the
same location as the shock front \citep{Markevitch2005,Govoni2004}, where a
sharp drop in radio surface brightness is detected. A faint tail of radio
emission is left behind in the cool region.  As already suggested by
\cite{Govoni2001} and \cite{Vacca2014}, this peculiar source could actually be
a peripheral source seen in projection.

\subsubsection{Region J}

The complex A508-A509 sits in the south-east corner of the full field of view and is shown in Fig.\,\ref{im7}. The two clusters are also visible in X-rays and at millimetre/sub-millimetre wavelengths but they are quite faint. Patches of
diffuse emission not coincident with point-like sources survive in the
residual image. Source J1 is located where an excess of radio emission is
observed in the NVSS image. The other patches are located at the edge of the
image and are therefore hard to discriminate if they are real or artefacts,
even though the SRT+NVSS image clearly shows a patch of diffuse emission
before compact-source subtraction.

\begin{table*}
        \caption{Diffuse large-scale emission properties of sources detected above 3$\sigma$}
\begin{center}
        \begin{tabular}{ccccccccc}
          \hline
          \hline
  Source     &   RA (J2000) & Dec (J2000)&   z &  $S_{\rm residual}$   & $P_{\rm1.4\,GHz}$& LLS & Class& Alternative name\\ 
             &   h:m:s      & d:m:s      & &    mJy             &  $10^{23}$W/Hz & Mpc &         &        \\
\hline
A1  & 04:59:08.81 & +08d48:52 & 0.104 & 45 $\pm$ 5  &13.3  & 1.3  & &Radio halo A523\\
A2  & 04:57:43.81 & +08d47:03 & 0.104 & 10 $\pm$ 3  &2.9   & 0.9  &&\\
A3  & 04:56:23.85 & +09d27:59 & 0.104 & 44 $\pm$ 7  &13.1  & 2.1  & &\\
B1  & 04:49:29.06 & +08d30:16 & 0.104 & 100$\pm$ 10 & 29.7 & 3.2  &&\\
B2  & 04:53:19.21 & +07d48:11 & 0.100 & 4  $\pm$ 2  & 1.2  & 0.5  &&\\
B3  & 04:51:39.15 & +07d15:01 & 0.100 & 5  $\pm$ 2  & 1.2  & 0.6  &&\\
C1  & 05:15:39.81 & +06d51:47 & 0.029 & 554$\pm$ 27 &11.6  & 2.6  &*&\\
C2  & 05:12:24.80 & +07d25:01 & 0.029 & 8  $\pm$ 3  &0.2   & 0.3  &&\\
C3  & 05:10:39.64 & +07d06:07 & 0.029 & 28 $\pm$ 5  &0.6   & 0.4  &&\\
C4  & 05:12:34.29 & +06d49:01 & 0.029 & 23 $\pm$ 5  &0.5   & 0.5  &&\\
C5  & 05:11:21.76 & +06d49:35 & 0.029 & 10 $\pm$ 3  &0.2   & 0.3  &&\\
C6  & 05:12:26.81 & +06d20:31 & 0.029 & 55 $\pm$ 7  &1.2   & 0.8  &*&\\
C7  & 05:07:44.04 & +06d26:13 & 0.100 & 16 $\pm$ 4  &4.3   & 0.8  &&\\
C8  & 05:06:57.73 & +06d21:59 & 0.100 & 21 $\pm$ 5  &5.7   & 1.5  &&\\
C9  & 05:05:57.34 & +06d14:45 & 0.100 & 4  $\pm$ 2  &1.1   & 0.5  &*&\\
C10 & 05:06:19.45 & +06d04:59 & 0.100& 7  $\pm$ 3  &2.1   & 0.7  &&\\
D1  & 05:05:00.00 & +06d44:00 & 0.100 & 6  $\pm$ 2  &1.8   & 0.7  &&\\
D2  & 05:01:52.93 & +06d06:57 & 0.100 & 17 $\pm$ 4  &4.5   & 1.2  &&\\
D3  & 05:00:19.57 & +05d44:24 & 0.100 & 14 $\pm$ 4  &3.9   & 1.4  &&\\
E1  & 04:57:26.67 & +06d52:01 & 0.405 & 109$\pm$ 10 &684.7 & 8.6  &&\\
E2  & 04:55:05.24 & +06d17:21 & 0.100 & 12 $\pm$ 3  &3.4   & 0.8  &*&\\
E3  & 04:57:10.28 & +06d04:15 & 0.100 & 4  $\pm$ 2  &1.1   & 0.5  &&\\
F1  & 05:11:24.89 & +03d46:42 & 0.100 & 7  $\pm$ 3  &12.0  & 0.7  &&\\
G1  & 05:02:21.28 & +05d26:12 & 0.100 & 6  $\pm$ 2  &1.6   & 0.6  &&\\
G2  & 04:55:03.01 & +05d33:20 & 0.100 & 14 $\pm$ 4  &3.8   & 0.9  &&\\
G3  & 05:00:28.93 & +05d03:38 & 0.100 & 6  $\pm$ 3  &1.7   & 0.7  &&\\
G4  & 04:59:12.63 & +05d01:05 & 0.100 & 13 $\pm$ 4  &3.6   & 1.7  &&\\
G5  & 04:57:59.36 & +04d58:01 & 0.100 & 8  $\pm$ 3  &2.3   & 0.9  &&\\
G6  & 04:58:34.65 & +04d42:47 & 0.100 & 15 $\pm$ 4  &4.0   & 1.2  &*&\\
H1  & 04:49:56.16 & +04d48:46 & 0.100 & 7  $\pm$ 3  &1.8   & 0.6  &&\\
H2  & 04:49:28.39 & +04d31:12 & 0.100 & 10 $\pm$ 3  &2.8   & 0.8  &&\\
I1  & 04:54:06.90 & +02d55:46 & 0.199 & 12 $\pm$ 3  &14.9  & 1.3  && Radio halo A520\\
I2  & 04:55:06.23 & +02d33:02 & 0.199 & 5  $\pm$ 2  &6.7   & 1.8  &&\\
I3  & 04:52:45.95 & +02d30:33 & 0.199 & 14 $\pm$ 4  &17.0  & 2.2  &&\\
J1  & 04:48:37.81 & +03d00:55 & 0.100 & 8  $\pm$ 3  & 2.1  & 0.6  &&\\

            \hline
            \hline
        \end{tabular}
        \label{tab_diff}\\
\end{center}

Col 1: Source label; Col 2 \& 3: source coordinates; Col 4: redshift; 
Col 5: Flux density at 1.4\,GHz of the residual diffuse emission from the SRT+NVSS combined image after the point source-subtraction; Col 6: Radio power at 1.4\,GHz of the diffuse emission; Col 7: Largest linear size of the diffuse emission; Col 8: An asterisk indicates that the source could be a leftover of compact sources after the subtraction process or an artefact; Col 9: Alternative name of the source.\\

 \end{table*}

\section{Discussion}
\label{Discussion}

After subtracting the point-source emission, we identified 35 patches of
diffuse synchrotron emission in the residual SRT+NVSS image. Two of them are
the radio halos in the galaxy clusters A520 and A523, and five are probably
artefacts or the leftover of point-like sources after the point-source
subtraction process. The remaining 28 are potentially new real
large-scale diffuse synchrotron sources, possibly associated with the
large-scale structure of the cosmic web. The bottom right panel of
Fig.\,\ref{srt_all} shows that most of the new sources presented in this
paper lie along the filament connecting the clusters from A523-A525 in the
north to A508-A509 in the south, with an over-density at the centre of the
field of view, at the location of RXC\,J0503.1+0607, A529 and A526. In the
following, we restrict our analysis to the properties of the 28 new detections
and the two radio halos already known. We excluded sources that could be
artefacts or a remnant of compact sources after the subtraction procedure. To
better investigate their properties, we measured their radio powers and sizes
and their X-ray {luminosities}, summarized in Table\,\ref{tab_diff} and in
Table\,\ref{tab_diff_xray} respectively. The X-ray fluxes have been estimated
by following the procedure described in \S\,\ref{x-ray}. Only eleven sources
have a significance of the X-ray count rate above 1$\sigma$. These sources are
A1 (diffuse emission in A523), A2, A3, C3, C8, E3, G3, G4, G5, I1 (diffuse
emission in A520), and I3.

Some of the sources presented here are potentially interesting and complex
systems.  For example, source A2 is a faint large-scale source 10$^{\prime}$
west of the radio halo in A523 (source A1). It has a radio flux at 1.4\,GHz
$S_{\rm residual}=(9\pm 3)$\,mJy ($P_{\rm 1.4\,GHz}=2.8\times 10^{23}$\,W/Hz),
a LLS of 1.3\,Mpc, and an X-ray luminosity $L_{\rm X 0.1-2.4\,keV}=(0.7\pm
0.6)\times 10^{44}$\,erg/s. In the south-west of A2, another diffuse patch of
synchrotron emission is detected at 2$\sigma$ significance and at the same
spatial location of an X-ray signal (see \S\,\ref{RegionA}). These three sources could form an
arc-shaped filament of diffuse synchrotron emission.
If A2 and the source south-west of it are interpreted
to be radio halos, Source A1 - Source A2 and this third source could represent
the first case of a triple radio halo, which would 
be particularly interesting because of its association with an 
under-luminous X-ray system. To, date only one other multiple system is known, namely the double
radio halo discovered by \cite{Murgia2010} in the galaxy cluster couple
A399-A401. In that case the two clusters are about 2.4\,Mpc apart, while in
this case the distance between these sources is about 1.4\,Mpc if all of them are at the same redshift of Source A1 (z=0.104). A follow up at the X-ray and optical
frequencies is needed to understand if at this location other galaxy clusters
are present and shed light on the nature of these sources.

An even more mysterious source is the emission we label E1. This source
appears to be exactly along the long filament connecting A523-A525 in the
north to A509-A508 in the south-west, and in the same direction of the radio
galaxy 4C\,+06.21, embedded in the diffuse emission and sitting at redshift
z=0.405. There is no indication in the literature of a galaxy cluster at the same
location in the sky and the closest galaxy clusters to the source are
MACS\,J0455.2+065 ($\approx 25^{\prime}$) and A529 (at about $50^{\prime}$).
Therefore, Source E1 could be either associated with the filament or a diffuse
large-scale structure in the background, as suggested by the redshift of the
closest radio source.  Source E1 is the most luminous and most extended source in
the sample, with a flux at 1.4\,GHz $S_{\rm residual}=(109\pm 10)$\,mJy
($P_{\rm 1.4\,GHz}=6.8\times 10^{25}$\,W/Hz), a LLS of 8.7\,Mpc, assuming it
is at the same redshift of 4C\,+06.21 (z=0.405). On the other hand, if we
assume that the source is at the same redshift of the filament
(z$\approx$0.1), we derive a radio power $P_{\rm 1.4\,GHz}=3.0\times
10^{24}$\,W/Hz and an LLS=3\,Mpc, still among the brightest and largest
sources of the sample. Nevertheless, the source is very faint in the X-ray, for
which we derive a luminosity $L_{\rm X ,0.1-2.4\,keV}<14.5\times
10^{44}$\,erg/s (2$\sigma$) if a redshift z=0.405 is taken.  A firm
classification is not possible without further information about the system
over a wide range of frequencies, from the X-ray to optical and radio bands.

Overall, the nature of the above sources remains unclear.  To investigate if
they are similar to diffuse cluster sources \citep[i.e. radio halos and
  relics, e.g.,][]{Feretti2012} or rather they represent a different population, we
compare the radio power, radio size and X-ray luminosity of these new sources
with those of known diffuse cluster sources.  Hereafter, we adopt for known
cluster sources the values given by \cite{Feretti2012} without applying any
correction for the cosmology.  In Fig.\,\ref{halos}, we show in the left panel
the radio power at 1.4\,GHz $P_{\rm 1.4\,GHz}$ versus the X-ray luminosity
$L_{\rm X, 0.1-2.4\,keV}$ in the energy band 0.1--2.4\,keV and in the right
panel the radio power at 1.4\,GHz $P_{\rm 1.4\,GHz}$ versus the largest linear
scale LLS at 1.4\,GHz, for cluster sources (radio halos and radio relics) from
\cite{Feretti2012} and for the new sources presented in this work.  For
the diffuse emission in A523 and A520 (Sources A1 and I1), we plot both the
values we derive and the values found in the literature in order to have a basis for comparison. Our values are slightly different than the values available in
the literature\footnote{We note that the discrepancy in the radio power of the
  halo in A520 is due to the fact that \cite{Feretti2012} report the value
  given by \cite{Govoni2001}. A re-analysis of the radio halo properties were
  performed by \cite{Vacca2014} that find a radio power consistent with our
  measurement, as described in \S\,\ref{A520}.}, as discussed in
\S\,\ref{diff}, but still follow the correlation between radio and X-ray
properties for known diffuse radio sources. Among the eleven sources in our
catalogue with a significance of the X-ray count rate above 1$\sigma$, two are the
radio halos in A520 and A523, one (G4) follows the correlation observed for
radio halos and relics, and the remaining eight show a X-ray luminosity
between 10 and 100 times lower than expected from their radio power given the
correlation observed for cluster sources, with an average X-ray luminosity of
$L_{\rm X, 0.1-2.4\,keV}$=0.77$\times 10^{44}$\,erg/s. They populate a new
region of the ($L_{\rm X, 0.1-2.4\,keV}$, $P_{\rm 1.4\,GHz}$) plane that was
previously unsampled. 
The remaining sources are very faint. They show a radio power comparable to
that of radio halos, but with fainter X-ray emission ($\lesssim 10^{43}$\,erg/s)
and larger size.  Their nature is quite mysterious, since they are located in
a region of the sky where no galaxy cluster has been identified in the literature.

The radio power versus largest linear size diagram (right panel of
Fig.\,\ref{halos}) is interesting, as extended diffuse radio sources in
clusters are known to follow the correlation
\begin{equation}
P_{\rm 1.4\,GHz}\propto LLS^{\rm a}
\label{corr}
\end{equation}
between their radio power $P_{\rm 1.4\,GHz}$ and their largest linear size 
\citep[see e.g.][]{Feretti2012}. The measurements for A520 and A523 sit in
the correlation for cluster radio sources, as expected, while the remaining
sources are located on the right of this correlation. The source G4 shows a
size larger than expected from the correlation observed for radio halos and
relics, despite its X-ray luminosity being comparable to that of these
sources. The mean power and the mean largest linear size of the sample
(excluding the radio halos in A520 and A523) are respectively $P_{\rm
  1.4\,GHz}=2.9\times 10^{24}$\,W/Hz and LLS=1.3\,Mpc.  The radio power of the
new sources presented in this work correlates with the largest linar size as
well. However, this correlation differs from that of radio halos and relics.  A linear fit in logarithmic scale
of Eq.\,\ref{corr} gives a slope $a=1.54\pm0.03$ for cluster sources and
$a=2.04\pm0.02$ for the new sources. The
values of LLS, $P_{\rm 1.4\,GHz}$ and $L_{\rm X, 0.1-2.4\,keV}$ of the sources
presented here have been derived under the assumption that the redshift for
these sources is known. As already discussed, the identification of the
redshift of these sources is not straightforward and, therefore, we have adopted
the values of the closest radio galaxy or galaxy cluster to the system, or
alternatively a redshift z=0.100, if no association was possible (see
Table\,\ref{tab_diff} and Table\,\ref{tab_diff_xray}).  However, we note that
a different redshift would shift the radio power and LLS to lower/higher
values but the correlation would remain.  In addition, we compare the mean
emissivity $\langle J\rangle$ of the new sources to those of the cluster sources, by
assuming that they have a spherical symmetry and radius $R=0.5\times LLS$. The
result is shown in Fig.\,\ref{emissivity}.  The histogram reveals two distinct
distributions with only a partial overlap. The mean value of the emissivity
for cluster sources is 2.7$\times 10^{-41}\,$erg/s/Hz/cm$^3$ (3.0$\times
10^{-42}\,$erg/s/Hz/cm$^3$ when only radio halos are considered), while for
the new sources we find a mean value of 3.1$\times 10^{-43}\,$erg/s/Hz/cm$^3$.
The distribution observed for cluster diffuse radio sources is consistent with
the findings of \cite{Murgia2009}, who assume a radius given by the
e-folding radius of the exponential fit of the radio source brightness
profile.

In order to investigate if the new candidate sources are randomly distributed in the sky or instead show a real connection with the filament of galaxy clusters located in the same area (A523,
A525, RXC\,J0503.1+0607, A529, A515, A526, RXC\,J0506.9+0257, A508, and A509), we compute the conditional nearest neighbor distance $D_{\rm CNN}$ \citep{Okabe1984}, defined as
\begin{equation}
D_{\rm CNN}=\frac{1}{N_{\rm s}+N_{\rm c}}\left(\sum_{i=1}^{N_{\rm s}}d_{{\rm s}i}+\sum_{j=1}^{N_{\rm c}}d_{{\rm c}j} \right)
\end{equation}
where $N_{\rm s}$ is the total number of new candidate sources, $N_{\rm c}$ is the total number of clusters, $d_{{\rm s}i}$ the distance of the i-source from the closest cluster, and $d_{{\rm c}j}$ the distance of the j-cluster from the closest new candidate source. We obtain $D_{\rm CNN}=1.09^{\circ}$. 
\begin{figure}
\begin{center}
\includegraphics[width=0.45\textwidth, angle=0]{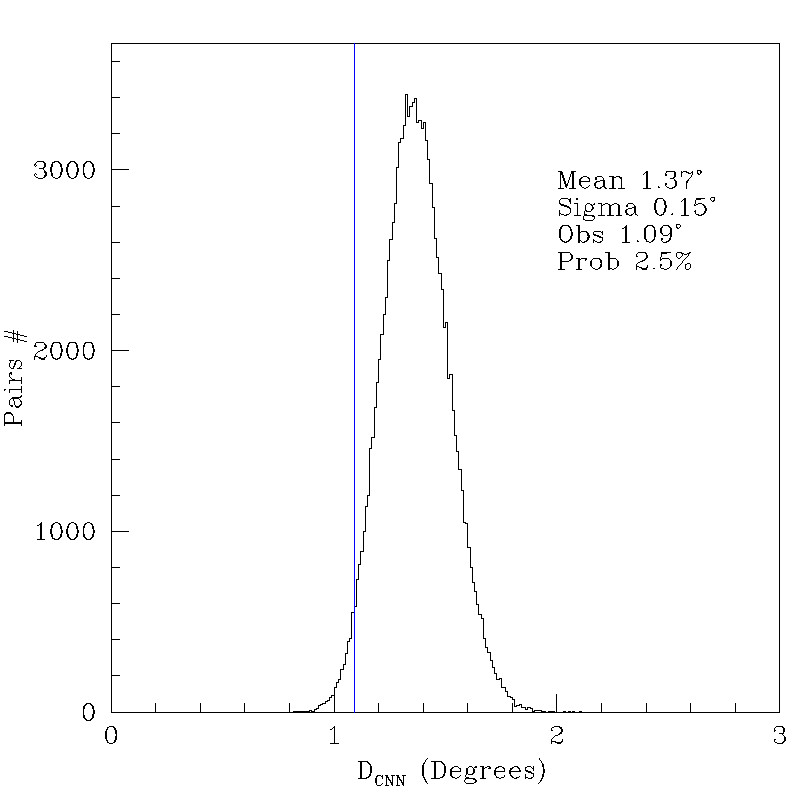}
\end{center}
\caption{Distribution of $D_{\rm CNN}$ for the random sets of sources (black histogram). The histogram is well described by a Gaussian distribution with mean 1.37$^{\circ}$ and standard deviation 0.15$^{\circ}$. The $D_{\rm CNN}$ for our sample is shown by the blue line. }
\label{chamfer}
 \end{figure}
As a comparison we compute $D_{\rm CNN}$ for about 130000 sets of random new candidate sources, while keeping the cluster coordinates fixed. The coordinates of the random sources have been extracted from uniform distributions in Right Ascension and Declination. In Fig.\,\ref{chamfer}, we show the distribution of $D_{\rm CNN}$ for the random sets of sources (black histogram) and for our sample (blue line). The probability of a random association between our new sources and the galaxy cluster filament is less than 2.5\%. This means that the number of spurious sources related to noise, or to the Galactic foreground, is small. We verified this point with the help of mock SRT observations that we ran through the same imaging pipeline as the real ones. We find that gain fluctuations 
are strongly reduced in our images after the application of wavelet and denoising techniques. However, the 18-20\% of the diffuse large-scale detected sources could be Galactic.
A detailed description of the simulations and of the results is given in Appendix\,\ref{fnoise}.

If confirmed, our results reveal a new population of sources, very luminous
and extended in radio, but very elusive in X-rays.  These sources could be
associated with the filaments of the cosmic web. The emission from these
structures is believed to be very faint at all wavelengths. Indeed, we are
able to detect only a fraction of them in X-rays and with low significance
(1$\sigma$), while in radio we begin to detect them thanks to the high
sensitivity to surface brightness of single-dish observations. The fact that
most of our sample is offset to larger sizes (at the same radio power), that the average emissivity is 10-100 smaller than radio halos and relics respectively, and
that the X-ray luminosity of the detected sources is about 10-100 weaker than
cluster sources, are evidence that we have discovered a new population of sources
at lower surface brightness.  This is consistent with the thermodynamic
properties of the gas in simulated cosmological filaments, whose typical X-ray
emissivity is expected to be $\sim 10-100$ smaller than the emissivity of
galaxy clusters with the same mass \citep[][]{Gheller2016}.

\begin{figure*}
\begin{center}
\includegraphics[width=0.9\textwidth, angle=0]{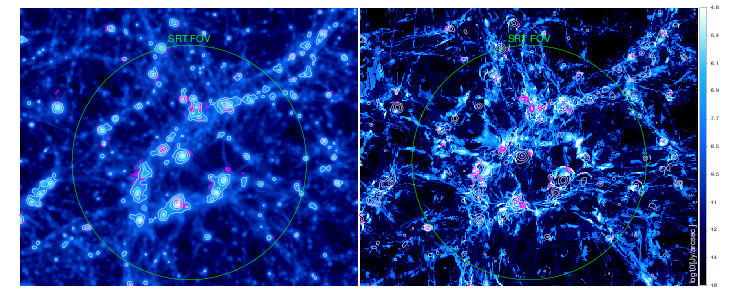}
\end{center}
\caption{\emph{Left panel}: estimated virial volume of all halos in the box (based on their total matter density, white contours) and  detectable radio emission in the SRT observing configuration presented in this paper (purple contours) overlaid on the projected gas density (colours). \emph{Right panel}: the
  colours show the projected full radio emission at the nominal resolution of
  the simulation (83\,kpc/cell) and assuming this field is at $z=0.1$. The contours are the same as in the
  left panel. The
  green circles in both images represent the field of view of SRT if the
  simulated volume is located at $z=0.1$.}
\label{fil_vs_sim_images}
 \end{figure*}

\begin{table*}
  \caption{X-ray fluxes from the regions associated with the diffuse sources in the field from RASS}
\begin{center}
        \begin{tabular}{cccccccc}
          \hline
          \hline
Source & z &$n_{\rm H}$ &$R_{200}$ &$S_{\rm X}$ & $D_{\rm L}$ & $L_{\rm X, 0.1-2.4\,keV}$ \\
& & $10^{20}$\,cm$^{-2}$&$^{\circ}$ & $10^{-12}$erg\,s$^{-1}$\,cm$^{-2}$ & Mpc & $10^{44}$erg\,s$^{-1}$ \\
\hline
A1 & 0.104 & 10.6 & 0.18 & 1.2 $\pm$ 0.3 &499& 5.4 $\pm$ 1.4\\
A2 & 0.104 & 10.5 & 0.118 & 0.2 $\pm$ 0.1 &499& 0.7 $\pm$ 0.6\\
A3 & 0.104 & 11.8 & 0.112 & 0.1 $\pm$ 0.1 &499& 0.6 $\pm$ 0.5\\
B1 & 0.104 & 10.3 & < 0.132 & < 0.3 &499& < 1.2\\
B2 & 0.1 & 9.01 & < 0.112 & < 0.1 & 478 & < 0.5\\
B3 & 0.1 & 8.43 & < 0.118 & < 0.2 & 478 & < 0.6\\
C1 & 0.029 & 10.7 & < 0.292 & < 0.5 & 132& < 0.2\\
C2 & 0.029 & 11.1 & < 0.293 & < 0.5 & 132& < 0.2\\
C3 & 0.029 & 10.7 & 0.323 & 0.9 $\pm$ 0.4 & 132& 0.3 $\pm$ 0.1\\
C4 & 0.029 & 10.8 & < 0.293 & < 0.5 & 132& < 0.2\\
C5 & 0.029 & 10.8 & < 0.288 & < 0.5 & 132& < 0.2\\
C6 & 0.029 & 11.2 & 0.268 & 0.4 $\pm$ 0.3 & 132& 0.1 $\pm$ 0.1\\
C7 & 0.1 & 10.2 & < 0.133 & < 0.3&478 & < 1.1\\
C8 & 0.1 & 9.81 & 0.116 & 0.1 $\pm$ 0.1 &478& 0.6 $\pm$ 0.5\\
C9 & 0.1 & 9.43 & < 0.13 & < 0.2 &478& < 1.0\\
C10 & 0.1 & 9.29 & < 0.119 & < 0.2 &478& < 0.7\\
D1 & 0.1 & 9.74 & < 0.126 & < 0.2 &478& < 0.9\\
D2 & 0.1 & 8.46 & < 0.118 & < 0.2 &478& < 0.6\\
D3 & 0.1 & 8.16 & < 0.112 & < 0.1 &478& < 0.5\\
E1 & 0.405 & 7.89 & < 0.064 & < 0.2 &2283& < 14.5\\
E2 & 0.1 & 7.58 & < 0.127 & < 0.2 & 478 & < 0.9\\
E3 & 0.1 & 7.42 & 0.117 & 0.1 $\pm$ 0.1 & 478 & 0.6 $\pm$ 0.5\\
F1 & 0.1 & 10.2 & < 0.125 & < 0.2 & 478 & < 0.8\\
G1 & 0.1 & 7.91 & < 0.123 & < 0.2 & 478 & < 0.8\\
G2 & 0.1 & 6.72 & < 0.119 & < 0.2 & 478 & < 0.6\\
G3 & 0.1 & 7.53 & 0.119 & 0.2 $\pm$ 0.1 & 478 & 0.7 $\pm$ 0.5\\
G4 & 0.1 & 7.38 & 0.243 & 4.6 $\pm$ 0.6& 478  & 18.5 $\pm$ 2.3\\
G5 & 0.1 & 7.06 & 0.12 & 0.2 $\pm$ 0.1 & 478 & 0.7 $\pm$ 0.4\\
G6 & 0.1 & 7.14 & < 0.124 & < 0.2 & 478 & < 0.8\\
H1 & 0.1 & 7.77 & < 0.127 & < 0.2& 478  & < 0.9\\
H2 & 0.1 & 7.67 & < 0.124 & < 0.2 & 478 & < 0.8\\
I1 & 0.199 & 5.66 & 0.162 & 3.3 $\pm$ 0.4 &1011& 55.8 $\pm$ 6.9\\
I2 & 0.199 & 5.64 & < 0.086 & < 0.2 &1011& < 2.8\\
I3 & 0.199 & 5.76 & 0.085 & 0.1 $\pm$ 0.1 &1011& 2.6 $\pm$ 1.8\\
J1 & 0.1 & 7.41 & < 0.125 & < 0.2 & 478 & < 0.8\\

            \hline
            \hline
        \end{tabular}
\label{tab_diff_xray}\\
\end{center}
Col 1: Source name; Col 2: redshift; Col 3: Hydrogen column density in the
direction of the source; Col 4: $R_{200}$ of the system; Col 5: X-ray flux in the 0.5-2\,keV energy band within $R_{200}$; Col 6: Luminosity distance; Col 7: X-ray luminosity in the 0.1-2.4\,keV energy band within $R_{200}$.    \\
\normalsize
\end{table*}

\subsection{Observations versus simulations}

\begin{figure}
\begin{center}
\includegraphics[width=0.4\textwidth, angle=0]{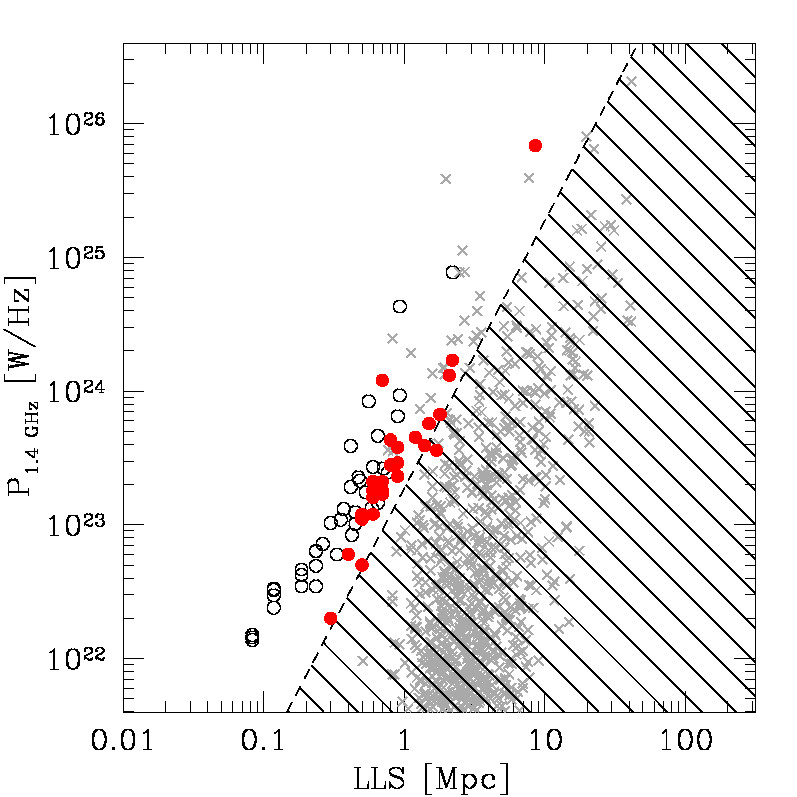}
\end{center}
\caption{Radio power at 1.4\,GHz versus the LLS  at 1.4\,GHz for the all simulated objects
  before the SRT observing parameters are considered (gray crosses), for
  simulated objects after the SRT noise and spatial resolution are applied
  (empty black dots), and for observed objects (full red dots). The shaded area indicates the region of the ($P_{\rm 1.4\,GHz}$, LLS) plane that
cannot be accessed considering the 2$\sigma$ ($\sigma$=2.5\,mJy/beam) sensitivity of the SRT+NVSS image after point source subtraction.}
\label{fil_vs_sim}
 \end{figure}

\begin{figure}
\begin{center}
\includegraphics[width=0.4\textwidth, angle=0]{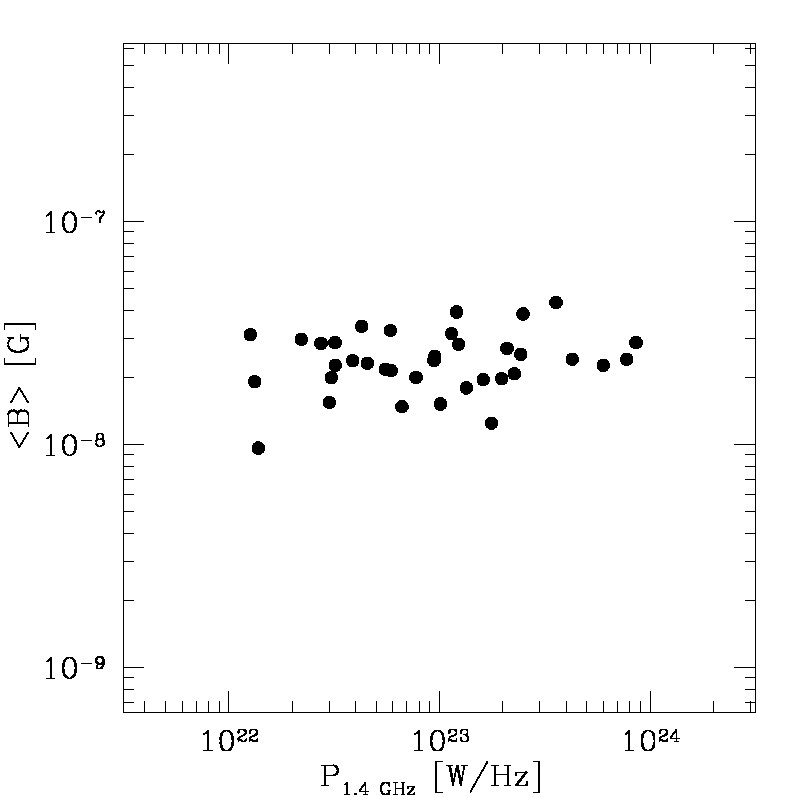}\\
\includegraphics[width=0.4\textwidth, angle=0]{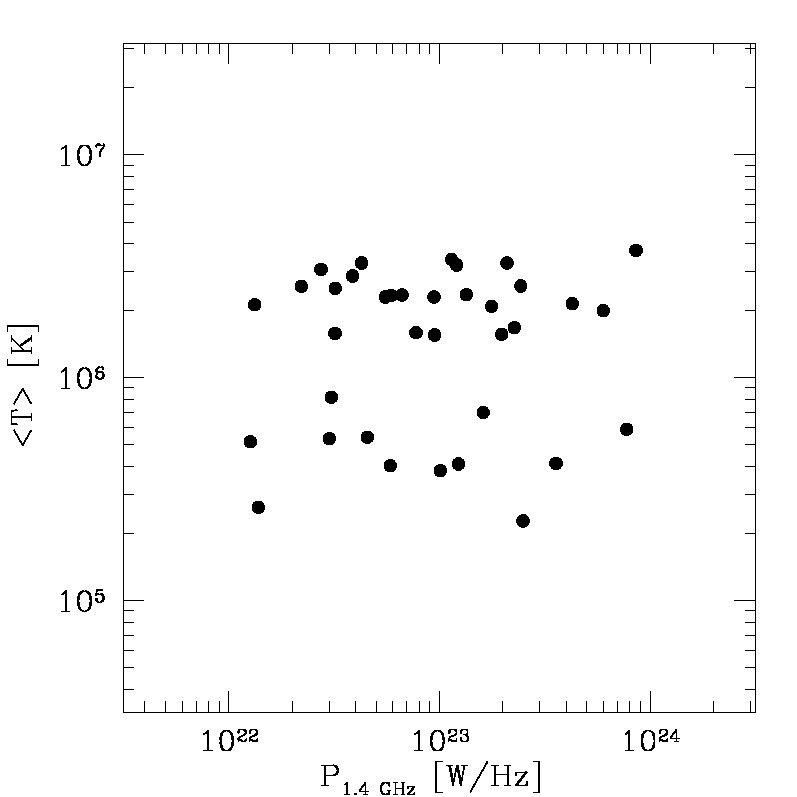}\\
\includegraphics[width=0.4\textwidth, angle=0]{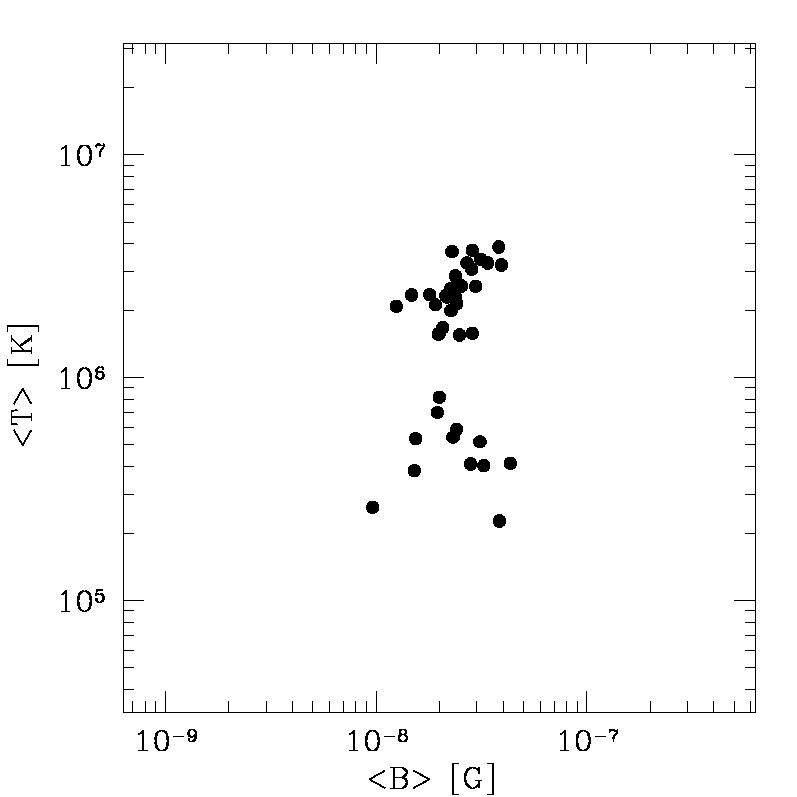}

\end{center}
\caption{\emph{Top panel:} mean volume-weighted magnetic field 
$\langle B\rangle$ versus $P_{\rm 1.4\,GHz}$. \emph{Middle panel:} mean temperature $\langle T\rangle$ versus $P_{\rm 1.4\,GHz}$. \emph{Bottom panel:} $\langle T\rangle$ versus $\langle B\rangle$.}
\label{fil_vs_sim_plot}
 \end{figure}

To interpret our results, we compared the observations with simulations from
the sample already introduced in \citet[][]{Vazza2015} and
\citet[][]{Gheller2016}, obtained with the cosmological grid code {\it Enzo}
\citep[][]{Bryan2014}.  In summary, we used a uniform $1200^3$ grid with a
resolution of 83\,kpc (co-moving) to simulate the evolution of a 100\,Mpc$^3$
region from $z=30$ to $z=0$, assuming a seed magnetic field of 1\,nG of
primordial origin\footnote{We note that in this work we re-normalize the
  magnetic field model of \citet{Gheller2016}, in which a 0.1\,nG initial seed
  field was assumed. This higher initial field is still within the bound of
  present constraints derived from the Cosmic Microwave Background
  \citep[e.g.][]{Subramanian2016} and it is legitimate because everywhere in
  the volume our simulated magnetic fields are far from saturation.}.  The
magnetic field in the magneto-hydro-dynamical (MHD) method uses the
conservative Dedner formulation \citep[][]{Dedner2002} which uses hyperbolic
divergence cleaning to keep $\nabla \cdot \vec{B}$ close to zero. Radiative
processes and feedback from star forming regions and/or active galactic nuclei
were not included in this run.  The assumed cosmology is the $\Lambda$CDM
cosmological model with density parameters $\Omega_0 = 1.0$, $\Omega_{\rm BM}
= 0.0455$, $\Omega_{\rm DM} =0.2265$ (BM and DM indicating the baryonic and
the dark matter respectively).

To simulate synchrotron radio emission from the cosmic web, we followed
\citet{Vazza2015} and assumed the diffusive shock acceleration of electrons at
cosmological shocks, relying on the formalism by \citet{Hoeft2007} to combine
simulated magnetic fields and the distribution of Mach numbers, $\mathcal{M}$, in order 
to estimate the level of synchrotron emission across the cosmic web, which is computed for each simulated cell as:
\begin{equation}
P_{\nu} \propto  n_u \xi(\mathcal{M}) \cdot M^3 c_s^3 S \cdot  B^2, 
\label{hb}
\end{equation}
where $n_u$ is the upstream gas density, $c_s$ is the sound speed in the
upstream gas, $B$ is the magnetic field in the shocked cell, and $S$ is the
shock surface. The acceleration efficiency of electrons, $\xi(\mathcal{M})$,
is taken from \citet{Hoeft2007} and (in the absence of seed relativistic
electrons to re-accelerate) is a steep function of $\mathcal{M}$ for weak
shocks, and rapidly saturates to $\xi_{\rm e} \approx 7 \cdot 10^{-4}$ for $M \gt
10$ shocks in our model.

Finally, in order to focus on the radio emission produced by filaments in our
volume and minimize the contamination by denser structures along the line of
sight, we relied on the \emph{filament finder} presented in \citep{Gheller2016},
which allows us to extract the 3-dimensional isosurfaces associated with the
mild over-densities associated with filaments, as well as to build a catalogue
of single filament objects, for which mean thermodynamic and magnetic
properties can be computed.

To compare with observation, we produced mock radio observation of the sky
model from cosmological simulations, with a procedure similar to
\citet{Vazza2015}.  In particular, we computed the radio emission model at 1.4
GHz (by locating our simulated box at a distance corresponding to $z=0.1$)
convolving the input sky model with a final beam of
3.5$^{\prime}\times$3.5$^{\prime}$ and considering a final sensitivity of 0.05
$\mu \rm Jy/arcsec^2$, the same as those of the image used for
our analysis.  The result is shown in Fig.\,\ref{fil_vs_sim_images}. colours in the left panel represent the projected density, while in the right panel, they show the projected full radio emission at the nominal resolution of the
simulation (83\,kpc/cell). White contours describe the
estimated virial volume of all halos in the box (based on their total matter
over-density). Purple contours show the detectable radio emission in the SRT
observing configuration presented in this paper.

Based on the 3-d catalogue of filaments obtained in \citet{Gheller2016}, we
computed the properties of filaments seen in projections in the entire volume,
i.e. their total intrinsic radio power at 1.4 GHz (prior to any observational
cut) and their largest linear scale. By considering the noise and resolution
of the SRT observations, only a very tiny fraction ($\leq 10^{-4}$) of the
radio emitting surface of the cosmic web in the simulation survives. We
recomputed the corresponding radio power and largest linear size of these
objects. In Fig.\,\ref{fil_vs_sim}, we show the radio power at 1.4\,GHz versus
the LLS for all the simulated objects before the SRT observing parameters are
considered (gray crosses), for simulated objects after the SRT noise and spatial
resolution are applied (empty black dots), and for observed objects (full red dots).  The crosses represent all the diffuse large-scale sources in the image before any
observing parameter (noise and convolution) is applied. After the SRT
noise and spatial resolution are taken into account, the detected
sources are only those marked by empty black circles. The
few detectable emission patches populate almost the same region of the ($LLS$,
$P_{\rm 1.4\,GHz}$) plane as the observed candidate sources presented in this
paper. In most cases, as suggested by the map in
Fig.\,\ref{fil_vs_sim_images}, these large diffuse emission regions are
associated with very peripheral shocks at the crossroad between the outer
virial volume of massive galaxy clusters and the filaments connecting them,
which are typically detectable only in very crowded over-dense regions,
similar to the high-density field observed in this work.

In Fig.\,\ref{fil_vs_sim_plot}, we show the mean magnetic field $\langle B\rangle$ and mean
temperature $\langle T\rangle$ versus $P_{\rm 1.4\,GHz}$ (top and middle panels), and
$\langle T\rangle$ versus $\langle B\rangle$ (bottom panel), for the simulated filaments that host emission patches
which should be detectable by our SRT observation. The average properties of
the host objects are typical of the most massive filaments in our simulations
\citep[][]{Gheller2016}, with $n/\langle n \rangle \sim 50-100$, $\langle B\rangle \sim 10-50$\,nG, $T \sim 2
\cdot 10^5- 5 \cdot 10^6$\,K), yet the detectable regions only cover a tiny
fraction (a few percent) of the filaments' projected area, and they tend to be
associated with the densest and hottest portions of filaments, connecting to
the surrounding clusters.  With the assumed prescription for extragalactic
magnetic fields and electron acceleration at shocks, the regions which are
within the range of detection in our SRT observation typically have an average
magnetic field along the line of sight of $\sim 0.02- 0.05\,\mu$G.

\section{Summary and conclusions}
\label{conclusions}
In this work, we report the detection of diffuse radio emission
which might be associated with a large-scale filament of the cosmic web
  covering a 8$^{\circ}\times$8$^{\circ}$ area in the sky, likely associated
  with a z$\approx$0.1 over-density traced by nine massive galaxy clusters. To investigate the presence of
large-scale diffuse synchrotron emission beyond the cluster periphery, we
observed this area with the Sardinia Radio Telescope.  These low spatial
resolution data have been combined with high resolution observations from the
NRAO Very Large Array Sky Survey, to permit separation of the diffuse large-scale synchrotron
emission from that of embedded discrete radio sources.

By inspecting the field of view, we identified 35 patches of diffuse emission
with significance above 3$\sigma$. Two are the cluster sources already known
in the direction of the galaxy clusters A523 (source A1) and A520 (source I1),
five sources (C1, C6, C9, E2, and G6) are probably the leftover of the compact
source subtraction process or artefacts, and the remaining 28 sources represent
diffuse synchrotron radio emission with no obvious interpretation.  To shed
light on the nature of these new sources, we studied their radio and
X-ray properties.  Only 
eleven sources have an X-ray count rate significantly above 1$\sigma$. These sources are A1 (diffuse emission in A523), A2, A3,
C3, C8, E3, G3, G4, G5, I1 (diffuse emission in A520), and I3. Apart from the
two radio halos in A520 and A523, one of the significant X-ray sources (G4)
sits in the $L_{\rm X, 0.1-2.4\,keV}$-$P_{\rm 1.4\,GHz}$ correlation observed
for radio halos and relics but has a larger size than expected for cluster
sources with this power. The remaining eight sources show a X-ray luminosity
between 10 and 100 times lower than expected from their radio power given the
correlation observed for cluster sources, with an average X-ray luminosity of
$L_{\rm X, 0.1-2.4\,keV}$=0.77$\times 10^{44}$\,erg/s. They populate a new
region of the ($L_{\rm X, 0.1-2.4\,keV}$, $P_{\rm 1.4\,GHz}$) plane that was
previously unsampled.

Some of these are particularly interesting. Source A2 in combination with
the source south-west of it and with Source A1 (the radio halo in A523) could
represent the first case of a triple
radio halo, particularly interesting because of its association with an under-luminous X-ray system. Source E1 is the largest and the more powerful diffuse radio source in the field. Its size is 25$^{\prime}$ and its location on the plane of the sky is in the same direction as the filament connecting A523-A525 to  A509-A508. Alternatively, could be located at higher redshift, if associated with 4C\,+06.21, the source embedded in it.  Whether this source
belongs to the filament or it is instead at higher redshift has an impact
on its size that ranges between 3 and 8.7\,Mpc, and in the latter case it
would represent the largest extragalactic diffuse synchrotron source known to
date.

Overall, these objects have
distinct properties from known radio halos and relics. Indeed, they show lower
X-ray luminosity in the 0.1--2.4\,keV band and a larger LLS than expected for
a given radio power from the correlation observed for cluster sources.  The
mean power and the mean largest linear size of the sample (excluding the radio
halos in A520 and in A523) are respectively $P_{\rm 1.4\,GHz}=2.9\times
10^{24}$\,W/Hz and LLS=1.3\,Mpc. The average radio emissivity and the X-ray
luminosity of these sources are about 10-100 weaker than those of cluster
diffuse sources. These characteristics are indicative of a very faint new
population of sources.  The conditional nearest neighbour distance of clusters and and candidate new sources indicates that the probability of a random association between our new sources and the galaxy cluster filament is less than 2.5\%. This means that it is unlikely spurious sources related to noise or to the Galactic foreground are present.
A comparison between simulations and observations
shows that the radio powers and radio sizes of these sources are comparable to those expected
by the brightest patches of diffuse synchrotron radio emission associated with the filaments of the cosmic web and correspond to an average
magnetic field along the line of sight of $\sim 0.02- 0.05\,\mu$G.

High-spatial resolution and deep radio, X-ray and optical
follow-ups are essential to shed light on the nature of these puzzling
sources.\\

\clearpage

\section*{Acknowledgements}
The Sardinia Radio Telescope \citep{Bolli2015,Prandoni2017} is funded by the Department of University and
Research (MIUR), Italian Space Agency (ASI), and the Autonomous Region of
Sardinia (RAS) and is operated as National Facility by the National Institute
for Astrophysics (INAF).  The development of the SARDARA back-end was
funded by the Autonomous Region of Sardinia (RAS) using resources from the
Regional Law 7/2007 ``Promotion of the scientific research and technological
innovation in Sardinia'' in the context of the research project CRP-49231
(year 2011, PI Possenti): ``High resolution sampling of the Universe in the
radio band: an unprecedented instrument to understand the fundamental laws of
the nature''.The National Radio Astronomy Observatory (NRAO) is a facility of
the National Science Foundation, operated under cooperative agreement by
Associated Universities, Inc. This research made use of the NASA/IPAC
Extragalactic Database (NED) which is operated by the Jet Propulsion
Laboratory, California Institute of Technology, under contract with the
National Aeronautics and Space Administration.  F. Loi gratefully acknowledges
Sardinia Regional Government for the financial support of her PhD scholarship
(P.O.R. Sardegna F.S.E. Operational Programme of the Autonomous Region of
Sardinia, European Social Fund 2007-2013 - Axis IV Human Resources, Objective
l.3, Line of Activity l.3.1.).  Basic research in radio astronomy at the Naval
Research Laboratory is funded by 6.1 Base funding. F.V. acknowledges the
usage of computational resources at CSCS (ETHZ in Lugano) under allocations
s701 and the financial support from the ERC Starting Grant "MAGCOW",
no.714196.








\appendix
\section{Compact sources in the field}
\label{pscat}

\begin{figure}
\begin{center}
\includegraphics[width=0.5\textwidth, angle=0]{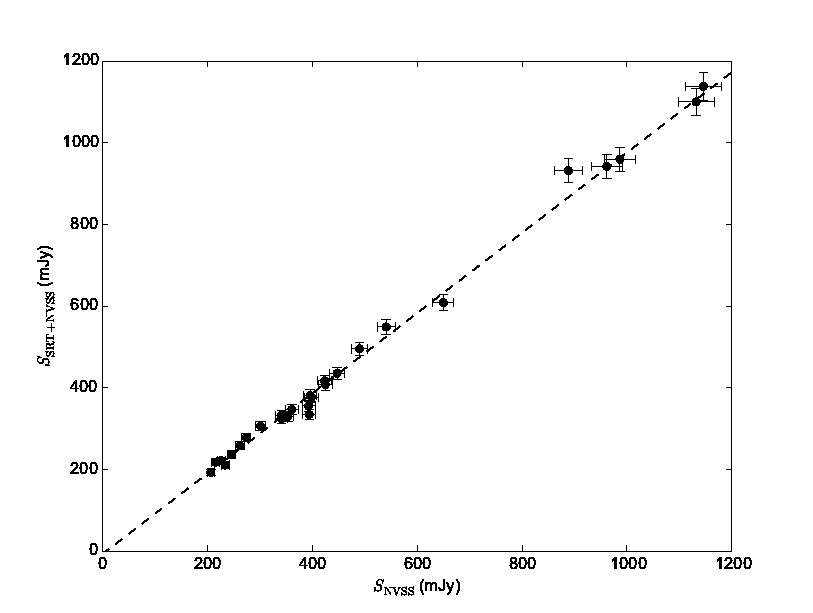}\\
\end{center}
\caption{Flux from NVSS mosaic image versus flux from the combined SRT+NVSS image for sources with fluxes $>200$\,mJy.}
\label{flux_check}
 \end{figure}

This field of view is rich in radio sources, with a total number of thousands
at 1.4\,GHz.  To further cross-check the flux density scale of our data, we measured
the flux density of a sample of sources that satisfies the following
requirements: peak brightness at 1.4\,GHz larger than 200\,mJy and flux larger
than the 3$\sigma$ simultaneously at all frequencies (with $\sigma$ being the
uncertainty in the flux measurement at the corresponding frequency).  For the
sake of comparison, we estimated the flux and its uncertainty 
over the frequency range 74\,MHz - 5\,GHz, by using the following observations
after convolving all the images at the resolution of the GB6
($3.6^{\prime}\times 3.4^{\prime}$):
\begin{itemize}
\addtolength{\itemindent}{0.5cm}

\item the VLA Low-Frequency Sky Survey Redux \citep[VLSSr,][]{Lane2014} at
  74\,MHz ($\sigma_{\rm VLSSr}=300$\,mJy);
\item the TIFR GMRT Sky Survey Alternative Data Release \citep[TGSS
  ADR,][]{Intema2017} at 150\,MHz ($\sigma_{\rm TGSS}=50$\,mJy);
\item the NVSS image, at 1.4\,GHz ($\sigma_{\rm NVSS}=2$\,mJy); 
\item the combined SRT and NVSS image, at 1.4\,GHz ($\sigma_{\rm SRT+NVSS}=3$\,mJy); 
\item the Green Bank 6-cm (GB6) survey \citep[GB6,][]{Gregory1996} at 4.85\,GHz ($\sigma_{\rm GB6}=5$\,mJy).
\end{itemize}
All these data are in the \cite{Baars1977} flux scale, apart from the TGSS ADR
and the VLLSr that have been calibrated according to respectively the
\cite{Scaife2012} and the \cite{Roger1973} flux density scale. Therefore, we corrected
the TGSS and VLSSr fluxes accordingly. Concerning the VLSSr, we additionally
increased the uncertainty in the flux measurement, as suggested by
\cite{Lane2014}.  The flux has been evaluated by applying a cut in brightness
in the NVSS image equal to 5$\sigma_{\rm NVSS}$ and by masking all the other
images accordingly.  This translates to a total of 24 sources.  In
Table\,\ref{tab:D}, we report the source ID, the RA and Dec coordinates
(J2000) of the peak flux in the combined SRT+NVSS image, and the peak flux as
measured from the VLSSr, TGSS, NVSS, combined SRT+NVSS, GB6, along with their
uncertainties that have been computed by adding in quadrature the statistic
and systematic (3\% of the flux) uncertainties.  The flux measurements from
the NVSS and the SRT+NVSS combined images have been compared to cross-check
the flux scale after the combination of the data.  In Fig.\,\ref{flux_check},
we show the flux of the SRT+NVSS image versus the NVSS flux (dots) along with a
linear fit (dashed line). The slope of the fit is 0.97, revealing a good
agreement of the two flux measurements within the errors.

\begin{figure*}
\begin{center}
\includegraphics[width=\textwidth, angle=0]{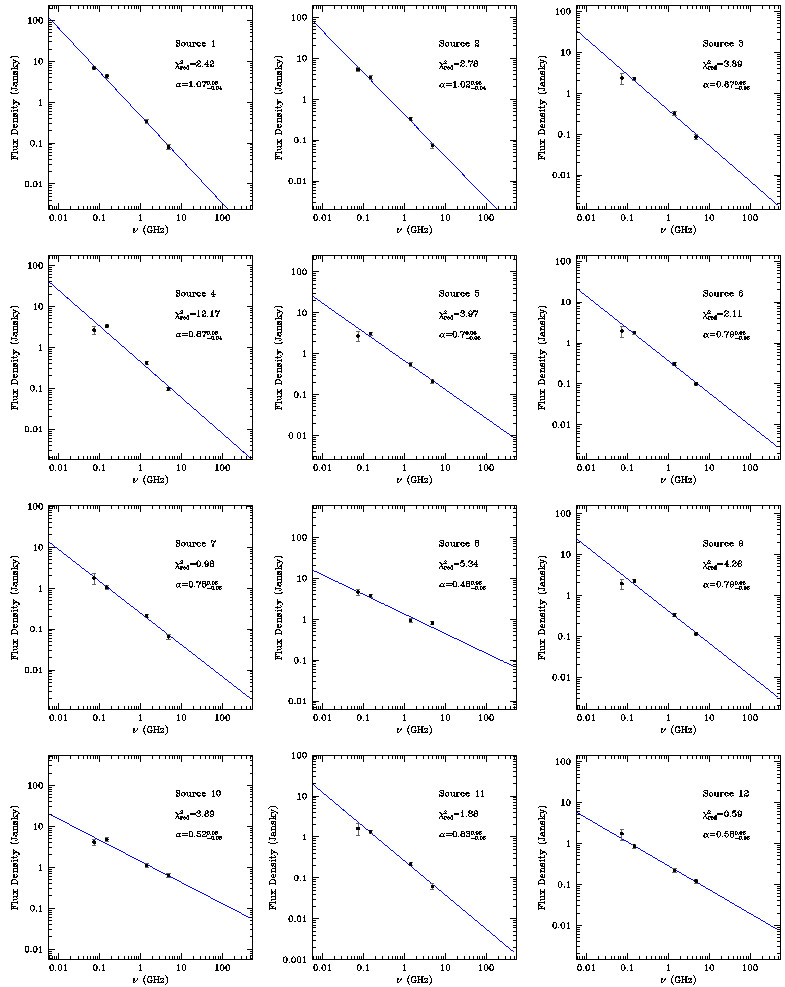}
\caption{Flux density versus frequency for the first 12 sources in the
  catalogue. For each source the flux density as measured by VLSSr, TGSS,
 combined SRT+NVSS, and GB6 images is shown along with the power law fit, the
 resulting spectral index $\alpha$ value and the $\chi^2_{\rm red}$ of the fit.}
\label{spixs1}
\end{center}
\end{figure*}
\begin{figure*}
\begin{center}
\includegraphics[width=\textwidth, angle=0]{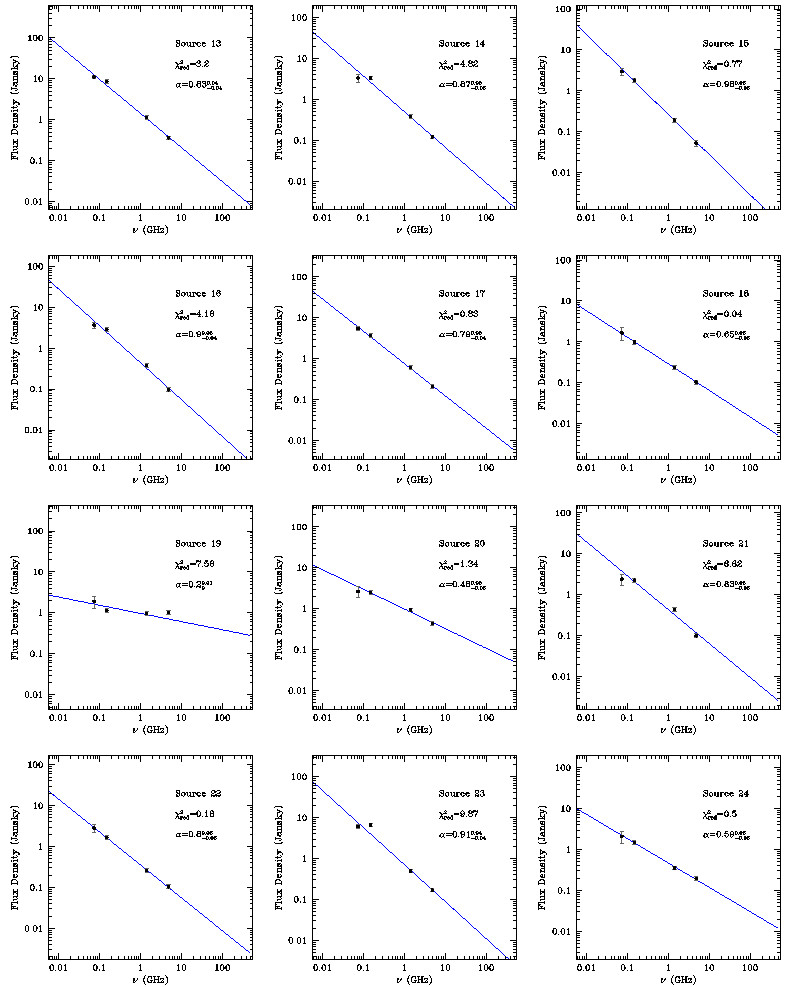}
\caption{Flux density versus frequency for the last 12 sources in the
  catalogue. For each source the flux density as measured by VLSSr, TGSS,
 combined SRT+NVSS, and GB6 images is shown along with the power law fit, the
 resulting spectral index $\alpha$ value and the $\chi^2_{\rm red}$ of the fit.}
\label{spixs2}
\end{center}
\end{figure*}
 
To compare the fluxes at 1.4\,GHz with the measurements at the other radio
frequencies, we fit the fluxes from VLSSr, TGSS, SRT+NVSS, and GB6 versus
frequency with a power law
\begin{equation}
S_{\nu}\propto\nu^{-\alpha},
\end{equation}
where $\nu$ is the observing frequency and $\alpha$ is the spectral index.
During the fitting procedure the relative errors were increased to 10\% if lower than this, to
take into account any systematic errors.  In Table\,\ref{tab:D}, we report the
spectral indices values, its error and the $\chi^2_{\rm red}$ from the fit. All
the sources have spectral index ranging from a minimum value of about 0.2 to
a maximum value of about 1.1, with a mean $\alpha$=0.76 and a standard
deviation $\sigma_{\alpha}$ =0.19. Plot of the flux densities versus
frequency and the corresponding fits are shown in Fig.\,\ref{spixs1} and
Fig.\,\ref{spixs2} for each source.

\begin{table*}
        \caption{Radio sources with peak brightness at 1.4\,GHz larger than 200\,mJy.}
\footnotesize
       \begin{tabular}{cccccccccc}
          \hline
          \hline

\# &  RA (J2000) & Dec (J2000) & $S_{\rm VLSSr}$ & $S_{\rm TGSS}$ & $S_{\rm NVSS}$ &  $S_{\rm SRT+NVSS}$&  $S_{\rm GB6}$ & $\alpha$ & $\chi^2_{\rm red}$\\
       &   hms      & $^{\circ}$:$^{\prime}$:$^{\prime\prime}$   & Jy & Jy & Jy &  Jy  & mJy & & \\
\hline
1 & 5:11:37& 9:31:21.40& 6.8$\pm$0.6 & 4.3$\pm$0.2 & 0.39$\pm$0.01  & 0.33$\pm$0.01 & 81$\pm$10 &  1.07$^{0.05}_{-0.05}$ & 2.42 \\
2 & 5:05:32& 9:26:01.79& 5.3$\pm$0.6 & 3.4$\pm$0.1 & 0.35$\pm$0.01  & 0.33$\pm$0.01 & 74$\pm$10 &  1.02$^{0.05}_{-0.05}$ & 2.78 \\
3 & 5:04:31& 8:27:00.24& 2.4$\pm$0.7 & 2.3$\pm$0.1 & 0.34$\pm$0.01  & 0.32$\pm$0.01 & 85$\pm$12 &  0.87$^{0.06}_{-0.06}$ & 3.89 \\
4 & 5:01:58& 6:50:57.71& 2.6$\pm$0.6 & 3.3$\pm$0.1 & 0.42$\pm$0.01  & 0.42$\pm$0.01 & 97$\pm$10 &  0.87$^{0.05}_{-0.05}$ & 12.17 \\
5 & 5:03:03& 6:09:58.82& 2.7$\pm$0.7 & 3.1$\pm$0.2 & 0.54$\pm$0.02  & 0.55$\pm$0.02 & 211$\pm$14 &  0.70$^{0.05}_{-0.05}$ & 3.97 \\
6 & 4:55:38& 7:52:08.42& 2.0$\pm$0.6 & 1.8$\pm$0.1 & 0.30$\pm$0.01  & 0.31$\pm$0.01 & 100$\pm$11 &  0.79$^{0.05}_{-0.06}$ & 2.11 \\
7 & 5:06:10& 5:22:28.42& 1.8$\pm$0.5 & 1.1$\pm$0.1 & 0.23$\pm$0.01  & 0.21$\pm$0.01 & 65$\pm$10 &  0.78$^{0.06}_{-0.06}$ & 0.98 \\
8 & 4:57:07& 6:45:29.63& 4.7$\pm$0.8 & 3.7$\pm$0.2 & 0.89$\pm$0.03  & 0.93$\pm$0.03 & 808$\pm$28 &  0.48$^{0.06}_{-0.06}$ & 5.34 \\
9 & 5:09:07& 3:48:03.21& 1.9$\pm$0.6 & 2.2$\pm$0.1 & 0.34$\pm$0.01  & 0.33$\pm$0.01 & 114$\pm$10 &  0.79$^{0.05}_{-0.05}$ & 4.26 \\
10 & 5:07:35& 3:08:00.44& 4.1$\pm$0.8 & 4.8$\pm$0.2 & 1.13$\pm$0.03  & 1.10$\pm$0.03 & 634$\pm$23 &  0.52$^{0.05}_{-0.05}$ & 3.89 \\
11 & 5:06:37& 2:37:11.02& 1.6$\pm$0.5 & 1.3$\pm$0.1 & 0.22$\pm$0.01  & 0.22$\pm$0.01 & 62$\pm$9 &  0.83$^{0.06}_{-0.06}$ & 1.88 \\
12 & 5:01:45& 2:24:36.45& 1.7$\pm$0.5 & 0.9$\pm$0.1 & 0.23$\pm$0.01  & 0.22$\pm$0.01 & 120$\pm$10 &  0.58$^{0.06}_{-0.07}$ & 0.59 \\
13 & 4:44:40& 2:48:01.22& 11.0$\pm$0.8 & 8.6$\pm$0.3 & 1.15$\pm$0.03  & 1.14$\pm$0.03 & 361$\pm$16 &  0.83$^{0.04}_{-0.04}$ & 3.20 \\
14 & 4:46:22& 5:41:03.56& 3.3$\pm$0.7 & 3.3$\pm$0.2 & 0.40$\pm$0.01  & 0.38$\pm$0.01 & 121$\pm$13 &  0.87$^{0.05}_{-0.05}$ & 4.82 \\
15 & 4:44:17& 5:01:41.76& 2.9$\pm$0.5 & 1.8$\pm$0.1 & 0.21$\pm$0.01  & 0.19$\pm$0.01 & 53$\pm$9 &  0.98$^{0.06}_{-0.06}$ & 0.77 \\
16 & 4:49:08& 1:52:58.99& 3.7$\pm$0.6 & 2.9$\pm$0.1 & 0.40$\pm$0.01  & 0.38$\pm$0.01 & 96$\pm$10 &  0.90$^{0.05}_{-0.05}$ & 4.16 \\
17 & 4:44:18& 1:54:45.61& 5.5$\pm$0.7 & 3.7$\pm$0.2 & 0.65$\pm$0.02  & 0.61$\pm$0.02 & 211$\pm$13 &  0.79$^{0.05}_{-0.05}$ & 0.83 \\
18 & 4:59:43& 5:20:06.76& 1.7$\pm$0.6 & 1.0$\pm$0.1 & 0.25$\pm$0.01  & 0.24$\pm$0.01 & 104$\pm$10 &  0.65$^{0.06}_{-0.06}$ & 0.04 \\
19 & 5:05:22& 4:59:57.84& 1.9$\pm$0.6 & 1.2$\pm$0.1 & 0.99$\pm$0.03  & 0.96$\pm$0.03 & 1022$\pm$33 &  0.20$^{0.01}_{-0.00}$ & 7.56 \\
20 & 4:44:38& 5:46:32.00& 2.7$\pm$0.8 & 2.5$\pm$0.2 & 0.96$\pm$0.03  & 0.94$\pm$0.03 & 434$\pm$19 &  0.48$^{0.05}_{-0.05}$ & 1.34 \\
21 & 4:55:38& 2:54:08.06& 2.4$\pm$0.7 & 2.3$\pm$0.1 & 0.45$\pm$0.01  & 0.44$\pm$0.01 & 100$\pm$12 &  0.83$^{0.05}_{-0.05}$ & 6.62 \\
22 & 5:15:31& 9:04:03.94& 2.8$\pm$0.6 & 1.7$\pm$0.1 & 0.26$\pm$0.01  & 0.26$\pm$0.01 & 104$\pm$12 &  0.80$^{0.06}_{-0.06}$ & 0.18 \\
23 & 5:13:51& 1:56:58.07& 6.1$\pm$0.8 & 6.6$\pm$0.2 & 0.49$\pm$0.02  & 0.50$\pm$0.02 & 170$\pm$14 &  0.91$^{0.04}_{-0.05}$ & 9.87 \\
24 & 5:11:34& 2:44:32.41& 2.1$\pm$0.7 & 1.5$\pm$0.1 & 0.39$\pm$0.01  & 0.36$\pm$0.01 & 195$\pm$13 &  0.59$^{0.06}_{-0.06}$ & 0.50 \\
25 & 4:52:48& 4:01:01.85& 3.2$\pm$0.7 & 2.4$\pm$0.1 & 0.36$\pm$0.01  & 0.35$\pm$0.01 & 96$\pm$12 &  0.87$^{0.06}_{-0.06}$ & 1.98 \\

            \hline
        \end{tabular}
        \label{tab:D}
\normalsize 
\newline
Col 1: Source name; Col 2 \& 3: source coordinates; Col 4: Flux density from
the VLSSr image at 74\,MHz; Col 5: Flux density from the TGSS image at
150\,MHz; Col 6: Flux density from the NVSS image at 1.4\,GHz; Col 7: Flux
density from the combined SRT+NVSS image at 1.4\,GHz; Col 8: Flux density from
the GB6 image at 4.85\,GHz; Col 9: Spectral index between 74\,MHz and 1.4\,GHz; Col
10: Reduced chi-squared.
    \end{table*}

\clearpage

\begin{figure}
\begin{center}
\includegraphics[width=0.4\textwidth, angle=0]{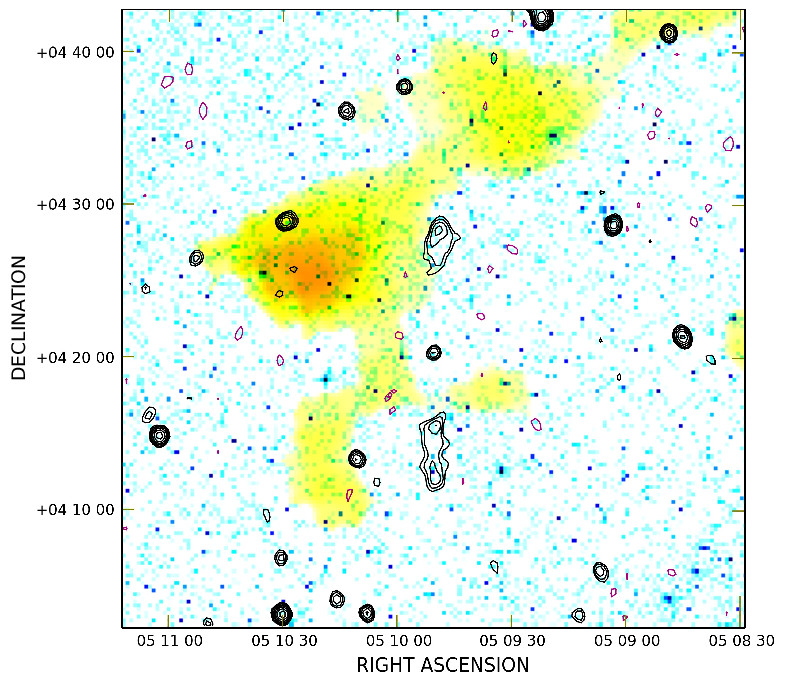}\\
\includegraphics[width=0.4\textwidth, angle=0]{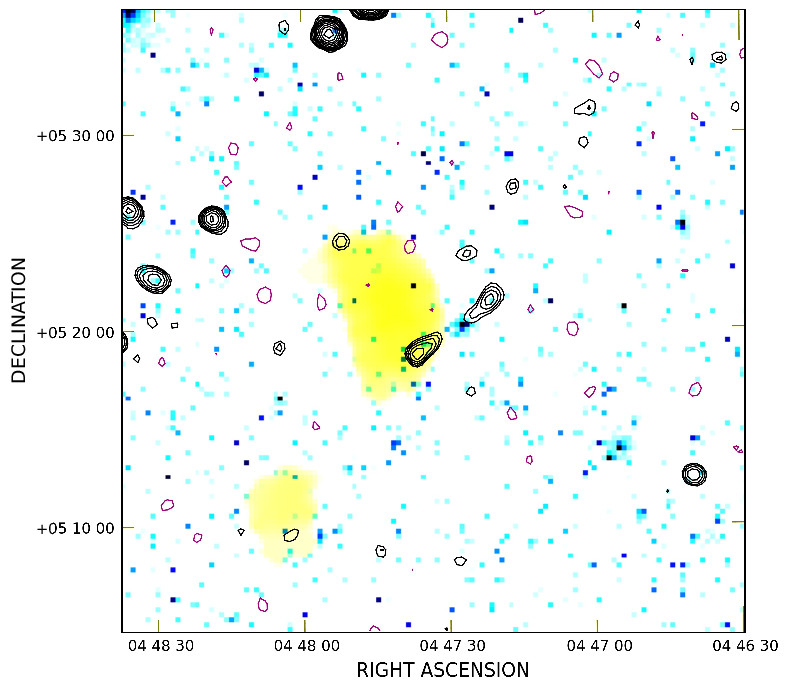}
\caption{SRT+NVSS contours overlaid on the \emph{Planck} Y-map and the optical DSS2 red image in colours (respectively yellow and blue), for the two radio galaxies (top from Region F and bottom from Region H) described in Appendix\,\ref{RRG}. The contour levels are -1.35\,mJy/beam, 1.35\,mJy/beam (with a beam of 45$^{\prime \prime}$) and the remaining increase by a factor $\sqrt{2}$ (negative in magenta and positive in black). }
\label{radio_gal}
\end{center}
 \end{figure}
\section{Other interesting sources}
\label{RRG}

In Fig.\,\ref{radio_gal}, we show the SRT+NVSS contours overlaid on the SZ
Y-map from the \emph{Planck} satellite and the optical emission from the
second Digitized Sky Survey (DSS2) red filter in the direction of two radio galaxies
we serendipitously found in Region F (Fig.\,\ref{imf}) and in Region H
(Fig.\,\ref{im9}), by inspecting the SRT+NVSS combined image.  The source in
the top panel of Fig.\,\ref{radio_gal} is a candidate giant radio galaxy. It
sits in the plane of the sky and is not present in the catalogues available in
the literature.  This radio source is located at RA (J2000) 05h09m50s and Dec
(J2000) +04$^{\circ}$20$^{\prime}$19$^{\prime\prime}$. Its flux density at 1.4\,GHz is
(46$\pm$3)\,mJy and its angular size is about 20$^{\prime}$, as measured by
the combined SRT+NVSS image. On the basis of its properties, we classify it as
a Fanaroff-Riley (FR) II radio galaxy with two bright hot spots and two
low-brightness jets.  The nucleus of this radio galaxy is clearly detected,
while the optical emission is below the noise level. Since radio galaxies are
usually giant elliptical galaxies with absolute magnitude of at least -19 and the
limiting apparent magnitude of the optical DSS2 red filter is 20.8, we derive
that the source distance is larger than 0.9\,Gpc. Considering that the angular
size has a minimum for z=1, we infer that the size of this source ranges
between 4 and 10\,Mpc. Interestingly, the SZ Y-map reveal hints of cavities
at the location of the lobes of the source.

The source in the bottom panel of Fig.\,\ref{radio_gal} is another interesting
source, located at RA (J2000) 04h47m23.9s Dec (J2000) +05d18m50s. It has a
flux density at 1.4\,GHz of (19$\pm$1)\,mJy and its angular size is about
6$^{\prime}$, as measured by the combined SRT+NVSS image. The closest systems
with optical identification have a redshift z$\approx$0.1. At this distance,
the source would have a linear size of about 700\,kpc. We classify it as
a FRII radio galaxy.

\section{Effects of gain fluctuations and Galactic foreground}
\label{fnoise}
Gain fluctuations of the receiving chain can affect the observed radio emission. To exclude that the new candidate sources are spurious patches of radio emission due to these gain fluctuations, we produced two 8$^{\circ}\times$8$^{\circ}$ RA-DEC mock SRT observations of the sky and run them through the same imaging pipeline as the observations. To reduce the computational burden, the simulations were run only in the same frequency ranges as the NVSS (1364.9-1414.9\,MHz and 1435.1-1485.1\,MHz) and then averaged.

\begin{figure}
\begin{center}
\includegraphics[width=0.4\textwidth, angle=0]{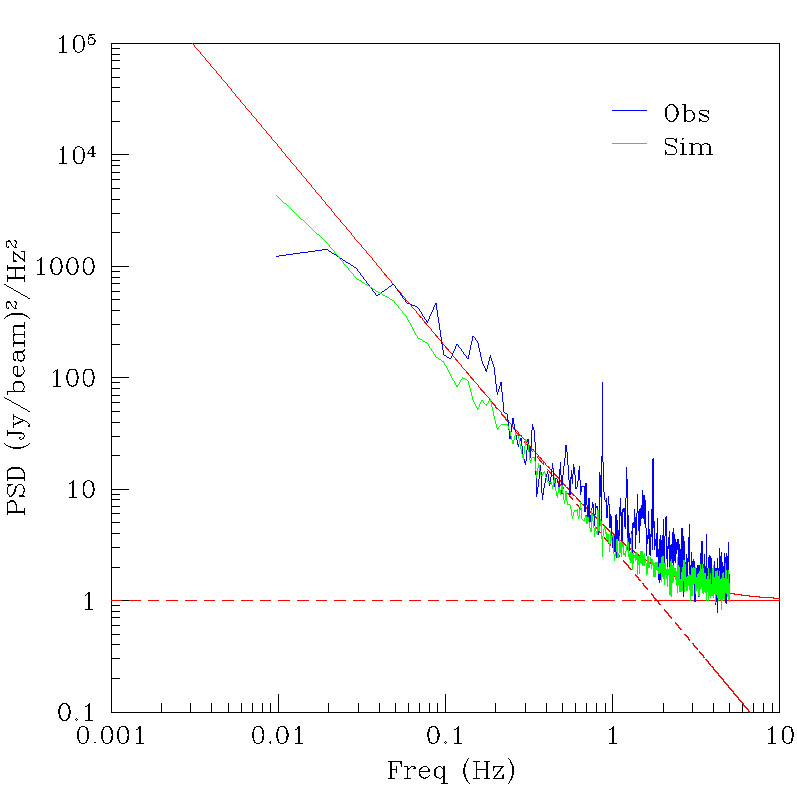}
\caption{Power spectrum of the noise derived from the observations (blue) and from the simulations (green).  The horizontal and diagonal red dashed lines represent respectively the white-noise and 1/f-noise components, while the red continuous line is the sum of white and 1/f-noise.}
\label{fnoise_fig}
\end{center}
 \end{figure}

As a first step, we characterize the noise properties of our images.
For each $8^{\circ}\times8^{\circ}$ RA-DEC observation and each polarization, we extracted 5 streams of data-points along the full sub-scan in cold regions of the sky. The total number of data points for each stream is about 850 and they have been acquired every 100\,ms corresponding to a sub-scan duration of about 85\,s. For each data stream we derived the power spectrum of the noise fluctuations in the frequency range 0.01-5\,Hz and all the power spectra were averaged, see Fig.\,\ref{fnoise_fig}. We interpreted the average observed noise spectrum as a combination of white noise with power spectrum $P_{\rm white}\simeq 1\,$(Jy/beam)$^2$/Hz$^2$ corresponding, according to the radiometer equation, to a system temperature of 33\,K, for an integration time of 100\,ms, and a bandwidth of 50\,MHz, and a 1/f-noise with power
\begin{equation}
P_{1/f}=P_{0}(f/f_{\rm 0})^{-s},
\end{equation}
where $P_{0}=3$(Jy/beam)$^2$/Hz$^2$, $f_{\rm 0}=1$\,Hz, and $s=1.8$.
In Fig.\,\ref{fnoise_fig}, we show the observed power spectrum of the noise (blue), the white-noise (horizontal red dashed) component, the f-noise (diagonal red dashed) component, and the sum of the white and 1/f noise (red continuous line).

To investigate whether these gain fluctuations can produce spurious large-scale radio emission, we produced mock observations with the same noise properties as the observations and a point-like source sky model. We assumed a system temperature $T_{\rm sys}=33$\,K, a gain $G=0.47$\,K/Jy, a forward efficiency $\eta_{\rm F}=1$, and a flat gain curve \citep[see][]{Bolli2015}. We adopted a ground temperature $T_{\rm ground}=303$\,K and a zenith opacity $\tau=0.006$, as registered at the time of our observations.
 The sky model has been obtained from the NVSS by applying a cut in brightness of 10$\sigma_{\rm NVSS}=4.5\,$mJy/beam. The image has been convolved to the SRT resolution at the NVSS frequency (13.66$^{\prime}$). 
In Fig.\,\ref{fnoise_fig}, we show for consistency the averaged power spectrum of the noise derived from the mock data in green. At high frequencies, the observed power spectral density is slightly higher than the mock one. This is probably due to the fact that the mock data do not include the confusion noise, naturally present in the observations.

\begin{figure*}
\begin{center}
\includegraphics[width=0.8\textwidth, angle=0]{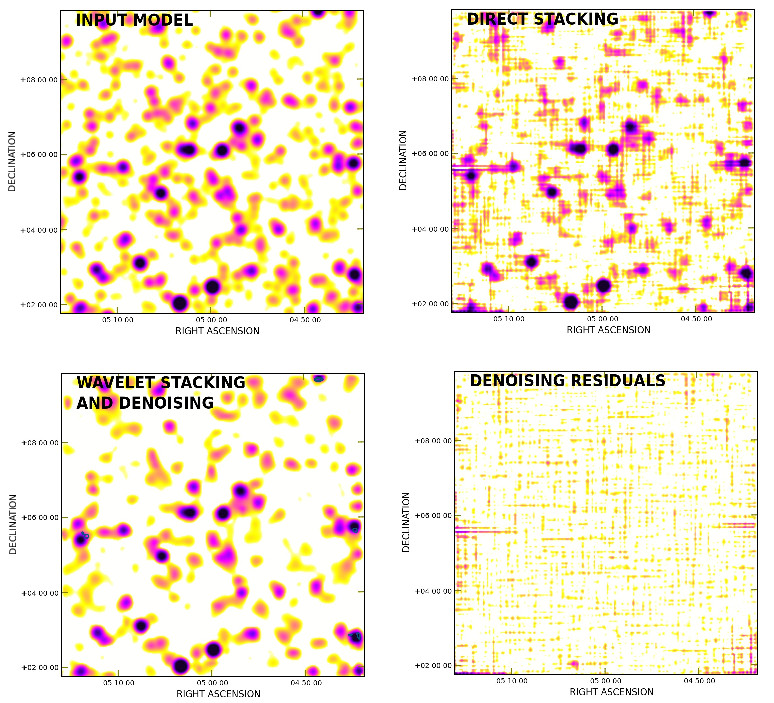}
\caption{Top-left panel: model of the sky including only the point-like sources. Top-right panel: the image derived by direct averaging all the RA and DEC  baseline subtracted scans. Bottom-left panel:  image after applying the wavelet stacking and denoising algorithms. The blue contours represent the difference between the wavelet stacking and denoising image and the input model and start from 3$\sigma$ of the first. Bottom-right panel: difference of the last two panels. }
\label{ptmodel}
\end{center}
 \end{figure*}

In Fig.\,\ref{ptmodel}, we compare the model to the images obtained after the pipeline: the model of the sky is shown in the top-left panel, the image derived by direct averaging all the RA and DEC baseline subtracted scans in the top-right panel, the image after mixing and denoising algorithms in the bottom-left panel, and the difference of the last two in the bottom-right panel. 
The top-right and bottom-left panels show that gain fluctuations are present along the sub-scan direction (with auto-correlation length of about 10\,s) but appear to be completely de-correlated between a sub-scan and those immediately before and after. After mixing and denoising, these small-scale fluctuations are strongly reduced. This analysis indicates that the new candidate sources are not the result of gain fluctuations of the receiving system.

\begin{figure*}
\begin{center}
\includegraphics[width=0.8\textwidth, angle=0]{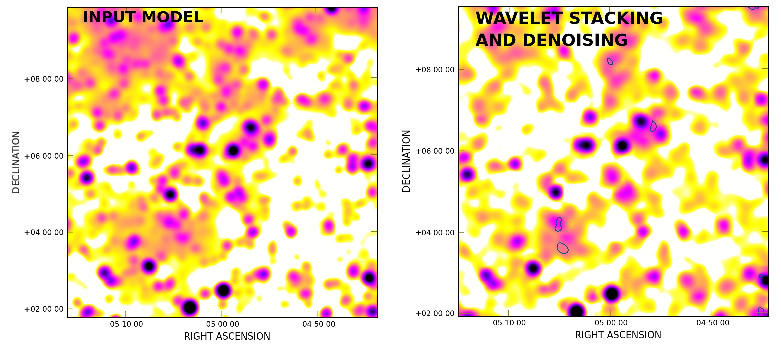}
\caption{Left panel: model of the sky including the point-like sources and a model of our Galaxy foreground. Right panel: The colours represent the radio emission obtained by applying the imaging pipeline to mock observations including a Galactic and a point-like source contribution. Blue contours show the diffuse synchrotron patches after the NVSS point-like source model has been subtracted and start from 3$\sigma$ of the  wavelet stacking and denoising image.}
\label{ptgalmodel}
\end{center}
 \end{figure*}

To investigate whether these sources could be of Galactic origin, we repeated the same procedure after including the Galactic contribution in the model. We produced the mock Galactic emission by using a power spectrum derived from the Stockert survey \citep{Reich1982} around the field of interest. The Stockert survey has a resolution of 35$^{\prime}$, coarser than the SRT. The power spectrum of the Galactic foreground estimated from the Stockert survey has been extrapolated down to the cell-size of the Stockert image (15$^{\prime}$, comparable to the SRT resolution). Moreover, it should be noted that most of the power is on large scales. The mock Galactic emission was added to the NVSS point-like source model described before.
In Fig.\,\ref{ptgalmodel}, we show on the left the model and on the right the image after using the same pipeline as for the observations in colours, while in contours the diffuse synchrotron patches after the NVSS point-like source model has been subtracted. Apart from structures at the edges of the image, five patches survive in the final image. Therefore, we conclude that only about 18-20\% of the new candidate sources we discover could be of Galactic origin.


\bsp	
\label{lastpage}
\end{document}